\documentclass[10pt,reqno,a4paper]{amsart}
\usepackage[british]{babel}
\usepackage{amssymb,amstext,amsthm,eucal,bbm,mathrsfs,amscd,bbold,pifont}
\usepackage[margin=1in]{geometry}
\usepackage{array}
\usepackage[svgnames,table]{xcolor}
\usepackage[T1]{fontenc}
\usepackage[utf8]{inputenc}
\usepackage{enumitem}
\usepackage[small]{eulervm}
\usepackage{tgpagella}
\usepackage{microtype}
\usepackage{booktabs,caption}
\usepackage{verbatim}
\usepackage{mdframed}
\usepackage{tikz,pgfplots}
\pgfplotsset{compat=1.14}
\usetikzlibrary{calc,arrows,cd}
\usepackage[unicode]{hyperref}
\hypersetup{%
  pdftitle   = {Geometry and BMS Lie algebras of spatially isotropic homogeneous spacetimes},
  pdfkeywords = {Lie algebras, kinematical, spacetime, homogeneous spaces},
  pdfauthor  = {José Figueroa-O'Farrill, Ross Grassie, Stefan Prohazka},
  pdfcreator = {\LaTeX\ with package \flqq hyperref\frqq},
  linkcolor=NavyBlue,
  citecolor=ForestGreen,
  urlcolor=OrangeRed,
  anchorcolor=OrangeRed,
  colorlinks=true
}
\theoremstyle{plain}
\newtheorem{lemma}{Lemma}

\newtheorem{theorem}[lemma]{Theorem}

\theoremstyle{definition}

\newcommand{\Kgr}{\mathcal{K}}
\newcommand{\Hgr}{\mathcal{H}}
\newcommand{\Sgr}{\mathcal{S}}
\newcommand{\Bgr}{\mathcal{B}}
\newcommand{\Sgrbar}{\overline{\mathcal{S}}}
\newcommand{\Bgrbar}{\overline{\mathcal{B}}}
\newcommand{\Khat}{\widehat{\mathcal{K}}}
\newcommand{\Hhat}{\widehat{\mathcal{H}}}

\newcommand{\h}{\mathfrak{h}}
\renewcommand{\a}{\mathfrak{a}}
\renewcommand{\b}{\mathfrak{b}}
\renewcommand{\c}{\mathfrak{c}}
\renewcommand{\d}{\partial}
\newcommand{\dbar}{\overline{\partial}}
\newcommand{\zbar}{\overline{z}}

\newcommand{\gl}{\mathfrak{gl}}

\newcommand{\e}{\mathfrak{e}}
\renewcommand{\r}{\mathfrak{r}}

\newcommand{\so}{\mathfrak{so}}

\renewcommand{\sl}{\mathfrak{sl}}

\renewcommand{\k}{\mathfrak{k}}

\newcommand{\s}{\mathfrak{s}}
\newcommand{\m}{\mathfrak{m}}
\newcommand{\eT}{\mathscr{T}}
\newcommand{\eX}{\mathscr{X}}
\renewcommand{\H}{H}
\newcommand{\x}{\boldsymbol{x}}
\newcommand{\y}{\boldsymbol{y}}
\renewcommand{\v}{\boldsymbol{v}}
\newcommand{\w}{\boldsymbol{w}}

\newcommand{\J}{\boldsymbol{J}}
\newcommand{\B}{\boldsymbol{B}}
\newcommand{\bbeta}{\boldsymbol{\beta}}
\newcommand{\bxi}{\boldsymbol{\xi}}

\newcommand{\bpi}{\boldsymbol{\pi}}
\renewcommand{\P}{\boldsymbol{P}}
\newcommand{\bzero}{\boldsymbol{0}}
\newcommand{\Pt}{\widetilde{\boldsymbol{P}}}
\newcommand{\Bt}{\widetilde{\boldsymbol{B}}}

\renewcommand{\L}{\mathcal{L}}

\newcommand{\csch}{\operatorname{csch}}
\newcommand{\sech}{\operatorname{sech}}
\newcommand{\Rad}{\operatorname{Rad}}
\newcommand{\rad}{\operatorname{rad}}
\renewcommand{\Re}{\operatorname{Re}}

\newcommand{\ad}{\operatorname{ad}}
\newcommand{\spn}[1]{\operatorname{span}\left\{#1\right\}}

\newcommand{\Pbar}{\overline{\P}}
\newcommand{\Hbar}{\overline{H}}
\newcommand{\Mbar}{\overline{M}}

\newcommand{\1}{\mathbb{1}}

\let\tensor=\otimes
\newcommand{\BB}{\mathbb{B}}
\newcommand{\PP}{\mathbb{P}}
\newcommand{\xx}{\mathbb{x}}
\newcommand{\xbar}{\overline{\mathbb{x}}}
\newcommand{\yy}{\mathbb{y}}
\newcommand{\ybar}{\overline{\mathbb{y}}}
\newcommand{\vv}{\mathbb{v}}
\newcommand{\vbar}{\overline{\mathbb{v}}}
\newcommand{\ww}{\mathbb{\beta}}
\newcommand{\wbar}{\overline{\mathbb{\beta}}}
\newcommand{\EE}{\mathbb{E}}
\newcommand{\MM}{\mathbb{M}}
\renewcommand{\SS}{\mathbb{S}}
\newcommand{\HH}{\mathbb{H}}
\newcommand{\RR}{\mathbb{R}}
\newcommand{\NN}{\mathbb{N}}
\newcommand{\ZZ}{\mathbb{Z}}

\newcommand{\CC}{\mathbb{C}}

\newcommand{\GL}{\operatorname{GL}}
\newcommand{\SO}{\operatorname{SO}}

\newcommand{\tG}{\widetilde{G}}
\newcommand{\tD}{\widetilde{D}}

\renewcommand{\div}{\operatorname{div}}
\newcommand{\Lbdl}{\mathscr{L}}
\newcommand{\mink}{S1}
\newcommand{\ds}{S2}
\newcommand{\ads}{S3}
\newcommand{\euc}{S4}
\newcommand{\sph}{S5}
\newcommand{\hyp}{S6}
\newcommand{\gal}{S7}
\newcommand{\dsg}{S8}
\newcommand{\tdsg}{S9}
\newcommand{\adsg}{S10}
\newcommand{\tadsg}{S11}
\newcommand{\twodgal}{S12}
\newcommand{\car}{S13}
\newcommand{\dsc}{S14}
\newcommand{\adsc}{S15}
\newcommand{\flc}{S16}
\newcommand{\xone}{S17}
\newcommand{\xtwo}{S18}
\newcommand{\xthree}{S19}
\newcommand{\xfour}{S20}
\newcommand{\st}{A21}
\newcommand{\tst}{A22}
\newcommand{\athree}{A23}
\newcommand{\twoda}{A24}
\newcommand{\zLC}{\mathsf{LC}}
\newcommand{\zAdSC}{\mathsf{AdSC}}
\newcommand{\zdSC}{\mathsf{dSC}}
\newcommand{\zdS}{\mathsf{dS}}
\newcommand{\zAdS}{\mathsf{AdS}}
\newcommand{\zC}{\mathsf{C}}
\newcommand{\zG}{\mathsf{G}}
\newcommand{\zS}{\mathsf{S}}
\newcommand{\zTS}{\mathsf{TS}}
\newcommand{\zAdSG}{\mathsf{AdSG}}
\newcommand{\zdSG}{\mathsf{dSG}}
\newcommand{\ztAdSG}{\mathsf{AdSG}}
\newcommand{\ztdSG}{\mathsf{dSG}}

\newcommand{\pd}{\partial}

\newcommand{\cm}{\checkmark}
\definecolor{gris}{rgb}{0.5,0.5,0.5}
\newcommand{\zero}{{\color{gris}0}}
\allowdisplaybreaks[0]
\begin{document}

\title{Geometry and BMS Lie algebras of spatially isotropic homogeneous spacetimes}

\author[Figueroa-O'Farrill]{José Figueroa-O'Farrill}
\author[Grassie]{Ross Grassie}
\author[Prohazka]{Stefan Prohazka}
\address[JMF,RG]{Maxwell Institute and School of Mathematics, The University
  of Edinburgh, James Clerk Maxwell Building, Peter Guthrie Tait Road,
  Edinburgh EH9 3FD, Scotland, United Kingdom}
\email[JMF]{\href{mailto:j.m.figueroa@ed.ac.uk}{j.m.figueroa@ed.ac.uk}}
\email[RG]{\href{mailto:s1131494@sms.ed.ac.uk}{s1131494@sms.ed.ac.uk}}
\address[SP]{Université Libre de Bruxelles and International Solvay Institutes,
  Physique Mathématique des Interactions Fondamentales, Campus Plaine
  - CP~231, B-1050 Bruxelles, Belgium, Europe}
\email[SP]{\href{mailto:stefan.prohazka@ulb.ac.be}{stefan.prohazka@ulb.ac.be}}
\begin{abstract}
  Simply-connected homogeneous spacetimes for kinematical and
  aristotelian Lie algebras (with space isotropy) have recently been
  classified in all dimensions. In this paper, we continue the study
  of these ``maximally symmetric'' spacetimes by investigating their
  local geometry. For each such spacetime and relative to exponential
  coordinates, we calculate the (infinitesimal) action of the
  kinematical symmetries, paying particular attention to the action of
  the boosts, showing in almost all cases that they act with generic
  non-compact orbits. We also calculate the soldering form, the
  associated vielbein and any invariant aristotelian, galilean or
  carrollian structures. The (conformal) symmetries of the galilean
  and carrollian structures we determine are typically
  infinite-dimensional and reminiscent of BMS Lie algebras. We also
  determine the space of invariant affine connections on each
  homogeneous spacetime and work out their torsion and curvature.
\end{abstract}
\dedicatory{Amelie Prohazka gewidmet}
\thanks{EMPG-18-14}
\maketitle
\tableofcontents

\section{Introduction}
\label{sec:introduction}

Half a century ago, Bacry and Lévy-Leblond \cite{Bacry:1968zf} asked
what were the possible kinematics. They provided an answer to this
question by classifying kinematical Lie algebras in $3+1$ dimensions
subject to the assumptions of invariance under parity and
time-reversal. They also showed that the kinematical Lie algebras in
their classification could be related by contractions. Moreover they
observed that each such Lie algebra acts transitively on some
($3+1$)-dimensional spatially isotropic homogeneous spacetime and that
the contractions could be interpreted as geometric limits of the
corresponding spacetimes. Physically, we can understand these limits
as approximations and this interpretation explains why these
particular spacetimes are relevant and continue to show up in
different corners of physics.

Indeed, most of the spacetimes in their work are known to play a
fundamental rôle in physics.  For example, the de Sitter spacetime is
important for cosmology, the anti de Sitter spacetime currently drives
much of our understanding of quantum gravity due to the AdS/CFT
correspondence~\cite{Maldacena:1997re}, and, in the limit 
where the cosmological constant goes to zero, Minkowski spacetime is
fundamental in particle physics.  Other important spacetimes of this
type include the galilean spacetime, which is the playing field for
condensed matter systems, and the carrollian spacetime, whose
relation to Bondi--Metzner--Sachs (BMS) symmetries, as shown
in~\cite{Duval:2014uva}, is leading to exciting progress in our
understanding of infrared physics in asymptotically flat spaces (for
reviews see~\cite{Strominger:2017zoo,Ashtekar:2018lor}).\footnote{%
  We refer to, e.g.,~\cite{Figueroa-OFarrill:2018ilb} for further
  motivation and a (non-exhaustive) list of further references. While
  this work was under completion the interesting
  work~\cite{Morand:2018tke} appeared which discusses similar aspects
  as this and our earlier work.  Recently, also further interesting
  works, which fall in the realm of the kinematical Lie algebras and
  spacetimes, have appeared, see, e.g.,~\cite{Parsa:2018kys,Harmark:2018cdl,
    Batlle:2018hoe, Campoleoni:2018ltl, Bagchi:2019xfx,Safari:2019zmc,
    Gibbons:2019zfs,Ozdemir:2019orp,Hansen:2019vqf,Bergshoeff:2019ctr}.}

Twenty years later, Bacry and Nuyts \cite{MR857383} dropped the ``by
no means compelling'' assumptions of parity and time-reversal
invariance and hence classified all kinematical Lie algebras in $3+1$
dimensions, observing that once again each such Lie algebra acts
transitively on some ($3+1$)-dimensional homogeneous spacetime.

Strictly speaking, what was shown in \cite{Bacry:1968zf, MR857383} is
that every kinematical Lie algebra $\k$ in their classification has a
Lie subalgebra $\h$ spanned by the infinitesimal generators of
rotations and boosts. This \emph{suggests} the existence of Lie groups
$\Hgr \subset \Kgr$ with Lie algebras $\h \subset \k$ and hence of a
homogeneous spacetime $\Kgr/\Hgr$. However the very existence of the
homogeneous spacetime and its precise relationship to the
infinitesimal description in terms of the Lie pair $(\k,\h)$ turns out
to be subtle. Furthermore, as mentioned already in \cite{Bacry:1968zf,
  MR857383}, a physically desirable property of a kinematical
spacetime is that orbits of the boost generators should be
non-compact. To the best of our knowledge, a proof of this fact did
not exist for many of the spacetimes in \cite{MR857383}. With this in
mind, and based on a recent deformation-theoretic classification of
kinematical Lie algebras \cite{Figueroa-OFarrill:2017ycu,
  Figueroa-OFarrill:2017tcy, Andrzejewski:2018gmz}, we revisited this
problem and in \cite{Figueroa-OFarrill:2018ilb} classified and showed
the existence of simply-connected spatially isotropic homogeneous
spacetimes in arbitrary dimension, making \emph{en passant} a small
correction to the ($3+1$)-dimensional classification in
\cite{MR857383}. Another novel aspect of
\cite{Figueroa-OFarrill:2018ilb} was the classification of
aristotelian spacetimes, which lack boost symmetry.  One way to
interpret this classification is as a generalisation of the
classification of maximally symmetric riemannian and lorentzian
spacetimes when we drop the requirement that there should exist an
invariant metric.

Another way is to understand this work as a generalisation of the work
of Bacry and Lévy-Leblond~\cite{Bacry:1968zf} when the assumption of
parity and time reversal invariance and the restriction to $3+1$
dimensions is dropped.  Simultaneously imposing parity and time
reversal invariance\footnote{This operation is $\sigma(H)=-H$ and
  $\sigma(\P)=-\P$ leaving the remaining generators unaltered.}
selects the symmetric spaces, leading to the omission of some
interesting spacetimes like, e.g., the non-reductive carrollian light
cone~$\hyperlink{S16}{\zLC}$ and the torsional galilean spacetimes. 

Let us emphasise that in identifying specific Lie algebra generators
as ``translations'' or ``boosts'' one is actually implicitly referring
to the homogeneous space.  Indeed, the Lie algebra itself does not
provide this interpretation.  For example, by inspecting Table
\ref{tab:spacetimes} one recognises that the Minkowski
($\hyperlink{S1}{\MM}$) and AdS carrollian ($\hyperlink{S15}{\zAdSC}$)
spacetimes share the same underlying Lie algebra. They are however
different homogeneous spacetimes and the precise relationship between
the kinematical Lie algebras and their spacetimes was also analysed in
\cite{Figueroa-OFarrill:2018ilb} and will be seen explicitly in the
following analysis.

The methods employed in \cite{Figueroa-OFarrill:2018ilb} are Lie
algebraic and this means that in that paper we concentrated on
geometrical properties which could be probed infinitesimally, such as
determining the characteristic invariant structures (in low rank) that
such a spacetime might possess, leaving the investigation of the
orbits of the boosts to the present paper. Indeed, we will prove that
the boosts do act with (generic) non-compact orbits in all spacetimes
with the unsurprising exceptions of the aristotelian spacetimes (which
have no boosts) and the riemannian symmetric spaces, where the
``boosts'' are actually rotations.\footnote{%
  Since some of these spacetimes are well studied, there is
  necessarily some overlap with existing work, like the original works
  \cite{Bacry:1968zf, MR857383} or more recent works that also discuss
  homogeneous spacetimes, e.g., \cite{SchaferNameki:2009xr,
    Jottar:2010vp, Bagchi:2010xw, Duval:2012qr,
    Bergshoeff:2015wma,Bekaert:2014bwa,
    Grosvenor:2017dfs,Morand:2018tke}.}

To those ends we introduce exponential coordinates for each of the
spacetimes in \cite{Figueroa-OFarrill:2018ilb}, relative to which we
write down the fundamental vector fields which generate the action of
the transitive Lie algebra. We also give explicit expressions for the
invariant structure (lorentzian, galilean, carrollian, aristotelian)
that the spacetime may possess. In addition, we determine the
invariant connections which the homogeneous spacetimes admit (if any)
and determine their torsion and curvature. We also pay particularly
close attention to the orbits of the boost generators and in most
cases show that the generic orbit is non-compact, as one would expect
to be the case for any reasonable spacetime.

Finally, using modified exponential coordinates, we determine the
infinitesimal (conformal) symmetries of the galilean and carrollian
structures of our spacetimes.  They are infinite-dimensional and
reminiscent of BMS algebras. Many of the results already appear in
\cite{Duval:2014uva,Duval:2014lpa}. Unobserved however was the close
relation of the conformal symmetries of the (anti) de Sitter
carrollian structure, belonging to null surfaces of (anti) de Sitter
spacetime, and BMS symmetries. Section \ref{sec:symm-spac-struct} can
be read in large parts independently.

The paper is organised as follows. In
Section~\ref{sec:homog-kinem-spac} we summarise the results of the
classification in~\cite{Figueroa-OFarrill:2018ilb}. In
Tables~\ref{tab:spacetimes} and~\ref{tab:aristotelian} we list the
simply-connected, spatially isotropic, homogeneous kinematical and
aristotelian spacetimes, respectively.  These are the spacetimes whose
geometry we study in this paper.  Figures~\ref{fig:generic-d-graph},
\ref{fig:d=3-graph}, and \ref{fig:d=2-graph} summarise the
relationships between these spacetimes. These relationships take the
form of limits which, in many cases, manifest themselves as
contractions of the corresponding kinematical Lie
algebras. Table~\ref{tab:spacetimes-props} summarises some of the
geometrical properties of the spacetimes in
Tables~\ref{tab:spacetimes} and \ref{tab:aristotelian}. The list of
spacetimes naturally breaks up into classes depending on which
invariant structures (if any) the spacetimes possess: lorentzian,
riemannian, galilean, carrollian and aristotelian. There are also
exotic two-dimensional spacetimes with no discernible invariant
structure. In Section~\ref{sec:geom-prop-homog} we briefly review the
basic notions of the local geometry of homogeneous spaces, tailored to
the case at hand and compute the action of the rotations and boosts on
the spacetimes. In Section~\ref{sec:nomizu} we discuss the space of
invariant connections for the reductive homogeneous spacetimes in
Tables~\ref{tab:spacetimes} and \ref{tab:aristotelian} and calculate
their torsion and curvature, paying particular attention to the
existence of flat and/or torsion-free connections. In
Section~\ref{sec:metric} we discuss the lorentzian and riemannian
homogeneous spaces and their limits. This leaves a few spacetimes
which are not obviously obtained in this way and we discuss them
separately: the torsional galilean homogeneous spacetimes are
discussed in Section~\ref{sec:galilean}, the carrollian light cone in
Section~\ref{sec:spacetime-flc}, the exotic two-dimensional spacetimes
in Section~\ref{sec:exotic}, and the aristotelian spacetimes in
Section~\ref{sec:aristotelian}. In Section~\ref{sec:symm-spac-struct}
we determine the infinitesimal (resp.\ conformal) symmetries of the
galilean and carrollian spacetimes; namely, the vector fields which
preserve (resp.\ rescale) the corresponding galilean and carrollian
structure.  The corresponding Lie algebras are typically
infinite-dimensional and reminiscent of the BMS algebras.  Finally, in
Section~\ref{sec:conclusions} we offer some conclusions. The paper
contains two appendices: In Appendix~\ref{app:modexp} we discuss the
carrollian and galilean spacetimes in terms of modified exponential
coordinates, which are the most convenient coordinates in order to
discuss their symmetries, and in Appendix~\ref{sec:conf-kill-vect} we
record for convenience the Lie algebras of conformal Killing vectors
on low-dimensional maximally symmetric riemannian manifolds.

\section{Homogeneous kinematical spacetimes}
\label{sec:homog-kinem-spac}

We use the notation of \cite{Figueroa-OFarrill:2018ilb}, which we now
review. Recall that a simply-connected homogeneous kinematical
spacetime is described infinitesimally by a Lie pair $(\k,\h)$. Here
$\k$ is a kinematical Lie algebra with $D$-dimensional space isotropy:
namely, a real $\tfrac{(D+2)(D+1)}{2}$-dimensional Lie algebra with
generators $J_{ab}$, $1\leq a<b\leq D$, spanning a Lie subalgebra
isomorphic to $\so(D)$, $B_a$ and $P_a$, for $1\leq a \leq D$,
transforming as vectors of $\so(D)$ and $H$ transforming as a scalar.
The Lie subalgebra $\h$ of $\k$ contains $\so(D)$ and an
$\so(D)$-vector representation, which is spanned by
$\alpha B_a + \beta P_a$, $1\leq a\leq D$, for some non-zero
$\alpha,\beta \in\RR$. We choose a basis for $\k$ such that $\h$ is
\emph{always} spanned by $J_{ab}$ and $B_a$. In this fashion, the Lie
brackets of $\k$ uniquely specify the Lie pair $(\k,\h)$.

Let us make a notational remark: we will refer to the generators $B_a$
as (infinitesimal) \emph{boosts}, even though in some cases (e.g., the
riemannian symmetric spaces) they act as rotations. A substantial part
of the work that went into this paper was devoted to determining when
the boosts really act like boosts and not, say, like rotations.

Notice that in writing down the Lie brackets of $\k$, it is only
necessary to list those brackets which do not involve $J_{ab}$ since
those involving $J_{ab}$ are common for all kinematical Lie algebras
and restate the fact that $J_{ab}$ span an $\so(D)$ subalgebra under
which $B_a$ and $P_a$ are vectors and $H$ is a scalar. Explicitly,
this reads \begin{equation}
  \label{eq:defkinbr}
  \begin{split}
    [J_{ab}, J_{cd}] &= \delta_{bc} J_{ad} - \delta_{ac} J_{bd} - \delta_{bd} J_{ac} + \delta_{ad} J_{bc},\\
    [J_{ab}, B_c] &= \delta_{bc} B_a - \delta_{ac} B_b,\\
    [J_{ab}, P_c] &= \delta_{bc} P_a - \delta_{ac} P_b,\\
    [J_{ab}, H ] &= 0;
  \end{split}
\end{equation}
although we will use an abbreviated notation in which we do not write
the $\so(D)$ indices explicitly. We write $\J$, $\B$, $\P$, and $H$
for the generators and rewrite the kinematical Lie brackets in
\eqref{eq:defkinbr} as
\begin{equation}
  \label{eq:kin}
  [\J,\J] = \J, \qquad [\J,\B] = \B, \qquad [\J, \P] = \P,
  \qquad\text{and}\qquad [\J, H] = 0.
\end{equation}
For $D\neq 2$, any other brackets can be reconstructed unambiguously
from the abbreviated expression since there is only one way to
reintroduce indices in an $\so(D)$-equivariant fashion.  For example,
\begin{equation}
  [H, \B] = \P \implies [H, B_a] = P_a \qquad\text{and}\qquad
  [\B,\P] = H + \J \implies [B_a, P_b] = \delta_{ab} H + J_{ab}.
\end{equation}
In $D=3$ we may also have brackets of the form
\begin{equation}
 [\P,\P] = \P \implies [P_a,P_b] = \epsilon_{abc} P_c.
\end{equation}
Similarly, for $D=2$, $\epsilon_{ab}$ is rotationally invariant and can
appear in Lie brackets.  So we will write, e.g.,
\begin{equation}
  [H, \B] = \B + \Pt  \qquad\text{for}\qquad  [H, B_a] = B_a +
  \epsilon_{ab} P_b.
\end{equation}

If the Lie subalgebra $\h$ contains an ideal $\b$ of $\k$, we say that
the Lie pair $(\k,\h)$ is not effective.  For a kinematical Lie
algebra $\k$, such an ideal is necessarily the one spanned by the
boosts, which act trivially on the homogeneous spacetime.  In such
cases, we quotient by $\b$ to arrive at an effective (by construction)
Lie pair $(\a,\r)$, where $\a = \k/\b$ is an aristotelian Lie algebra
and $\r \cong \so(D)$ is the Lie subalgebra of rotations.  The Lie
pair $(\a,\r)$ corresponds to an aristotelian spacetime.  Not all
aristotelian spacetimes arise in this way, and this justifies the
separate classification of aristotelian Lie algebras and their
corresponding spacetimes in \cite[App.~A]{Figueroa-OFarrill:2018ilb}.

\subsection{Classification}
\label{sec:classification}

We now summarise the results of \cite{Figueroa-OFarrill:2018ilb}.
Table~\ref{tab:spacetimes} lists the (isomorphism classes of)
simply-connected, spatially isotropic, homogeneous spacetimes.  We
shall refer to them as ``simply-connected homogeneous kinematical
spacetimes'' from now on.  These are described by Lie pairs $(\k,\h)$,
where $\k$ is a kinematical Lie algebra with generators $\J,\B,\P, H$
and $\h$ is the Lie subalgebra spanned by $\J,\B$.  The first column
is the label given in \cite{Figueroa-OFarrill:2018ilb}.  The second
column specifies the value of $D$, where the dimension of the
spacetime is $D+1$.  The middle columns are the Lie brackets of $\k$
in addition to the common kinematical Lie brackets in
equation~\eqref{eq:kin}.  It is tacitly assumed that when $D=1$, we
set $\J=0$ whenever it appears.  The final column contains any
relevant comments, including the name of the spacetime if known.  The
table is divided by horizontal rules into five sections, from top to
bottom:
\begin{description}
\item[Lorentzian] These are the homogeneous kinematical spacetimes
  admitting an invariant lorentzian metric, which due to the dimension
  of the symmetry algebra must be maximally symmetric:
  \begin{itemize}
  \item Minkowski spacetime ($\hyperlink{S1}{\MM}$),
  \item de~Sitter spacetime ($\hyperlink{S2}{\zdS}$), and
  \item anti de~Sitter spacetime ($\hyperlink{S3}{\zAdS}$).
  \end{itemize}
\item[Riemannian] These are the homogeneous kinematical ``spacetimes''
  admitting an invariant riemannian metric, which again must be
  maximally symmetric by dimension:
  \begin{itemize}
  \item euclidean space ($\hyperlink{S4}{\EE}$),
  \item round sphere ($\hyperlink{S5}{\SS}$), and
  \item hyperbolic space ($\hyperlink{S6}{\HH}$).
  \end{itemize}
\item[Galilean] These are the homogeneous kinematical spacetimes
  admitting an invariant galilean structure:
  \begin{itemize}
  \item galilean spacetime ($\hyperlink{S7}{\zG}$),
  \item galilean de~Sitter spacetime ($\hyperlink{S8}{\zdSG} = \ztdSG_{-1}$), 
  \item torsional galilean de~Sitter spacetime ($\hyperlink{S9}{\ztdSG_\gamma}$,
    $\gamma \in (-1,1]$),
  \item galilean anti de~Sitter spacetime ($\hyperlink{S10}{\zAdSG} = \ztAdSG_0$), and 
  \item torsional galilean anti de~Sitter spacetime ($\hyperlink{S11}{\ztAdSG_\chi}$,
    $\chi> 0$),
  \item a two-parameter family
    (\hyperlink{S12}{$\text{\twodgal}_{\gamma,\chi}$}) of three-dimensional
    galilean spacetimes interpolating between the torsional galilean (anti)
    de~Sitter spacetimes.
  \end{itemize}
\item[Carrollian] These are the homogeneous kinematical spacetimes
  admitting an invariant carrollian structure:
  \begin{itemize}
  \item carrollian spacetime ($\hyperlink{S13}{\zC}$),
  \item carrollian de~Sitter spacetime ($\hyperlink{S14}{\zdSC}$),
  \item carrollian anti de~Sitter spacetime ($\hyperlink{S15}{\zAdSC}$), and
  \item carrollian light cone ($\hyperlink{S16}{\zLC}$).
  \end{itemize}
  These spacetimes are identifiable as null hypersurfaces in
  homogeneous, lorentzian kinematical spacetimes in one dimension
  higher: $\hyperlink{S1}{\MM}$ for $\hyperlink{S13}{\zC}$ and
  $\hyperlink{S16}{\zLC}$, $\hyperlink{S3}{\zAdS}$ for
  $\hyperlink{S15}{\zAdSC}$ and $\hyperlink{S2}{\zdS}$
  for $\hyperlink{S14}{\zdSC}$.  In particular, the image of the embedding $\zLC
  \subset \MM$ is the future light cone.\footnote{Strictly speaking, it
    is the future light cone if $D>1$ and its universal cover if
    $D=1$, a fact that was initially glossed over in
    \cite{Figueroa-OFarrill:2018ilb}.}
\item[Exotic] These are two-dimensional kinematical spacetimes
  without any discernible invariant structures.
\end{description}

Since, in two dimensions, it is largely a matter of convention what
one calls space and time \footnote{%
  While true when discussing the geometry of homogeneous spacetimes,
  there is of course a physical distinction between space and time:
  time translations are generated by the hamiltonian, whose spectrum
  one often requires to be bounded from below, whereas the spectrum of
  spatial translations is not subject to such a requirement.}, some of
the spacetimes become accidentally pairwise isomorphic when $D=1$:
namely, $\zC \cong \zG$, $\zdS \cong \zAdS$, $\zdSC \cong \zAdSG$ and
$\zAdSC \cong \zdSG$.  In order to arrive at a one-to-one
correspondence between the rows of the table and the isomorphism class
of simply-connected homogeneous spacetimes, we write $D\geq2$ for
$\hyperlink{S2}{\zdS}$, $\hyperlink{S13}{\zC}$,
$\hyperlink{S14}{\zdSC}$ and $\hyperlink{S15}{\zAdSC}$.

\begin{table}[h!]
  \centering
  \caption{Simply-connected homogeneous ($D+1$)-dimensional kinematical spacetimes}
  \label{tab:spacetimes}
  \rowcolors{2}{blue!10}{white}
  \resizebox{\textwidth}{!}{
    \begin{tabular}{l|>{$}c<{$}|*{5}{>{$}l<{$}}|l}\toprule
      \multicolumn{1}{c|}{Label} & D & \multicolumn{5}{c|}{Non-zero Lie brackets in addition to $[\J,\J] = \J$, $[\J, \B] = \B$, $[\J,\P] = \P$} & \multicolumn{1}{c}{Comments}\\\midrule
      \hypertarget{S1}{\mink} & \geq 1 & [H,\B] = -\P & & [\B,\B] = \J & [\B,\P] = H & & $\MM$\\
      \hypertarget{S2}{\ds} & \geq 2 & [H,\B] = -\P & [H,\P] = -\B & [\B,\B]= \J & [\B,\P] = H & [\P,\P]= - \J & $\zdS$\\
      \hypertarget{S3}{\ads} & \geq 1 & [H,\B] = -\P & [H,\P] = \B & [\B,\B]= \J & [\B,\P] = H & [\P,\P] = \J & $\zAdS$\\\midrule
      \hypertarget{S4}{\euc} & \geq 1 & [H,\B] = \P & & [\B,\B] = -\J & [\B,\P] = H & & $\EE$\\
      \hypertarget{S5}{\sph} & \geq 1 & [H,\B] = \P & [H,\P] = -\B & [\B,\B]= -\J & [\B,\P] = H & [\P,\P]= - \J & $\SS$\\
      \hypertarget{S6}{\hyp} & \geq 1 & [H,\B] = \P & [H,\P] = \B & [\B,\B]= -\J & [\B,\P] = H & [\P,\P] = \J & $\HH$\\\midrule
      \hypertarget{S7}{\gal} & \geq 1 & [H,\B] = -\P & & & & & $\zG$\\
      \hypertarget{S8}{\dsg} & \geq 1 & [H,\B] = -\P & [H,\P] = -\B & & & & $\zdSG = \ztdSG_{\gamma=-1}$\\
      \hypertarget{S9}{\tdsg$_\gamma$} & \geq 1 & [H,\B] = -\P & [H,\P] = \gamma\B + (1+\gamma)\P & & & & $\ztdSG_{\gamma\in (-1,1]}$\\
      \hypertarget{S10}{\adsg} & \geq 1 & [H,\B] =  -\P & [H,\P] = \B & & & & $\zAdSG = \ztAdSG_{\chi=0}$\\
      \hypertarget{S11}{\tadsg$_\chi$} & \geq 1 & [H,\B] = -\P & [H,\P] = (1+\chi^2) \B + 2\chi \P & & & & $\ztAdSG_{\chi>0}$ \\
      \hypertarget{S12}{$\text{\twodgal}_{\gamma,\chi}$} & 2 & [H,\B] = -\P & [H,\P] = (1+\gamma) \P - \chi\Pt + \gamma \B - \chi \Bt & & & &  $\gamma\in [-1,1), \chi >0$\\\midrule
      \hypertarget{S13}{\car} & \geq 2 & & & & [\B,\P] = H & & $\zC$\\
      \hypertarget{S14}{\dsc} & \geq 2 & & [H,\P] = -\B & & [\B,\P] = H & [\P,\P] = -\J & $\zdSC$ \\
      \hypertarget{S15}{\adsc} & \geq 2 & & [H,\P] = \B & & [\B,\P] = H & [\P,\P] = \J & $\zAdSC$ \\
      \hypertarget{S16}{\flc} & \geq 1 & [H,\B] = \B & [H,\P] = -\P & & [\B,\P] = H + \J & & $\zLC$ \\\midrule
      \hypertarget{S17}{\xone} & 1 & [H,B] = -P & & & [B,P] = - H - 2 P & & \\
      \hypertarget{S18}{\xtwo} & 1 & [H,B] = H & & & [B,P] = -P & & \\
      \hypertarget{S19}{\xthree$_\chi$} & 1 & [H,B] = (1+\chi) H & & & [B,P]= (1-\chi)P & & $\chi>0$\\
      \hypertarget{S20}{\xfour$_\chi$} & 1 & [H,B] = -P & & & [B,P] = -(1+\chi^2) H - 2\chi P & & $\chi>0$ \\ \bottomrule
    \end{tabular}
  }
  \caption*{The horizontal rules separate the lorentzian, riemannian,
    galilean, carrollian and exotic spacetimes. For further properties
    see Table \ref{tab:spacetimes-props}.}
\end{table}

Table~\ref{tab:aristotelian} lists the isomorphism classes of
simply-connected aristotelian spacetimes.  Homogeneous aristotelian
spacetimes are always reductive, and they admit simultaneously
invariant galilean and carrollian structures.  We label them as A\# as
opposed to S\#, for mnemonic reasons:
\begin{itemize}
\item A21 is the static aristotelian spacetime
  (\hyperlink{A21}{$\zS$}),
\item A22 is the torsional static aristotelian spacetime
  (\hyperlink{A22}{$\zTS$}),
\item A23$_\varepsilon$ are the Einstein static spacetime
  \hyperlink{A23p}{$\RR \times \SS^D$} for $\varepsilon=+1$ and the
  hyperbolic version \hyperlink{A23m}{$\RR \times \HH^D$} for
  $\varepsilon=-1$, and
\item \hyperlink{A24}{A24} is a three-dimensional static spacetime
  with underlying manifold the Heisenberg Lie group.
\end{itemize}

\begin{table}[h!]
  \centering
  \caption{Simply-connected homogeneous ($D+1$)-dimensional aristotelian spacetimes}
  \label{tab:aristotelian}
  \rowcolors{2}{blue!10}{white}
  \begin{tabular}{l|>{$}c<{$}|*{2}{>{$}l<{$}}|l}\toprule
    \multicolumn{1}{c|}{Label} & D & \multicolumn{2}{c|}{Non-zero Lie brackets in addition to $[\J,\J] = \J $ and $[\J,\P] = \P$}& \multicolumn{1}{c}{Comments}\\\midrule
    \hypertarget{A21}{\st} & \geq 0 & & & $\zS$\\
    \hypertarget{A22}{\tst} & \geq 1 & [H,\P] = \P & & $\zTS$\\
    \hypertarget{A23p}{\athree$_{+1}$} & \geq 2 & & [\P,\P] = - \J & $\RR \times \SS^D$\\
    \hypertarget{A23m}{\athree$_{-1}$} & \geq 2 & & [\P,\P] = \J & $\RR \times \HH^D$\\
    \hypertarget{A24}{\twoda} & 2 & & [\P,\P] = H & \\\bottomrule
  \end{tabular}
  \caption*{For further properties see Table \ref{tab:spacetimes-props}.}
\end{table}

\subsection{Geometric limits}
\label{sec:geometric-limits}

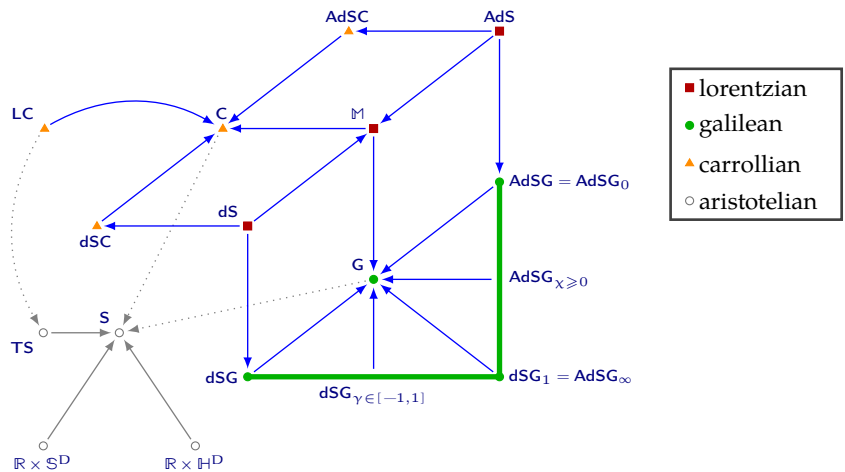
\begin{figure}[h!]
  \centering
  \begin{tikzpicture}[scale=1,>=latex, shorten >=3pt, shorten <=3pt, x=1.0cm,y=1.0cm]
    %
    %
    %
    %
    \coordinate [label=above left:{\hyperlink{S2}{\tiny $\zdS$}}] (ds) at (5.688048519056286,-0.5838170592960186);
    \coordinate [label=left:{\hyperlink{S8}{\tiny $\zdSG$}}] (dsg) at (5.688048519056286, -2.5838170592960186);
    \coordinate [label=right:{\hyperlink{S9}{\tiny $\ztdSG_1 =\ztAdSG_\infty$}}] (dsgone) at (9, -2.5838170592960186);
    \coordinate [label=below:{\hyperlink{S14}{\tiny $\zdSC$}}] (dsc) at  (3.688048519056286,-0.5838170592960186);
    \coordinate [label=above:{\hyperlink{S13}{\tiny $\zC$}}] (c) at (5.344024259528143, 0.7080914703519907);
    \coordinate [label=above left:{\hyperlink{S1}{\tiny $\MM$}}] (m) at (7.344024259528143,0.7080914703519907);
    \coordinate [label=above left:{\hyperlink{S7}{\tiny $\zG$}}] (g) at  (7.344024259528143, -1.2919085296480093);
    \coordinate [label=above:{\hyperlink{S3}{\tiny $\zAdS$}}] (ads) at (9,2);
    \coordinate [label=right:{\hyperlink{S10}{\tiny $\zAdSG =\ztAdSG_0$}}] (adsg) at (9,0);
    \coordinate [label=above:{\hyperlink{S15}{\tiny $\zAdSC$}}]  (adsc) at (7,2);
    \coordinate [label=above left:{\hyperlink{A21}{\tiny $\zS$}}] (s) at (4, -2);
    \coordinate [label=above left:{\hyperlink{S16}{\tiny $\zLC$}}]  (flc) at (3, 0.7080914703519907);
    \coordinate [label=below left:{\hyperlink{A22}{\tiny $\zTS$}}]  (ts) at (3, -2); 
    \coordinate [label=below:{\hyperlink{A23p}{\tiny $\RR\times\SS^D$}}]  (esu) at (3, -3.5); 
    \coordinate [label=below:{\hyperlink{A23m}{\tiny $\RR\times\HH^D$}}]  (hesu) at (5, -3.5); 
    %
    %
    \coordinate [label=below:{\hyperlink{S9}{\tiny $\ztdSG_{\gamma\in[-1,1]}$}}] (tdsg) at (7.344024259528143, -2.5838170592960186);
    \coordinate [label=right:{\hyperlink{S11}{\tiny $\ztAdSG_{\chi\geq0}$}}] (tadsg) at (9, -1.2919085296480093);
    %
    %
    \draw [->,line width=0.5pt,dotted,color=gray] (c) -- (s);
    \draw [->,line width=0.5pt,dotted,color=gray] (g) -- (s);
    \draw [->,line width=0.5pt,color=blue] (dsgone) -- (g);
    \draw [->,line width=0.5pt,color=blue] (7.344024259528143,-2.5838170592960186) -- (g);
    \draw [->,line width=0.5pt,color=blue] (9, -1.2919085296480093) -- (g);
    %
    %
    \draw [->,line width=0.5pt,color=blue] (adsc) -- (c);
    \draw [->,line width=0.5pt,color=blue] (dsc) -- (c);
    \draw [->,line width=0.5pt,color=blue] (ads) -- (m);
    \draw [->,line width=0.5pt,color=blue] (adsg) -- (g);
    \draw [->,line width=0.5pt,color=blue] (dsg) -- (g);
    \draw [->,line width=0.5pt,color=blue] (ds) -- (m);
    \draw [->,line width=0.5pt,color=blue] (ds) -- (dsc);
    \draw [->,line width=0.5pt,color=blue] (ds) -- (dsg);
    \draw [->,line width=0.5pt,color=blue] (m) -- (c);
    \draw [->,line width=0.5pt,color=blue] (m) -- (g);
    \draw [->,line width=0.5pt,color=blue] (ads) -- (adsc);
    \draw [->,line width=0.5pt,color=blue] (ads) -- (adsg);
    %
    %
    \draw [->,line width=0.5pt,color=blue] (flc) to [out=30,in=150] (c);
    \draw [->,line width=0.5pt,dotted,color=gray] (flc) to [out=240,in=120] (ts);
    \draw [->,line width=0.5pt,color=gray] (ts) to (s); 
    \draw [->,line width=0.5pt,color=gray] (esu) to (s); 
    \draw [->,line width=0.5pt,color=gray] (hesu) to (s); 
    %
    %
    \begin{scope}[>=latex, shorten >=0pt, shorten <=0pt, line width=2pt, color=green!70!black]
      \draw (adsg) --(dsgone);
      \draw (dsg) -- (dsgone);
    \end{scope}
    \foreach \point in {g,adsg,dsg,dsgone}
    \filldraw [color=green!70!black,fill=green!70!black] (\point) circle (1.5pt);
    \foreach \point in {ads,ds,m}
    \filldraw [color=red!70!black,fill=red!70!black] (\point) ++(-1.5pt,-1.5pt) rectangle ++(3pt,3pt);
    \foreach \point in {adsc,dsc,flc,c}
    \filldraw [color=DarkOrange,fill=DarkOrange] (\point) ++(-1pt,-1pt) -- ++(3pt,0pt) -- ++(-1.5pt,2.6pt) -- cycle;
    \foreach \point in {s,ts,esu,hesu}
    \draw [color=gray!90!black] (\point) circle (1.5pt);
    %
    %
    \begin{scope}[xshift=0.5cm]
    \draw [line width=1pt,color=gray!50!black] (10.75,-0.5) rectangle (13,1.5);
    \filldraw [color=red!70!black,fill=red!70!black] (11,1.25) ++(-1.5pt,-1.5pt) rectangle ++(3pt,3pt) ; 
    \draw (11,1.25) node[color=black,anchor=west] {\small lorentzian}; 
    \filldraw [color=green!70!black,fill=green!70!black] (11,0.75) circle (1.5pt) node[color=black,anchor=west] {\small galilean};
    \filldraw [color=DarkOrange,fill=DarkOrange] (11,0.25) ++(-1.5pt,-1pt) -- ++(3pt,0pt) -- ++(-1.5pt,2.6pt) -- cycle;
    \draw (11,0.25) node[color=black,anchor=west] {\small carrollian};
    \draw [color=gray!90!black] (11,-0.25) circle (1.5pt) node[color=black,anchor=west] {\small aristotelian};       
    \end{scope}
  \end{tikzpicture}
  \caption{Homogeneous spacetimes in dimension $D+1 \geq 4$ and their limits.}
  \label{fig:generic-d-graph}
\end{figure}

Many of the above spacetimes are connected by geometric limits, some
of which manifest themselves as contractions of the kinematical Lie
algebras.  Figure~\ref{fig:generic-d-graph} illustrates these limits
for generic $D\geq 3$.  For $D\leq 2$, the picture is modified  in a way
that will be explained below.

There are several types of limits displayed in
Figure~\ref{fig:generic-d-graph}:
\begin{itemize}
\item \emph{flat limits} in which the curvature of the canonical
  connection goes to zero: $\zAdS \to \MM$, $\zdS \to \MM$, $\zAdSC \to
  \zC$, $\zdSC \to \zC$, $\zAdSG \to \zG$ and $\zdSG \to \zG$;
\item \emph{non-relativistic limits} in which the speed of light goes
  to infinity (morally speaking): $\MM \to \zG$, $\zAdS \to \zAdSG$
  and $\zdS \to \zdSG$;

  In this limit there is still the notion of relativity,
  it just differs from the standard lorentzian one.  Therefore,
  although it might be more appropriate to call it the ``galilean
  limit'',  we will conform to the literature and call it the
  non-relativistic limit.
\item \emph{ultra-relativistic limits} in which the speed of light
  goes to zero (again, morally speaking): $\MM \to \zC$, $\zAdS \to
  \zAdSC$ and $\zdS \to \zdSC$.
\item limits to non-effective Lie pairs which, after quotienting by the
  ideal generated by the boosts, result in an aristotelian spacetime:
  the dotted arrows $\zLC \to  \zTS$, $\zC \to \zS$ and $\zG \to \zS$;
\item $\zLC \to \zC$, which is a contraction of $\so(D+1,1)$;
\item $\ztdSG_\gamma \to \zG$ and $\ztAdSG_\gamma \to \zG$, which are
  contractions of the corresponding kinematical Lie algebras;
\item limits between aristotelian spacetimes $\zTS \to \zS$, $\RR
  \times \SS^D \to \zS$ and $\RR \times \HH^D \to \zS$; and
\item a limit $\lim_{\chi \to \infty} \zAdSG_\chi = \zdSG_1$, which is
  not due to a contraction of the kinematical Lie algebras.
\end{itemize}
We can compose these limits like arrows in a commutative diagram, and 
therefore we do not show all the possible limits. 
All these limits are explained in~\cite{Figueroa-OFarrill:2018ilb}.

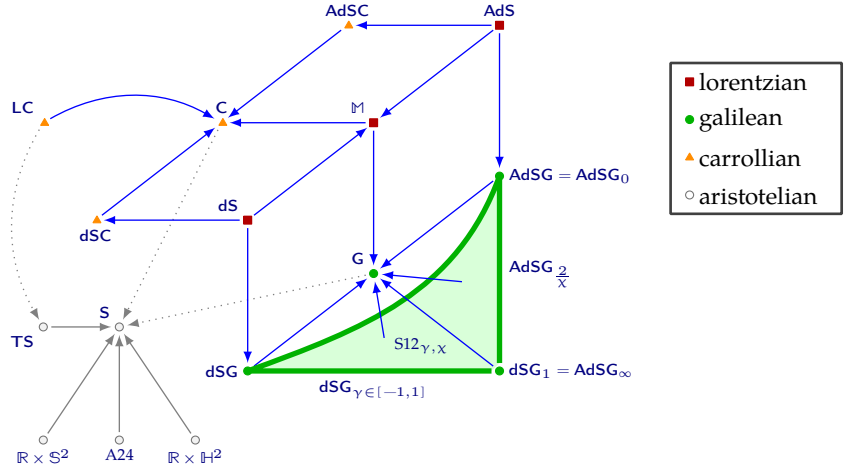
\begin{figure}[h!]
  \centering
  \begin{tikzpicture}[scale=1,>=latex, shorten >=3pt, shorten <=3pt,
    x=1.0cm,y=1.0cm,line join=bevel]
    %
    %
    %
    %
    \coordinate [label=above left:{\hyperlink{S2}{\tiny $\zdS$}}] (ds) at (5.688048519056286,-0.5838170592960186);
    \coordinate [label=left:{\hyperlink{S8}{\tiny $\zdSG$}}] (dsg) at (5.688048519056286, -2.5838170592960186);
    \coordinate [label=right:{\hyperlink{S9}{\tiny $\ztdSG_1 = \ztAdSG_\infty$}}] (dsgone) at (9, -2.5838170592960186);
    \coordinate [label=below:{\hyperlink{S14}{\tiny $\zdSC$}}] (dsc) at  (3.688048519056286,-0.5838170592960186);
    \coordinate [label=above:{\hyperlink{S13}{\tiny $\zC$}}] (c) at (5.344024259528143, 0.7080914703519907);
    \coordinate [label=above left:{\hyperlink{S1}{\tiny $\MM$}}] (m) at (7.344024259528143,0.7080914703519907);
    \coordinate [label=above left:{\hyperlink{S7}{\tiny $\zG$}}] (g) at  (7.344024259528143, -1.2919085296480093);
    \coordinate [label=above:{\hyperlink{S3}{\tiny $\zAdS$}}] (ads) at (9,2);
    \coordinate [label=right:{\hyperlink{S10}{\tiny $\zAdSG = \ztAdSG_0$}}] (adsg) at (9,0);
    \coordinate [label=above:{\hyperlink{S15}{\tiny $\zAdSC$}}]  (adsc) at (7,2);
    \coordinate [label=above left:{\hyperlink{A21}{\tiny $\zS$}}] (s) at (4, -2);
    \coordinate [label=above left:{\hyperlink{S16}{\tiny $\zLC$}}]  (flc) at (3, 0.7080914703519907);
    \coordinate [label=below left:{\hyperlink{A22}{\tiny $\zTS$}}]  (ts) at (3, -2);
    \coordinate [label=below:{\hyperlink{A23p}{\tiny $\RR\times\SS^2$}}]  (esu) at (3, -3.5); 
    \coordinate [label=below:{\hyperlink{A23m}{\tiny $\RR\times\HH^2$}}]  (hesu) at (5, -3.5); 
    \coordinate [label=below:{\hyperlink{A24}{\tiny \twoda}}]  (twoda) at (4, -3.5); 
    %
    %
    \coordinate [label=below:{\hyperlink{S9}{\tiny $\ztdSG_{\gamma\in[-1,1]}$}}] (tdsg) at (7.344024259528143, -2.5838170592960186);
    \coordinate [label=right: {\hyperlink{S11}{\tiny $\ztAdSG_{\frac{2}{\chi}}$}}] (tadsg) at (9, -1.2919085296480093);
    %
    %
    \draw [->,line width=0.5pt,dotted,color=gray] (c) -- (s);
    \draw [->,line width=0.5pt,dotted,color=gray] (g) -- (s);
    %
    %
    \draw [->,line width=0.5pt,color=blue] (adsc) -- (c);
    \draw [->,line width=0.5pt,color=blue] (dsc) -- (c);
    \draw [->,line width=0.5pt,color=blue] (ads) -- (m);
    \draw [->,line width=0.5pt,color=blue] (adsg) -- (g);
    \draw [->,line width=0.5pt,color=blue] (dsg) -- (g);
    \draw [->,line width=0.5pt,color=blue] (ds) -- (m);
    \draw [->,line width=0.5pt,color=blue] (ds) -- (dsc);
    \draw [->,line width=0.5pt,color=blue] (ds) -- (dsg);
    \draw [->,line width=0.5pt,color=blue] (m) -- (c);
    \draw [->,line width=0.5pt,color=blue] (m) -- (g);
    \draw [->,line width=0.5pt,color=blue] (ads) -- (adsc);
    \draw [->,line width=0.5pt,color=blue] (ads) -- (adsg);
    %
    %
    \draw [->,line width=0.5pt,color=blue] (flc) to [out=30,in=150] (c);
    \draw [->,line width=0.5pt,dotted,color=gray] (flc) to [out=240,in=120] (ts);
    \draw [->,line width=0.5pt,color=gray] (ts) to (s);
    \draw [->,line width=0.5pt,color=gray] (esu) to (s); 
    \draw [->,line width=0.5pt,color=gray] (hesu) to (s); 
    \draw [->,line width=0.5pt,color=gray] (twoda) to (s); 
    %
    %
    \begin{scope}[line width=2pt, color=green!70!black]
      \filldraw [color=green!70!black, fill=green!15!white] (dsgone)
      -- (dsg) to [out=20,in=250] (adsg) -- (dsgone);
    \end{scope}
    \coordinate [label=right:{\hyperlink{S12}{\tiny $\text{\twodgal}_{\gamma, \chi}$}}] (tdg) at (7.5,-2.25); 
    \draw [->, line width=0.5pt,color=blue] (dsgone) -- (g);
    \draw [->, line width=0.5pt,color=blue] (tdg) -- (g);
    \draw [->, line width=0.5pt,color=blue] (8.61128, -1.4101) -- (g);
    \foreach \point in {g,adsg,dsg,dsgone}
    \filldraw [color=green!70!black,fill=green!70!black] (\point) circle (1.5pt);
    \foreach \point in {ads,ds,m}
    \filldraw [color=red!70!black,fill=red!70!black] (\point) ++(-1.5pt,-1.5pt) rectangle ++(3pt,3pt);
    \foreach \point in {adsc,dsc,flc,c}
    \filldraw [color=DarkOrange,fill=DarkOrange] (\point) ++(-1pt,-1pt) -- ++(3pt,0pt) -- ++(-1.5pt,2.6pt) -- cycle;
    \foreach \point in {s,ts,esu,hesu,twoda}
    \filldraw [color=gray!90!black,fill=gray!10!white] (\point) circle (1.5pt);
    %
    %
    \begin{scope}[xshift=0.5cm]
    \draw [line width=1pt,color=gray!50!black] (10.75,-0.5) rectangle (13,1.5);
    \filldraw [color=red!70!black,fill=red!70!black] (11,1.25) ++(-1.5pt,-1.5pt) rectangle ++(3pt,3pt) ; 
    \draw (11,1.25) node[color=black,anchor=west] {\small lorentzian}; 
    \filldraw [color=green!70!black,fill=green!70!black] (11,0.75) circle (1.5pt) node[color=black,anchor=west] {\small galilean};
    \filldraw [color=DarkOrange,fill=DarkOrange] (11,0.25) ++(-1.5pt,-1pt) -- ++(3pt,0pt) -- ++(-1.5pt,2.6pt) -- cycle;
    \draw (11,0.25) node[color=black,anchor=west] {\small carrollian};
    \draw [color=gray!90!black] (11,-0.25) circle (1.5pt) node[color=black,anchor=west] {\small aristotelian};
    \end{scope}
  \end{tikzpicture}
  \caption{Three-dimensional homogeneous spacetimes and their limits.}
  \label{fig:d=3-graph}
\end{figure}

The situation in $D\leq 2$ is slightly different. As can be seen in
Tables~\ref{tab:spacetimes} and \ref{tab:aristotelian}, there are two
classes of spacetimes which are unique to $D=2$: a two-parameter
family of galilean spacetimes
(\hyperlink{S12}{$\text{\twodgal}_{\gamma,\chi}$}, for
$\gamma \in [-1,1)$ and $\chi>0$) and the aristotelian spacetime
\hyperlink{A24}{\twoda}. We can understand this latter spacetime as
the group manifold of the three-dimensional Heisenberg group. The
former two-parameter family interpolates between the torsional
galilean (anti) de Sitter spacetimes. As shown in
Figure~\ref{fig:d=3-graph}, the limit $\gamma \to 1$ of
\hyperlink{S12}{$\text{\twodgal}_{\gamma,\chi}$} is
$\hyperlink{S11}{\ztAdSG_{2/\chi}}$, so that if we then take
$\chi \to 0$, we arrive at $\hyperlink{S9}{\ztdSG_1}$. More generally,
the limit $\chi \to 0$ of
$\hyperlink{S12}{\text{\twodgal}_{\gamma,\chi}}$ is
$\hyperlink{S9}{\ztdSG_\gamma}$, whereas the limit $\chi \to \infty$
is independent of $\gamma$ and given by $\hyperlink{S10}{\zAdSG}$.

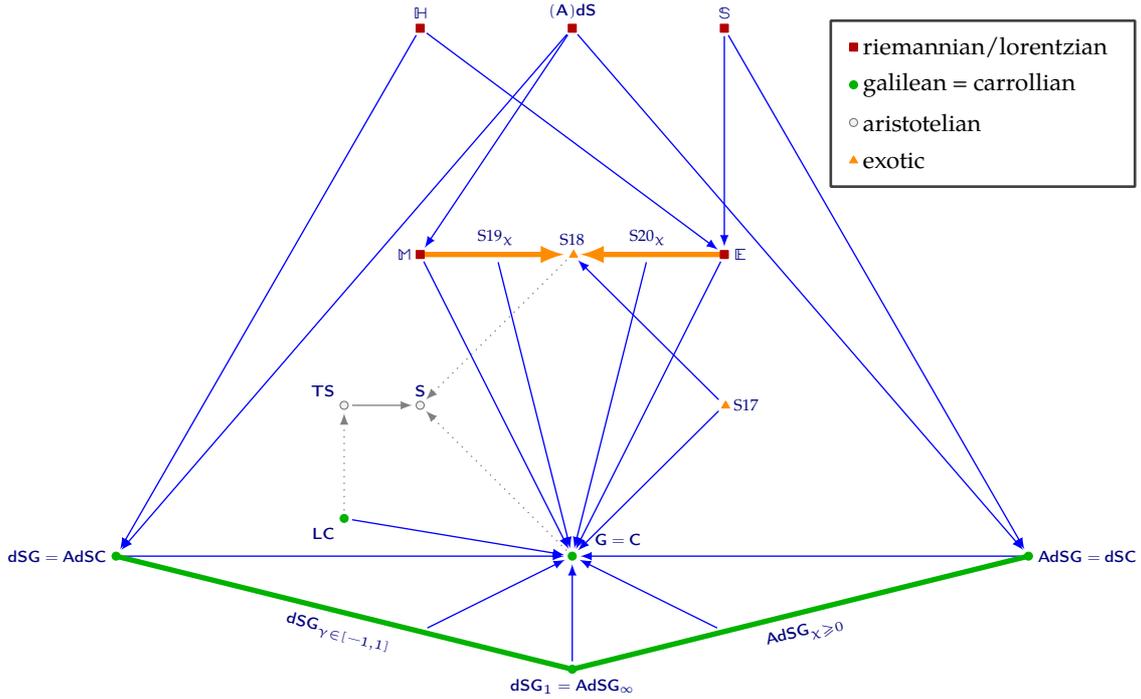
\begin{figure}[h!]
  \centering
  \begin{tikzpicture}[scale=1,>=latex, shorten >=3pt, shorten <=3pt,
    x=1.0cm,y=1.0cm,line join=bevel]
    %
    %
    %
    %
    \coordinate [label=below:{\hyperlink{S9}{\tiny $\ztdSG_1 = \ztAdSG_\infty$}}] (dsgone) at (0,1.5);
    \coordinate [label=left:{\hyperlink{S8}{\tiny $\zdSG = \zAdSC$}}]  (dsg) at (-6,3); 
    \coordinate [label=right:{\hyperlink{S10}{\tiny $\zAdSG = \zdSC$}}] (adsg) at (6,3); 
    \coordinate [label={[shift={(0.6,0.02)}]{\hyperlink{S7}{\tiny $\zG=\zC$}}}] (g) at  (0,3); 
    \coordinate [label=above:{\hyperlink{S18}{\tiny \xtwo}}] (xtwo) at (0,7);    
    \coordinate [label=above:{\hyperlink{S3}{\tiny $\mathsf{(A)dS}$}}] (ads) at (0,10); 
    \coordinate [label=above:{\hyperlink{A21}{\tiny $\zS$}}] (s) at (-2, 5); 
    \coordinate [label=below left:{\hyperlink{S16}{\tiny $\zLC$}}]  (flc) at (-3,3.5); 
    \coordinate [label=above left:{\hyperlink{A22}{\tiny $\zTS$}}]  (ts) at (-3, 5); 
    \coordinate [label=right:{\hyperlink{S17}{\tiny \xone}}] (xone) at (2,5); 
    \coordinate [label=left:{\hyperlink{S1}{\tiny $\MM$}}] (m) at (-2,7); 
    \coordinate [label=right:{\hyperlink{S4}{\tiny $\EE$}}] (e) at (2,7);     
    \coordinate [label=above:{\hyperlink{S5}{\tiny $\SS$}}] (sph) at (2,10);
    \coordinate [label=above:{\hyperlink{S6}{\tiny $\HH$}}] (hyp) at (-2,10);
    \coordinate [label=above:{\hyperlink{S19}{\tiny  \xthree$_\chi$}}] (xthree) at (-1,7); 
    \coordinate [label=above:{\hyperlink{S20}{\tiny  \xfour$_\chi$}}] (xfour) at (1,7); 
    \draw [->,line width=0.5pt,dotted,color=gray] (g) -- (s);
    \draw [->,line width=0.5pt,color=blue] (xone) -- (g); 
    \draw [->,line width=0.5pt,color=blue] (xone) -- (xtwo); 
    \draw [->,line width=0.5pt,color=blue] (m) -- (g);
    \draw [->,line width=0.5pt,color=blue] (e) -- (g);
    \draw [->,line width=0.5pt,color=blue] (adsg) -- (g);
    \draw [->,line width=0.5pt,color=blue] (dsg) -- (g);
    \draw [->,line width=0.5pt,color=blue] (ads) -- (m);
    \draw [->,line width=0.5pt,color=blue] (ads) -- (dsg);
    \draw [->,line width=0.5pt,color=blue] (ads) -- (adsg); 
    \draw [->,line width=0.5pt,color=blue] (sph) -- (e); 
    \draw [->,line width=0.5pt,color=blue] (hyp) -- (e); 
    \draw [->,line width=0.5pt,color=blue] (sph) -- (adsg); 
    \draw [->,line width=0.5pt,color=blue] (hyp) -- (dsg); 
    \draw [->,line width=0.5pt,color=blue] (flc) to (g);
    \draw [->, line width=0.5pt,dotted,color=gray] (flc) to (ts);
    \draw [->,line width=0.5pt,color=gray] (ts) to (s);
    \draw [->, line width=0.5pt,dotted,color=gray] (xtwo) to (s);
    \begin{scope}[->,>=latex, shorten >=3pt, shorten <=0pt, line width=2pt, color=DarkOrange]
      \draw (e) --(xtwo); 
      \draw (m) --(xtwo); 
    \end{scope}
    \draw [->,line width=0.5pt,color=blue] (-1,7) -- (g);
    \draw [->,line width=0.5pt,color=blue] (1,7) -- (g);
    \foreach \point in {xone,xtwo}
    \filldraw [color=DarkOrange,fill=DarkOrange] (\point) ++(-1pt,-1pt) -- ++(3pt,0pt) -- ++(-1.5pt,2.6pt) -- cycle;
    \begin{scope}[shorten >=0pt, shorten <=0pt, line width=2pt, color=green!70!black]
      \draw (dsg) -- (dsgone) node [midway, below, sloped] (tdsg) {\tiny \hyperlink{S9}{$\ztdSG_{\gamma\in[-1,1]}$}};
      \draw (dsgone) -- (adsg) node [midway, below, sloped] (tadsg) {\tiny \hyperlink{S11}{$\ztAdSG_{\chi\geq0}$}};
    \end{scope}
    \draw [->,line width=0.5pt,color=blue] (dsgone) -- (g); 
    \draw [->,line width=0.5pt,color=blue] (-2,2) -- (g); 
    \draw [->,line width=0.5pt,color=blue] (2,2) -- (g); 
    \foreach \point in {g,adsg,dsg,flc,dsgone}
    \filldraw [color=green!70!black,fill=green!70!black] (\point) circle (1.5pt);
    \foreach \point in {e,m,sph,hyp,ads}
    \filldraw [color=red!70!black,fill=red!70!black] (\point) ++(-1.5pt,-1.5pt) rectangle ++(3pt,3pt);
    \foreach \point in {s,ts}
    \filldraw [color=gray!90!black,fill=gray!10!white] (\point) circle (1.5pt);
    \begin{scope}[shift={(0.2cm,6.5cm)}]
      \draw [line width=1pt,color=gray!50!black] (3.2,1.4) rectangle  (7.2,3.6);
      \filldraw [color=red!70!black,fill=red!70!black] (3.5,3.25) ++(-1.5pt,-1.5pt) rectangle ++(3pt,3pt);
      \draw (3.5,3.25) node[color=black,anchor=west] {\small riemannian/lorentzian};
      \filldraw [color=green!70!black,fill=green!70!black] (3.5,2.75) circle (1.5pt) node[color=black,anchor=west] {\small galilean = carrollian};
      \filldraw [color=gray!90!black,fill=gray!10!white] (3.5,2.25) circle (1.5pt) node[color=black,anchor=west] {\small aristotelian};
      \filldraw [color=DarkOrange,fill=DarkOrange] (3.5,1.75) ++(-1.5pt,-1pt) -- ++(3pt,0pt) -- ++(-1.5pt,2.6pt) -- cycle;
      \draw (3.5,1.75) node[color=black,anchor=west] {\small exotic};
    \end{scope}
  \end{tikzpicture}
  \caption{Two-dimensional homogeneous spacetimes and their limits}
  \label{fig:d=2-graph}
\end{figure}

Table~\ref{tab:spacetimes} shows that there are four classes of
two-dimensional spacetimes unique to $D=1$.  These spacetimes are
affine but have no discernible structure.  In \cite{Figueroa-OFarrill:2018ilb}
we describe a number of limits involving these two-dimensional
spacetimes.  Figure~\ref{fig:d=2-graph} illustrates the relationship
between the two-dimensional spacetimes. This figure includes the
riemannian maximally symmetric spaces which are missing from
Figures~\ref{fig:generic-d-graph} and \ref{fig:d=3-graph}.

\subsection{Geometrical properties}
\label{sec:geom-prop}

In Table~\ref{tab:spacetimes-props} we summarise the basic properties
of the homogeneous kinematical spacetimes in
Table~\ref{tab:spacetimes} and aristotelian spacetimes in
Table~\ref{tab:aristotelian}. The first column is our label in this
paper, the second column specifies the value of $D$, where the
dimension of the spacetime is $D+1$. The columns labelled ``R'',
``S'', and ``A'' indicate whether or not the spacetime is reductive,
symmetric, or affine, respectively. A $\cm$ indicates that it is. A
$(\cm)$ in the affine column reflects the existence of an invariant
connection (other than the canonical connection) with vanishing
torsion and curvature. The columns labelled ``L'', ``E'', ``G'', and
``C'' indicate the kind of invariant structures the spacetime
possesses: lorentzian, riemannian (``euclidean''), galilean, and
carrollian, respectively. Again a $\cm$ indicates that the spacetime
possesses that structure. The columns ``P'', ``T'', and ``PT''
indicate whether the spacetime is invariant under parity, time
reversal or their combination, respectively, with $\cm$ signalling
that they do. The column ``B'' summarises results of the current paper
(to be found below) and indicates whether the boosts act with
non-compact orbits in a kinematical spacetime. The columns
``$\Theta$'' and ``$\Omega$'' tell us, respectively, about the torsion
and curvature of the canonical invariant connection for the reductive
spacetimes (that is, all but $\hyperlink{S16}{\zLC}$). A ``$\neq 0$'' indicates the presence
of torsion, curvature, or both torsion and curvature. Its absence
indicates that the connection is torsion-free, flat, or both. The
final column contains any relevant comments, including, when known,
the name of the spacetime.

The table is divided into six sections. The first four correspond to
lorentzian, euclidean, galilean and carrollian spacetimes. The fifth
section contains two-dimensional spacetimes with no invariant
structure of these kinds. The sixth and last section contains the
aristotelian spacetimes. Some of the spacetimes which exist for all
$D\geq 1$ become accidentally pairwise isomorphic in $D=1$: namely,
$\zC \cong \zG$, $\zdS \cong \zAdS$, $\zdSC \cong \zAdSG$ and
$\zAdSC \cong \zdSG$. These accidental isomorphisms explain why we
write $D\geq2$ for carrollian, de~Sitter, and carrollian (anti)
de~Sitter. In this way no two rows are isomorphic, and hence every row
in the table specifies a unique simply-connected homogeneous
spacetime, up to isomorphism.

\begin{table}[h!]
  \centering
  \caption{Properties of simply-connected homogeneous spacetimes}
  \label{tab:spacetimes-props}
  \rowcolors{2}{blue!10}{white}
    \begin{tabular}{l|>{$}c<{$}|*{3}{>{$}c<{$}}|*{4}{>{$}c<{$}}|*{3}{>{$}c<{$}}|>{$}c<{$}|*{2}{>{$}r<{$}}|l} \toprule
      \multicolumn{1}{c|}{Label} & \multicolumn{1}{c|}{$D$} & R & S & A & L & E & G & C & P & T & PT & B & \multicolumn{1}{c}{$\Theta$} &  \multicolumn{1}{c|}{$\Omega$} & \multicolumn{1}{c}{Comments}\\\midrule
      \hyperlink{S1}{\mink} & \geq 1 & \cm & \cm & \cm & \cm &  &  &  & \cm & \cm & \cm & \cm & & & $\MM$\\
      \hyperlink{S2}{\ds} & \geq 2 & \cm & \cm &  & \cm &  &  &  & \cm & \cm & \cm & \cm & & \neq0 & $\zdS$\\
      \hyperlink{S3}{\ads} & \geq 1 & \cm & \cm &  & \cm &  &  &  & \cm & \cm & \cm & \cm & & \neq0 & $\zAdS$\\\midrule
      \hyperlink{S4}{\euc} & \geq 1 & \cm & \cm & \cm &  & \cm &  &  & \cm & \cm & \cm & & & & $\EE$\\
      \hyperlink{S5}{\sph} & \geq 1 & \cm & \cm &  &  & \cm &  &  & \cm & \cm & \cm & & & \neq0 & $\SS$\\
      \hyperlink{S6}{\hyp} & \geq 1 & \cm & \cm &  &  & \cm &  &  & \cm & \cm & \cm & & & \neq0 & $\HH$\\\midrule
      \hyperlink{S7}{\gal} & \geq 1 & \cm & \cm & \cm &  &  & \cm &  & \cm & \cm & \cm & \cm & & & $\zG$\\
      \hyperlink{S8}{\dsg} & \geq 1 & \cm & \cm & (\cm) &  &  & \cm &  & \cm & \cm & \cm & \cm & & \neq0 & $\zdSG$\\
      \hyperlink{S9}{\tdsg$_{\gamma\neq 0}$} & \geq 1 & \cm &  & (\cm) &  &  & \cm &  & \cm & &  & \cm & \neq0 & \neq0 & $\ztdSG_\gamma$, $0\neq \gamma \in (-1,1]$\\
      \hyperlink{S9}{\tdsg$_0$} & \geq 1 & \cm &  & (\cm) &  &  & \cm &  & \cm &  &  & \cm & \neq0 & & $\ztdSG_0$\\
      \hyperlink{S10}{\adsg} & \geq 1 & \cm & \cm &  &  &  & \cm &  & \cm & \cm & \cm & \cm & & \neq0 & $\zAdSG$\\
      \hyperlink{S11}{\tadsg$_\chi$} & \geq 1 & \cm &  &  &  &  & \cm &  & \cm &  &  & \cm & \neq0 & \neq0 & $\ztAdSG_\chi$, $\chi>0$\\
      \hyperlink{S12}{$\text{\twodgal}_{\gamma,\chi}$} & 2 & \cm &  &  &  &  & \cm &  & \cm &  &  & \cm & \neq0 & \neq0 & $\gamma\in [-1,1)$, $\chi>0$\\\midrule
      \hyperlink{S13}{\car} & \geq 2 & \cm & \cm & \cm &  &  &  & \cm & \cm & \cm & \cm & \cm & & & $\zC$\\
      \hyperlink{S14}{\dsc} & \geq 2 & \cm & \cm & & &  &  & \cm & \cm & \cm & \cm & \cm & & \neq0 & $\zdSC$\\
      \hyperlink{S15}{\adsc} & \geq 2 &  \cm & \cm & &  &  &  & \cm & \cm & \cm & \cm & \cm & & \neq0 & $\zAdSC$\\
      \hyperlink{S16}{\flc} & \geq 1 &  &  & (\cm)_{D=1} &  &  &  & \cm & \cm &  &  & \cm & & & $\zLC$\\\midrule
      \hyperlink{S17}{\xone} & 1 & \cm & \cm & \cm &  &  &  &  &  &  & \cm & \cm & & & \\
      \hyperlink{S18}{\xtwo} & 1 & \cm & \cm & \cm &  &  &  &  &  & \cm &  & \cm & & & \\
      \hyperlink{S19}{\xthree$_\chi$} & 1 & \cm & \cm & \cm &  &  &  &  &  & \cm &  & \cm & & & $\chi>0$\\
      \hyperlink{S20}{\xfour$_\chi$} & 1 & \cm & \cm & \cm &  &  &  &  &  &  & \cm & \cm & & & $\chi>0$\\\midrule
      \hyperlink{A21}{\st} & \geq 0 & \cm & \cm & \cm & \cm & \cm & \cm & \cm & \cm & \cm & \cm & & & & $\zS$\\
      \hyperlink{A22}{\tst} & \geq 1 & \cm &  & (\cm) & \cm & \cm & \cm & \cm & \cm &  &  & & \neq0 & & $\zTS$\\
      \hyperlink{A23p}{\athree$_{+1}$} & \geq 2 & \cm & \cm &  & \cm & \cm & \cm & \cm & \cm & \cm & \cm & & & \neq0 & $\RR \times \SS^D$ \\
      \hyperlink{A23m}{\athree$_{-1}$} & \geq 2 & \cm & \cm &  & \cm & \cm & \cm & \cm & \cm & \cm & \cm & & & \neq0 & $\RR \times \HH^D$\\
      \hyperlink{A24}{\twoda} & 2 & \cm &  & (\cm) & \cm & \cm & \cm & \cm & \cm &  & &  & \neq0 & & \\ \bottomrule
    \end{tabular}
    \\[10pt]
    \caption*{This table describes if a $D+1$ dimensional kinematical
      spacetime (Table \ref{tab:spacetimes}) or aristotelian spacetime
      (Table \ref{tab:aristotelian}) is reductive (R), symmetric (S)
      or affine (A). A spacetime might exhibit a lorentzian (L),
      riemannian (E), galilean (G) or carrollian (C) structure, and be
      invariant under parity (P), time reversal (T) or their
      combination (PT). The boosts (B) may act with non-compact
      orbits. Furthermore the canonical connection of a reductive
      spacetime might be have torsion ($\Theta$) and/or curvature
      ($\Omega$).}
\end{table}

\section{Local geometry of homogeneous spacetimes}
\label{sec:geom-prop-homog}

In this section, we review some basic properties of homogeneous
spaces, tailored to the cases of interest. We discuss exponential
coordinates, the fundamental vector fields, the group action, the
action of rotations and boosts, the soldering form, and the vielbein.
In addition, we discuss the invariant connections on a reductive
homogeneous space.

\subsection{Exponential coordinates}
\label{sec:expon-coord}

Let $M = \Kgr/\Hgr$ be a kinematical spacetime with associated Lie
pair $(\k,\h)$ in which $\k$ is a kinematical Lie algebra and $\h$ is
the Lie subalgebra spanned by the rotations $J_{ab}$ and the boosts
$B_a$. The identification of $M$ with the coset manifold $\Kgr/\Hgr$
singles out a point $o \in M$ corresponding to the identity coset. We
call it the \emph{origin} of $M$. Any other point in $M$ would be
equally valid as an ``origin'', but that choice would induce an
identification with a different coset manifold since the new origin
typically has a different, but of course conjugate, stabiliser
subgroup.

The action of $\Kgr$ on $M$ is induced by left multiplication on
$\Kgr$.  If we let $\varpi : \Kgr \to M= \Kgr/\Hgr$ denote the
canonical surjection, then for all $g \in \Kgr$, we have that
\begin{equation} 
  g \cdot \varpi(k) = \varpi (g k).
\end{equation}
This is well defined because if $\varpi(k) = \varpi(k')$, then there
is some $h \in \Hgr$ such that $k'= k h$ and by associativity of the
group multiplication $g k' = g(k h) = (gk) h$, so that $\varpi(gk) =
\varpi(gk')$.

Now consider acting with $g \in \Kgr$ on the origin.  If $g \in \Hgr$,
$g \cdot o = o$, so this suggests the following.  Let $\m =
\spn{P_a,H}$ denote a vector space complement of $\h$ in $\k$ and
define $\exp_o : \m \to M$ by
\begin{equation}
  \exp_o(X) = \exp(X) \cdot o \qquad\text{for all $X \in \m$.}
\end{equation}
This map defines a local diffeomorphism from a neighbourhood of $0$ in
$\m$ and a neighbourhood of $o$ in $M$, and hence it defines
\emph{exponential coordinates} near $o$ via
$\sigma : \RR^{D+1} \to M$, where
$\sigma(t,\x) = \exp_o(t H + \x \cdot \P)$. This coordinate chart has
an origin $o\in M$, which is the point with coordinates
$(t,\x) = (0,\bzero)$. We may translate this coordinate chart from the
origin to any other point of $M$ via the action of the group and in
this way arrive at an exponential coordinate atlas for $M$. It is not
the only natural coordinate system associated with a choice of basis
for $\m$. Indeed, it is often more convenient computationally to use
modified exponential coordinates via products of exponentials, say,
$\sigma'(t,\x) = \exp(t H) \exp(\x\cdot\P) \cdot o$. For most of this
work we have opted to use strict exponential coordinates in our
calculations for uniformity and to ease comparison: the exception
being the determination of the symmetries, where modified exponential
coordinates (as described in Appendix~\ref{app:modexp}) allow for a
more uniform description.

There are some natural questions one can ask about the local
diffeomorphism $\exp_o : \m \to M$ or, equivalently, the local
diffeomorphism $\sigma: \RR^{D+1} \to M$.  One can ask how much of $M$
is covered by the image of $\exp_o$.  We say that $M$ is
\textbf{exponential} if $M = \exp_o(\m)$ and \textbf{weakly
  exponential} if $M = \overline{\exp_o(\m)}$, where the bar denotes
topological closure.  Similarly, we can ask about the domain of
validity of exponential coordinates: namely, the subspace of
$\RR^{D+1}$ where $\sigma$ remains injective.  In particular, if
$\sigma$ is everywhere injective, does it follow that $\sigma$ is also
surjective?  We know very little about these
questions for general homogeneous spaces, even in the reductive case.
However, there are some general theorems for the case of $M$ a
symmetric space.

\begin{theorem}[Voglaire~\cite{MR3273068}]\label{thm:voglaire}
  Let $M = \Kgr/\Hgr$ be a connected symmetric space with symmetric
  decomposition $\k = \h \oplus \m$ and define $\exp_o: \m \to M$.  
  Then the following are equivalent:
  \begin{enumerate}
  \item $\exp_o: \m \to M$ is injective
  \item $\exp_o: \m \to M$ is a global diffeomorphism
  \item $M$ is simply connected and for no $X \in \m$, does $\ad_X :
    \k \to \k$ have purely imaginary eigenvalues.
  \end{enumerate}
\end{theorem}

Since our homogeneous spaces are by assumption simply-connected, the
last criterion in the theorem is infinitesimal and, therefore, easily
checked from the Lie algebra. This result makes it a relatively simple
task to inspect Table~\ref{tab:spacetimes} and determine for which of
the symmetric spaces the last criterion holds by studying the
eigenvalues of $\ad_H$ and $\ad_{P_a}$ on $\k$. Inspection of
Table~\ref{tab:spacetimes} shows that $\hyperlink{S1}{\MM}$,
$\hyperlink{S4}{\EE}$, $\hyperlink{S6}{\HH}$, $\hyperlink{S7}{\zG}$,
$\hyperlink{S8}{\zdSG}$, $\hyperlink{S13}{\zC}$ and
$\hyperlink{S15}{\zAdSC}$ satisfy criterion (3) above and hence that
the exponential coordinates define a diffeomorphism to $\RR^{D+1}$ for
these spaces. It also follows by inspection that
$\hyperlink{S2}{\zdS}$, $\hyperlink{S3}{\zAdS}$,
$\hyperlink{S5}{\SS}$, $\hyperlink{S10}{\zAdSG}$ and
$\hyperlink{S14}{\zdSC}$ do not satisfy criterion (3) above and hence
the exponential coordinates do not give us a global chart. We will be
able to confirm this directly when we calculate the soldering form for
these symmetric spaces.

Concerning the (weak) exponentiality of symmetric spaces, we will make
use of the following result.

\begin{theorem}[Rozanov~\cite{MR2503866}]\label{thm:rozanov}
  Let $M = \Kgr/\Hgr$ be a symmetric space with $\Kgr$ connected.
  Then
  \begin{enumerate}
  \item If $\Kgr$ is solvable, then $M$ is weakly exponential.
  \item $M$ is weakly exponential if and only if $\widehat{M} =
    \Khat/\Hhat$ is weakly exponential, where $\Khat =
    \Kgr/\Rad(\Kgr)$ and similarly for $\Hhat$, where the
    \textbf{radical} $\Rad(\Kgr)$ is the maximal connected solvable
    normal subgroup of $\Kgr$.
  \end{enumerate}
\end{theorem}

The Lie algebra of $\Rad(\Kgr)$ is the radical of the Lie algebra
$\k$, which is the maximal solvable ideal, and can be calculated
efficiently via the identification $\rad\k = [\k,\k]^\perp$, namely,
the radical is the perpendicular subspace (relative to the Killing
form, which may be degenerate) of the first derived ideal.

It will follow from Theorem~\ref{thm:rozanov} that
$\hyperlink{S10}{\zAdSG}$ is weakly exponential.

\subsection{The group action and the fundamental vector fields}
\label{sec:group-acti-fund}

The action of the group $\Kgr$ on $M$ is induced by left
multiplication on the group.  Indeed, we have a commuting square
\begin{equation}
  \begin{tikzcd}
    \Kgr \arrow[d, "\varpi"'] \arrow[r, "L_g"] & \Kgr \arrow[d,
    "\varpi"] \\
    M \arrow[r, "\tau_g"] & M
  \end{tikzcd} \qquad\qquad
  \tau_g \circ \varpi = \varpi \circ L_g,
\end{equation}
where $L_g$ is the 
diffeomorphism of $\Kgr$ given by left multiplication by $g \in \Kgr$
and $\tau_g$ is the diffeomorphism of $M$ given by acting with $g$.
In terms of exponential coordinates, we have $g \cdot (t,\x) =
(t',\x')$ where
\begin{equation}
  g \exp(t H + \x \cdot \P) = \exp(t'H + \x'\cdot \P) h,
\end{equation}
for some $h \in \Hgr$ which typically depends on $g$, $t$, and $\x$.

If $g = \exp(X)$ with $X \in \h$ and if $A = t H + \x \cdot \P \in
\m$, the following identity will be useful:
\begin{equation}
  \label{eq:Hact}
  \exp(X)\exp(A)= \exp\left(\exp(\ad_X)A\right) \exp(X).
\end{equation}
If $M$ is reductive, so that $[\h,\m] \subset \m$, which is the case
for all but one of the kinematical spacetimes, then $\ad_X A \in
\m$ and, since $\m$ is a finite-dimensional vector space and hence
topologically complete, $\exp(\ad_X) A \in \m$ as well.  In this case,
we may act on the origin $o \in M$, which is stabilised by $\Hgr$, to
rewrite equation~\eqref{eq:Hact} as
\begin{equation}
  \exp(X)\exp_o(A) = \exp_o\left(\exp(\ad_X)A\right),
\end{equation}
or, in terms of $\sigma$,
\begin{equation}
  \exp(X) \sigma(t,\x) = \sigma(\exp(\ad_X)(t H + \x \cdot \P)) = \sigma(t',\x').
\end{equation}
This latter way of writing the equation shows the action of $\exp(X)$
on the exponential coordinates $(t,\x)$, namely
\begin{equation}\label{eq:exp-coord-action}
  (t,\x) \mapsto (t',\x') \qquad\text{where}\qquad t' H + \x' \cdot \P
  := \exp(\ad_X) (t H + \x \cdot \P).
\end{equation}

As we will show below, the rotations act in the usual way: they leave
$t$ invariant and rotate $\x$, so we will normally concentrate on the
action of the boosts and translations. This requires calculating, for
example,
\begin{equation}
  \exp(v^aP_a) \sigma(t,\x) = \sigma(t',\x') h.
\end{equation}
In some cases, e.g., the non-flat spacetimes, this calculation is not
practical and instead we may take $\v$ to be very small and work out
$t'$ and $\x'$ to first order in $\v$. This approximation then gives
the vector field $\xi_{P_a}$ generating the infinitesimal action of
$P_a$. To be more concrete, let $X \in \k$ and consider
\begin{equation}
  \exp(s X) \sigma(t,\x) = \sigma(t',\x') h
\end{equation}
for $s$ small.  Since for $s=0$, $t'=t$, $\x'=\x$, and $h = 1$, we may
write (up to $O(s^2)$)
\begin{equation}
  \exp(s X) \sigma(t,\x) = \sigma(t + s \tau, \x + s \y) \exp(Y(s)),
\end{equation}
for some $Y(s) \in \h$ with $Y(0) = 0$, and where $\tau$ and $\y$ do
not depend on $s$.  Equivalently,
\begin{equation}
  \label{eq:inf-action}
  \exp(s X) \sigma(t,\x) \exp(-Y(s)) = \sigma(t + s \tau, \x + s \y),
\end{equation}
again up to terms in $O(s^2)$.  We now differentiate this equation
with respect to $s$ at $s=0$.  Since the equation holds up to
$O(s^2)$, the differentiated equation is exact.

To calculate the derivative, we recall the expression for
the differential of the exponential map (see, e.g., \cite[§1.2,Thm.~5]{MR1889121})
\begin{equation}
\label{eq:d-exp}
  \left.\frac{d}{ds} \exp(X(s))\right|_{s=0} = \exp(X(0))
  D(\ad_{X(0)}) X'(0)~,
\end{equation}
where $D$ is the Maclaurin series corresponding to the analytic function
\begin{equation}
  \label{eq:function-d}
  D(z) = \frac{1-e^{-z}}{z} = 1 - \tfrac12 z + O(z^2).
\end{equation}
(We have abused notation slightly and written equations as if we were
working in a matrix group.  This is only for clarity of exposition:
the results are general.)

Let $A = t H + \x \cdot \P$. Differentiating
equation~\eqref{eq:inf-action}, we find
\begin{equation}
  X \exp(A) - \exp(A) Y'(0) = \exp(A) D(\ad_A) (\tau H + \y \cdot \P),
\end{equation}
and multiplying through by $\exp(-A)$ and using that $D(z)$ is
invertible as a power series with inverse the Maclaurin series
corresponding to the analytic function $F(z) = z/(1-e^{-z})$, we find
\begin{equation}
  \label{eq:master}
  G(\ad_A) X - F(\ad_A) Y'(0) = \tau H + \y \cdot \P,
\end{equation}
where we have introduced $G(z) = e^{-z} F(z) = z/(e^z-1)$.  It is a
useful observation that the analytic functions $F$ and $G$ satisfy the
following relations:
\begin{equation}
  \label{eq:f-and-g}
  F(z) = K(z^2) + \frac{z}{2} \qquad\text{and}\qquad G(z) = K(z^2) - \frac{z}{2}~,
\end{equation}
for some analytic function $K(\zeta) = 1 + \tfrac1{12}\zeta +
O(\zeta^2)$.  To see this, simply notice that $F(z) - G(z) = z$ and
that the analytic function $F(z) + G(z)$ is invariant under $z \mapsto
-z$.

Equation~\eqref{eq:master} can now be solved for $\tau$ and $\y$ on a
case by case basis.  To do this, we need to compute $G(\ad_A)$ and $F(\ad_A)$
on Lie algebra elements.  Often a pattern emerges which allows us to
write down the result.  If this fails, one can bring $\ad_A$ into
Jordan normal form and then apply the usual techniques from operator
calculus.  A good check of our calculations is that the linear map
$\k \to \eX(M)$, sending $X$ to the vector field
\begin{equation}
  \xi_X = \tau \frac{\d}{\d t} + y^a \frac{\d}{\d x^a},
\end{equation}
should be a Lie algebra \emph{anti}-homomorphism: namely,
\begin{equation}
  [\xi_X, \xi_Y] = - \xi_{[X,Y]}.
\end{equation}
We have an anti-homomorphism since the action of $\k$ on $M$ is
induced from the vector fields which generate left translations on
$\Kgr$ and these are right-invariant, hence obeying the opposite Lie
algebra.

\subsection{The action of the rotations}
\label{sec:rotations}

In this section, we illustrate the preceding discussion for the case
of rotations. Here, of course, $D\geq 2$. We will see rotations act in
the way we may naively expect on the exponential coordinates: namely,
$t$ is a scalar and $x^a$ is a vector.

The infinitesimal action of the rotational generators $J_{ab}$ on the
exponential coordinates can be deduced from
\begin{equation}
  [J_{ab}, H ] = 0 \qquad\text{and}\qquad [J_{ab}, P_c] = \delta_{bc}
  P_a - \delta_{ac} P_b.
\end{equation}
To be concrete, consider $J_{12}$, which rotates $P_1$ and $P_2$ into
each other:
\begin{equation}
  [J_{12}, P_1] = - P_2 \qquad\text{and}\qquad [J_{12},P_2] = P_1,
\end{equation}
but leaves $H$ and $P_3,\cdots,P_D$ inert.  We see that
$\ad^2_{J_{12}} P_a = - P_a$ for $a = 1,2$, so that exponentiating,
\begin{equation}
  \begin{split}
    \exp(\theta \ad_{J_{12}}) (t H + \x \cdot \P) &= t H + x^1 (\cos\theta
  P_1 - \sin\theta P_2) + x^2 (\cos\theta P_2 + \sin\theta P_1) +
  x^3P_3 + \cdots x^D P_D\\
  &= t H + (x^1 \cos\theta + x^2 \sin\theta) P_1 + (x^2\cos\theta -
  x^1\sin\theta) P_2 + x^3 P_3 + \cdots + x^D P_D.
  \end{split}
\end{equation}
Restricting attention to the $(x^1,x^2)$ plane, we see that the orbit
of $(x^1_0, x^2_0)$ under the one-parameter subgroup
$\exp(\theta J_{12})$ of rotations is
\begin{equation}
  \label{eq:rot}
    \begin{pmatrix}
    x^1(\theta) \\
    x^2(\theta)
  \end{pmatrix} = 
  \begin{pmatrix}
    \cos\theta &  \sin\theta \\
    - \sin \theta & \cos\theta 
  \end{pmatrix} \cdot
  \begin{pmatrix}
    x^1_0\\
    x^2_0
  \end{pmatrix}.
\end{equation}
Differentiating $(x^1(\theta),x^1(\theta))$ with respect to
$\theta$ yields
\begin{equation}
  \frac{d x^1}{d \theta} = x^2 \qquad\text{and}\qquad   \frac{d x^2}{d \theta} = -x^1,
\end{equation}
so that
\begin{equation}
  \xi_{J_{12}} = x^2 \frac{\d}{\d x^1} - x^1 \frac{\d}{\d x^2}.
\end{equation}
In the general case, and in the same way, we find
\begin{equation}\label{eq:fvf-rot}
 \xi_{J_{ab}} = x^b \frac{\d}{\d x^a} - x^a \frac{\d}{\d x^b},
\end{equation}
which can be checked to obey the opposite Lie algebra
\begin{equation}
  [\xi_{J_{ab}}, \xi_{J_{cd}}] = - \delta_{bc} \xi_{J_{ad}} +
  \delta_{bd} \xi_{J_{ac}} + \delta_{ac} \xi_{J_{bd}} - \delta_{ad}
  \xi_{J_{bc}} = -\xi_{[J_{ab},J_{cd}]}.
\end{equation}

\subsection{The action of the boosts}
\label{sec:boosts}

For a homogeneous space $M=\Kgr/\Hgr$ of a kinematical Lie group
$\Kgr$ to admit a physical interpretation as a genuine spacetime, one
would seem to require that the boosts act with non-compact
orbits~\cite{Bacry:1968zf}. Otherwise, it would be more suitable to
interpret them as (additional) rotations. In other words, if $(\k,\h)$
is the Lie pair describing the homogeneous spacetime, with $\h$ the
subalgebra spanned by the rotations and the boosts, then a desirable
geometrical property of $M$ is that for all $X = w^a B_a \in \h$ the
orbit of the one-parameter subgroup $\Bgr_X \subset \Hgr$ generated by
$X$ should be homeomorphic to the real line. Of course, this
requirement is strictly speaking never satisfied: the ``origin'' of
$M$ is fixed by $\Hgr$ and, in particular, by any one-parameter
subgroup of $\Hgr$, so its orbit under any $\Bgr_X$ consists of just
one point. Therefore the correct requirement is that the
\emph{generic} orbits be non-compact. It is interesting to note that
we impose no such requirements on the space and time translations.

With the exception of the carrollian light cone
$\hyperlink{S16}{\zLC}$, which will have to be studied separately, the
action of the boosts are uniform in each class of spacetimes:
lorentzian, riemannian, galilean and carrollian. (There are no boosts
in aristotelian spacetimes.) We can read the action of the boosts
(infinitesimally) from the Lie brackets:
\begin{itemize}
\item \emph{lorentzian}:
  \begin{equation}
    [\B, H] = \P, \qquad [\B, \P] = H \qquad\text{and}\qquad [\B,\B] = \J;
  \end{equation}
\item \emph{riemannian}:
  \begin{equation}
    [\B, H] = -\P, \qquad [\B, \P] = H \qquad\text{and}\qquad [\B,\B] = -\J;
  \end{equation}
\item \emph{galilean}:
  \begin{equation}
    [\B, H] = \P;
  \end{equation}
\item \emph{(reductive) carrollian}:
  \begin{equation}
    [\B, \P] = H;
  \end{equation}
\item and \emph{carrollian light cone ($\hyperlink{S16}{\zLC}$)}:
  \begin{equation}
    [\B, H] = -\B \qquad\text{and}\qquad [\B,\P] = H + \J.
  \end{equation}
\end{itemize}
Below we will calculate the action of the boosts for all spacetimes
except for the carrollian light cone and the exotic two-dimensional
spacetimes (\hyperlink{S17}{\xone}, \hyperlink{S18}{\xtwo},
\hyperlink{S19}{\xthree$_\chi$} and \hyperlink{S20}{\xfour$_\chi$})
which will be studied case by case.

In order to simplify the calculation, it is convenient to introduce
two parameters $\varsigma$ and $c$ and write the infinitesimal action
of the boosts as
\begin{equation}\label{eq:inf-boosts}
  [B_a,H] = -\varsigma P_a \qquad\text{and}\qquad [B_a,P_b] =
  \frac1{c^2} \delta_{ab} H.
\end{equation}
Then $(\varsigma,c^{-1}) = (-1,1)$ for lorentzian,
$(\varsigma,c^{-1}) = (1,1)$ for riemannian,
$(\varsigma,c^{-1}) = (-1,0)$ for galilean and
$(\varsigma,c^{-1}) = (0,1)$ for (reductive) carrollian spacetimes.

The action of the boosts on the exponential coordinates, as
described in Section~\ref{sec:group-acti-fund}, is given by
equation~\eqref{eq:exp-coord-action}, which in this case becomes
\begin{equation}
  t H + \x \cdot \P \mapsto \exp(\ad_{\w \cdot \B}) ( t H + \x \cdot \P ).
\end{equation}
From equation \eqref{eq:inf-boosts}, we see that
\begin{equation}
  \begin{aligned}[m]
    \ad_{\w\cdot \B} H &= -\varsigma \w \cdot \P \\
    \ad^2_{\w\cdot \B} H &= -\frac1{c^2} \varsigma w^2 H, \\
  \end{aligned}
  \qquad\text{and}\qquad
  \begin{aligned}[m]
    \ad_{\w\cdot \B} \P &= \frac1{c^2} \w H \\
    \ad^2_{\w\cdot \B} \P &= -\frac1{c^2} \varsigma \w (\w\cdot\P),\\
  \end{aligned}
\end{equation}
so that in all cases $\ad^3_{\w\cdot \B} = -\frac1{c^2} \varsigma w^2
\ad_{\w \cdot \B}$.  This allows us to exponentiate $\ad_{\w\cdot\B}$ easily:
\begin{equation}
  \exp(\ad_{\w\cdot\B}) = 1 + \frac{\sinh z}{z} \ad_{\w \cdot \B}
  + \frac{\cosh z - 1}{z^2} \ad^2_{\w \cdot \B},
\end{equation}
where $z^2 = -\frac1{c^2} \varsigma w^2$, and hence
\begin{equation}
  \begin{split}
    \exp(\ad_{\w\cdot\B}) t H &= t \cosh z H - \varsigma t \frac{\sinh z}{z} \w \cdot \P,\\
    \exp(\ad_{\w\cdot\B}) \x \cdot \P &= \x \cdot \P +  \frac1{c^2}
    \frac{\sinh z}{z} \x \cdot \w H + \frac{\cosh z - 1}{w^2}
    (\x \cdot \w) \w \cdot \P.
  \end{split}
\end{equation}
Therefore, the orbit of $(t_0,\x_0)$ under $\exp(s\w\cdot \B)$ is given
by
\begin{equation}\label{eq:boost-orbit}
  \begin{split}
    t(s) &= t_0 \cosh(s z) + \frac1{c^2} \frac{\sinh(s z)}{z}\x_0 \cdot \w,\\
    \x(s) &= \x_0^\perp - \varsigma t_0 \frac{\sinh(s z)}{z} \w +
    \frac{\cosh(s z)}{w^2} (\x_0 \cdot \w) \w,
  \end{split}
\end{equation}
where we have introduced
$\x_0^\perp := \x_0 - \frac{\x_0 \cdot \w}{w^2}\w$ to be the component
of $\x_0$ perpendicular to $\w$. It follows from this expression that
$\x^\perp(s) = \x_0^\perp$, so that the orbit lies in a plane spanned
by $\w$ and the time direction.

Differentiating these expressions with respect to $s$, we arrive at
the fundamental vector field $\xi_{B_a}$.  Indeed, differentiating
$(t(s),\x(s))$ with respect to $s$ at $s=0$, we obtain the value of
$\xi_{\w \cdot \B}$ at the point $(t_0,x_0)$.  Letting $(t_0,x_0)$
vary we obtain that
\begin{equation}\label{eq:fvf-boost}
  \xi_{B_a} = \frac{1}{c^2} x^a \frac{\d}{\d t} - \varsigma t \frac{\d}{\d x^a}.
\end{equation}

In particular, notice that one of the virtues of the exponential
coordinates, is that the fundamental vector fields of the stabiliser
$\h$ -- that is, of the rotations and the boosts -- are linear and, in
particular, they are complete. This will be useful in determining
whether or not the generic orbits of one-parameter subgroup of boosts
are compact.

Let $\exp(s \w\cdot\B)$, $s \in \RR$, be a one-parameter subgroup
consisting of boosts.  Given any $p \in M$, its orbit under this
subgroup is the image of the map $c: \RR \to M$, where
$c(s) := \exp(s \w\cdot \B) \cdot p$.  As we just saw, in the
reductive examples (all but $\hyperlink{S16}{\zLC}$) the fundamental vector field
$\xi_{\w\cdot\B}$ is linear in the exponential coordinates, and hence
it is complete.  Therefore, its integral curves are one-dimensional
connected submanifolds of $M$ and hence either homeomorphic to the
real line (if non compact) or to the circle (if compact).  The compact
case occurs if and only if the map $c$ is periodic.

If the exponential coordinates define a global coordinate chart (which
means, in particular, that the homogeneous space is diffeomorphic to
$\RR^{D+1}$), then it is only a matter of solving a linear ODE to
determine whether or not $c$ is periodic. In any case, we can
determine whether or not this is the case in the exponential
coordinate chart centred at the origin. For the special case of
symmetric spaces, which are the spaces obtained via limits from the
riemannian and lorentzian maximally symmetric spaces, we may use
Theorem~\ref{thm:voglaire}, which gives an infinitesimal criterion for
when the exponential coordinates define a global chart. Recalling the
discussion in Section~\ref{sec:expon-coord}, we again state that
$\hyperlink{S1}{\MM}$, $\hyperlink{S4}{\EE}$, $\hyperlink{S6}{\HH}$,
$\hyperlink{S7}{\zG}$, $\hyperlink{S8}{\zdSG}$,
$\hyperlink{S13}{\zC}$, and $\hyperlink{S15}{\zAdSC}$ satisfy
criterion (3) in Theorem~\ref{thm:voglaire} and hence that the
exponential coordinates define a diffeomorphism $M \cong \RR^{D+1}$.
Using exponential coordinates, we will see that the orbits of boosts
in $\hyperlink{S4}{\EE}$ and $\hyperlink{S6}{\HH}$ are compact,
whereas the generic orbits of boosts in the other cases are
non-compact.

The remaining symmetric spacetimes $\hyperlink{S2}{\zdS}$,
$\hyperlink{S3}{\zAdS}$, $\hyperlink{S5}{\SS}$,
$\hyperlink{S10}{\zAdSG}$, and $\hyperlink{S14}{\zdSC}$ do not satisfy
the infinitesimal criterion (3) in Theorem~\ref{thm:voglaire}, and
hence the exponential coordinates are not a global chart. It may
nevertheless still be the case that the image of $\exp_o$ covers the
homogeneous spacetime (or a dense subset). It turns out that
$\hyperlink{S5}{\SS}$ is exponential and $\hyperlink{S10}{\zAdSG}$ is
weakly exponential. The result for $\hyperlink{S5}{\SS}$ is classical,
since the sphere is a compact riemannian symmetric space, and the case
of $\hyperlink{S10}{\zAdSG}$ follows from Theorem~\ref{thm:rozanov}.
If $D\leq 2$, then the kinematical Lie group for
$\hyperlink{S10}{\zAdSG}$ is solvable and hence
$\hyperlink{S10}{\zAdSG}$ is weakly exponential, whereas if $D\geq 3$,
the radicals $\rad\k = \spn{\B,\P,H}$ and $\rad\h = \spn{\B}$.
Therefore, $\k/\rad\k \cong \so(D) \cong \h/\rad\h$. Therefore, with
$\Khat := \Kgr/\Rad(\Kgr)$ and similarly for $\Hhat$, $\Khat/\Hhat$ is
trivially weakly exponential and hence, by Theorem~\ref{thm:rozanov},
so is $\Kgr/\Hgr$. We will see that boosts act with compact orbits in
$\hyperlink{S5}{\SS}$, but with non-compact orbits in
$\hyperlink{S10}{\zAdSG}$.

Among the symmetric spaces in Table~\ref{tab:spacetimes}, this leaves
$\hyperlink{S2}{\zdS}$, $\hyperlink{S3}{\zAdS}$, and
$\hyperlink{S14}{\zdSC}$. We treat those cases using the same
technique, which will also work for the non-symmetric
$\hyperlink{S16}{\zLC}$. Let $M$ be a simply-connected homogeneous
spacetime and $q: M \to \Mbar$ a covering map which is equivariant
under the action of (the universal covering group of) $\Kgr$. By
equivariance,
$q(\exp(s\w \cdot \B) \cdot o) =\exp(s\w\cdot\B) \cdot q(o)$, so the
orbit of $o \in M$ under the boost is sent by $q$ to the orbit of
$q(o) \in \Mbar$. Since $q$ is continuous it sends compact sets to
compact sets, so if the orbit of $q(o) \in \Mbar$ is \emph{not}
compact then neither is the orbit of $o \in M$. For $M$ one of
$\hyperlink{S2}{\zdS}$, $\hyperlink{S3}{\zAdS}$,
$\hyperlink{S14}{\zdSC}$, or $\hyperlink{S16}{\zLC}$, there is some
covering $q : M \to \Mbar$ such that we can equivariantly embed
$\Mbar$ as a hypersurface in some pseudo-euclidean space where $\Kgr$
acts linearly. It is a simple matter to work out the nature of the
orbits of the boosts in the ambient pseudo-euclidean space (and hence
on $\Mbar$), with the caveat that what is a boost in $\Mbar$ need not
be a boost in the ambient space. Having shown that the boost orbit is
non-compact on $\Mbar$ we deduce that the orbit is non-compact on $M$.
We will show in this way that the generic boost orbits are non-compact
for $\hyperlink{S2}{\zdS}$, $\hyperlink{S3}{\zAdS}$,
$\hyperlink{S14}{\zdSC}$, and $\hyperlink{S16}{\zLC}$.

Finally, this still leaves the torsional galilean spacetimes
$\hyperlink{S9}{\ztdSG_\gamma}$, $\hyperlink{S11}{\ztAdSG_\chi}$ and
$\hyperlink{S12}{\text{\twodgal}_{\gamma,\chi}}$, which require a
different argument to be explained when we discuss these spacetimes in
Section~\ref{sec:action-boosts-2}.

\subsection{Invariant connections}
\label{sec:invar-conn}

There is only one non-reductive homogeneous spacetime in
Table~\ref{tab:spacetimes} and \ref{tab:aristotelian}, namely
$\hyperlink{S16}{\zLC}$, and its invariant connections were already
determined in \cite{Figueroa-OFarrill:2018ilb}. There it is shown the
light cone for $D\geq 2$ admits no invariant connections, whereas for
$D=1$ there is a three-parameter family of invariant connections and a
unique torsion-free, flat connection. We will, therefore, restrict
ourselves to the remaining reductive homogeneous spaces in this
section.

Let $(\k,\h)$ be a Lie pair associated to a reductive homogeneous
space. We assume that $(\k,\h)$ is effective so that $\h$ does not
contain any non-zero ideals of $\k$. We let $\k = \h \oplus \m$ denote
a reductive split, where $[\h,\m] \subset \m$. This split makes $\m$ into an
$\h$-module relative to the \textbf{linear isotropy representation}
$\lambda: \h \to \gl(\m)$, where
\begin{equation}
  \lambda_X Y = [X,Y] \qquad\text{for all $X\in\h$ and $Y\in\m$.}
\end{equation}

As shown in \cite{MR0059050}, one can uniquely characterise 
the invariant affine connections on $(\k,\h)$ by their \textbf{Nomizu map}
$\alpha: \m \times \m \to \m$, an $\h$-equivariant bilinear map; that
is, such that for all $X \in \h$ and $Y,Z \in \m$,
\begin{equation} \label{eq:Nomizuinv}
  [X,\alpha(Y,Z)] = \alpha([X,Y],Z) + \alpha(Y,[X,Z]).
\end{equation}
The torsion and curvature of an invariant affine connection with
Nomizu map $\alpha$ are given, respectively, by the following
expressions for all $X,Y,Z \in \m$,
\begin{equation} \label{eq:reductive_tor_and_curv}
  \begin{split}
    \Theta(X,Y) &= \alpha(X,Y) - \alpha(Y,X) - [X,Y]_\m,\\
    \Omega(X,Y) Z &= \alpha(X,\alpha(Y,Z)) - \alpha(Y,\alpha(X,Z)) -
    \alpha([X,Y]_\m,Z) - [[X,Y]_\h, Z],
  \end{split}
\end{equation}
where $[X,Y] = [X,Y]_\h + [X,Y]_\m$ is the decomposition of $[X,Y] \in
\k = \h \oplus \m$.  In particular, for the canonical invariant
connection with zero Nomizu map, we have
\begin{equation} \label{eq:tor_and_curv_can}
  \Theta(X,Y) = - [X,Y]_\m \qquad\text{and}\qquad
  \Omega(X,Y) Z = - \lambda_{[X,Y]_\h} Z.
\end{equation}

For kinematical homogeneous spacetimes, we can determine the possible
Nomizu maps in a rather uniform way. Rotational invariance determines
the form of the Nomizu map up to a few parameters and then we need
only study the action of the boosts. From Table~\ref{tab:spacetimes}
it is clear that the action of the boosts is common to all spacetimes
within a given class: lorentzian, riemannian, galilean, and
carrollian; although the curvature and torsion of the invariant
connections of course do depend on the spacetime in question.

\subsection{The soldering form and the canonical connection}
\label{sec:sold-invar-conn}

Recall that on the Lie group $\Kgr$ there is a left-invariant
$\k$-valued one-form $\vartheta$: the (left-invariant)
Maurer--Cartan one-form.  It obeys the structure equation
\begin{equation}
  \label{eq:MCSE}
  d\vartheta = -\tfrac12 [\vartheta,\vartheta],
\end{equation}
where the notation hides the wedge product in the right-hand side.
Using exponential coordinates, we can pull back $\vartheta$ to a
neighbourhood of the origin on $M$. The following formula, which
follows from equation \eqref{eq:d-exp}, shows how to calculate it:
\begin{equation}
  \label{eq:MC-pullback}
  \sigma^* \vartheta = D(\ad_A) (dt H + d\x \cdot \P),
\end{equation}
where, as before, $A = tH + \x\cdot \P$ and $D$ is the Maclaurin series
corresponding to the analytic function in \eqref{eq:function-d}.

The pull-back $\sigma^* \vartheta$ is a one-form defined near the
origin on $M$ with values in the Lie algebra $\k$.  Let $\m$
be a vector space complement to $\h$ in $\k$ so that as a vector
space $\k = \h \oplus \m$.  This split allows us to write
\begin{equation}
  \sigma^* \vartheta = \theta + \omega~,
\end{equation}
where $\theta$ is $\m$-valued and and $\omega$ is $\h$-valued. If the
Lie pair $(\k,\h)$ is reductive and $\m$ is chosen to be an
$\h$-submodule of $\k$, then $\omega$ is the one-form corresponding to
the \textbf{canonical invariant connection} on $M$. The
\textbf{soldering form} is then given by $\theta$.

The torsion and curvature of $\omega$ are easy to calculate using the
fact that $\vartheta$ obeys the Maurer--Cartan structure
equation~\eqref{eq:MCSE}.\footnote{%
  Let us emphasise that in this work, curvature always refer to the
  curvature of an invariant affine connection and hence should not be
  confused with the curvature of the associated Cartan connection,
  which is always flat for the homogeneous spaces (also called Klein
  geometries in that context).}
Indeed, the torsion two-form $\Theta$ is given by
\begin{equation}
  \label{eq:torsion}
  \Theta = d\theta + [\omega,\theta] = - \tfrac12 [\theta,\theta]_\m
\end{equation}
and the curvature two-form $\Omega$ by
\begin{equation}
  \label{eq:curvature}
  \Omega = d\omega + \tfrac12 [\omega,\omega] = - \tfrac12 [\theta,\theta]_\h,
\end{equation}
which agree with the expressions in
equation~\eqref{eq:tor_and_curv_can}.

In the non-reductive case $\omega$ does not define a connection, but
we may still project the locally defined $\k$-valued one-form
$\sigma^*\vartheta$ to $\k/\h$. The resulting local one-form $\theta$
with values in $\k/\h$ is a soldering form which defines an
isomorphism $T_pM \to \k/\h$ for every $p\in M$ near the
origin. Wherever $\theta$ is invertible, the exponential coordinates
define an immersion, which may however fail to be an embedding or
indeed even injective.  In practice, it is not easy to determine
injectivity, but it is easy to determine where $\theta$ is invertible
by calculating the top exterior power of $\theta$ and checking that it
is non-zero. Provided that $\theta$ is invertible, the inverse
isomorphism is the vielbein $E$, where $E(p): \k/\h \to T_pM$ for
every $p \in M$ near the origin. The vielbein allows us to transport
tensors on $\k/\h$ to tensor fields on $M$ and, as we now recall, it
takes $\Hgr$-invariant tensors on $\k/\h$ to $\Kgr$-invariant tensor
fields on $M$.

\subsection{Invariant tensors}
\label{sec:invariant-tensors}

It is well-known that $\Kgr$-invariant tensor fields on $M=\Kgr/\Hgr$
are in one-to-one correspondence with $\Hgr$-invariant tensors on
$\k/\h$ and if $\Hgr$ is connected, with $\h$-invariant tensors on
$\k/\h$.  We may assume that $\Hgr$ is indeed connected, passing 
to the universal cover of $M$, if necessary.  In practice, given an
$(r,s)$-tensor $T$ on $\k/\h$---that is, an element of
$(\k/\h)^{\otimes r} \otimes ((\k/\h)^*)^{\otimes s}$---we can turn it
into an $(r,s)$-tensor field $\eT $ on $M$ by contracting with
soldering forms and vielbeins as appropriate to arrive, for every $p
\in M$, to $\eT(p) \in (T_pM)^{\otimes r} \otimes (T^*_pM)^{\otimes
  s}$. Moreover, if $T$ is $\Hgr$-invariant, $\eT$ is
$\Kgr$-invariant.

Our choice of basis for $\k$ is such that $\J$ and $\B$ span $\h$ 
and therefore $\Pbar := \P \mod \h$ and $\Hbar := H \mod \h$ span
$\k/\h$.  In the reductive case, $\k = \h \oplus \m$ and
$\m \cong \k/\h$ as $\h$-modules.  We will let $\eta$ and $\pi^a$
denote the canonical dual basis for $(\k/\h)^*$.

Invariant non-degenerate metrics are in one-to-one correspondence with
$\h$-invariant non-degenerate symmetric bilinear forms on $\k/\h$ and
characterise, depending on their signature, \textbf{lorentzian} or
\textbf{riemannian} spacetimes. On the other hand, invariant
\textbf{galilean} structures\footnote{%
  We will not distinguish notationally the $\Hgr$-invariant tensor
  from the $\Kgr$-invariant tensor field.}  consist of a pair
$(\tau,h)$, where $\tau \in (\k/\h)^*$ and $h \in S^2(\k/\h)$ are
$\h$-invariant, $h$ has co-rank $1$ and $h(\tau,-) = 0$, if we think
of $h$ as a symmetric bilinear form on $(\k/\h)^*$. On $M$, $\tau$
gives rise to an invariant clock one-form and $h$ to an invariant
spatial metric on one-forms. \textbf{Carrollian} structures are dual
to galilean structures and consist of a pair $(\kappa,b)$, where
$\kappa \in \k/\h$ defines an invariant vector field and
$b \in S^2(\k/\h)^*$ is an invariant symmetric bilinear form of
co-rank $1$ and such that $b(\kappa,-) = 0$.  Homogeneous
\textbf{aristotelian} spacetimes admit an invariant galilean structure
and an invariant carrollian structure simultaneously.

Invariance under $\h$ implies, in particular, invariance under the
rotational subalgebra, which is non-trivial for $D \geq 2$.  Assuming
that $D\geq 2$ for now, it is easy to write down the possible
rotationally invariant tensors and therefore we need only check
invariance under $\B$.  The action of $\B$ is induced by duality from
the action on $\k/\h$ which is given by
\begin{equation}
  \lambda_{B_a} (\Hbar) = \overline{[B_a,\H]} \qquad\text{and}\qquad
  \lambda_{B_a} (\overline{P}_b) = \overline{[B_a, P_b]},
\end{equation}
with the brackets being those of $\k$. In practice, we can determine
this from the explicit expression of the Lie brackets by computing the
brackets in $\k$ and simply dropping any $\B$ or $\J$ from the
right-hand side. The only possible invariants in $\k/\h$ are
proportional to $H$, which is invariant provided that
$[\B,H] = 0 \mod \h$. Dually, the only possible invariants in
$(\k/\h)^*$ are proportional to $\eta$, which is invariant provided
that there is no $X \in \k$ such that $H$ appears in $[\B,X]$.
Omitting the tensor product symbol, the only rotational invariants in
$S^2(\k/\h)$ are linear combinations of $H^2$ and $P^2 := \delta^{ab}
P_a P_b$, whereas in $S^2(\k/\h)^*$ are $\eta^2$ and $\pi^2 =
\delta_{ab}\pi^a\pi^b$.

In $D=1$ there are no rotations, so we need only concern ourselves with
the action of $\B$.  Possible invariants in $\k/\h$ are linear
combinations of $H$ and $\P$, whereas in $(\k/\h)^*$ they are linear
combinations of $\eta$ and $\pi$.  Similarly in the space of symmetric
tensors, we can have now linear combinations of $H^2$, $H\P$, and $\P^2$
in $S^2(\k/\h)$ and of $\eta^2$, $\eta\pi$, and $\pi^2$ in
$S^2(\k/\h)^*$.  These are again easy to determine from the Lie
bracket.

\section{Invariant connections, curvature, and torsion for reductive spacetimes}
\label{sec:nomizu}

In this section we determine the invariant affine connections for the
reductive spacetimes in Tables~\ref{tab:spacetimes} and
\ref{tab:aristotelian}. This is equivalent to determining the space of
Nomizu maps which, as explained above, can be done uniformly, a class
at a time.  We also calculate the curvature and torsion of the
invariant connections.

For reductive homogeneous spaces there always exists, besides the
canonical connection with vanishing Nomizu map, another interesting
connection.  It is given by the torsion-free connection
defined\footnote{%
  It is the unique Nomizu map with $\alpha(X,X)=0$ for all $X \in \m$
  and vanishing torsion and called ``canonical affine connection of
  the first kind'' in~\cite{MR0059050}.}  by
$\alpha(X,Y)= \frac{1}{2} [X,Y]_{\m}$.  The canonical and the natural
torsion-free connections have the same geodesics and, as one can easily
observe below, the connections coincide for symmetric spaces.

For any spacetime the Nomizu maps needs to be rotationally invariant
which gives us
\begin{equation}\label{eq:nomizu-rotations}
  \begin{aligned}[m]
    \alpha(H,H) &=
    \begin{cases}
      \mu H & D>1\\
      \mu H + \mu' P & D=1
    \end{cases}\\
    \alpha(P_a,P_b) &=
  \begin{cases}
    \zeta \delta_{ab} H & D>3\\
    \zeta \delta_{ab} H + \zeta' \epsilon_{abc} P_c & D=3\\
    \zeta \delta_{ab} H + \zeta' \epsilon_{ab} H & D=2\\
    \zeta H + \zeta' P & D=1
  \end{cases}
  \end{aligned}
  \qquad\qquad
  \begin{aligned}[m]
  \alpha(H,P_a) &=
  \begin{cases}
    \nu P_a & D>2\\
    \nu P_a + \nu' \epsilon_{ab} P_b & D=2\\
    \nu P + \nu' H & D=1
  \end{cases}\\
  \alpha(P_a,H) &=
  \begin{cases}
    \xi P_a & D>2\\
    \xi P_a + \xi' \epsilon_{ab} P_b & D=2\\
    \xi P + \xi' H & D=1,
  \end{cases}
  \end{aligned}
\end{equation}
for some real parameters $\mu,\mu',\nu,\nu',\zeta,\zeta',\xi,\xi'$.
Now we simply impose invariance under $B_a$.

\subsection{Nomizu maps for lorentzian spacetimes}
\label{sec:nomizu-maps-lorentz}

The lorentzian spacetimes in Table~\ref{tab:spacetimes} all share the
same action of the boosts:
\begin{equation}
  \lambda_{B_a} H = P_a \qquad\text{and}\qquad \lambda_{B_a} P_b =
  \delta_{ab} H.
\end{equation}
  We will impose invariance explicitly in this case to illustrate the
calculation and only state the results in all other cases.

\subsubsection{$D\geq 4$}
\label{sec:dgeq-4-lor}

We calculate (remember \eqref{eq:Nomizuinv})
\begin{equation}
  (\lambda_{B_c}\alpha)(P_a,P_b) = \zeta \delta_{ab} P_c - \nu
  \delta_{ac} P_b - \xi \delta_{bc} P_a,
\end{equation}
whose vanishing requires $\zeta = \nu = \xi = 0$, as can be seen by
considering $a=b\neq c$, $a=c \neq b$, and $b=c\neq a$ in turn.
Finally,
\begin{equation}
  (\lambda_{B_c}\alpha)(H,H) = \mu P_c,
\end{equation}
whose vanishing imposes $\mu = 0$ and hence the only invariant Nomizu
map is the zero map.

\subsubsection{$D=3$}
\label{sec:deq-3-lor}

The only change here is an additional term $\zeta'\epsilon_{abc}P_c$ in
$\alpha(P_a,P_b)$.  This results in
\begin{equation}
  (\lambda_{B_c}\alpha)(P_a,P_b) = \zeta \delta_{ab} P_c + \zeta'
  \epsilon_{abc} H - \nu \delta_{ac} P_b - \xi \delta_{bc} P_a,
\end{equation}
whose vanishing again requires $\zeta = \zeta'= \nu = \xi = 0$.  Hence
continuing as in $D\geq 4$, we find that the only invariant Nomizu map
is the zero map.

\subsubsection{$D=2$}
\label{sec:deq-2-lor}

The $2+1$ dimensional case differs with respect to its higher
dimensional counterparts. We start by calculating
\begin{equation}
  (\lambda_{B_c}\alpha)(P_a,P_b)
  = \zeta \delta_{ab} P_c + \zeta' \epsilon_{ab} P_c
  - \delta_{ac} (\nu P_b + \nu' \epsilon_{bd} P_d)
  - \delta_{bc} ( \xi  P_a + \xi'  \epsilon_{ad} P_d).
\end{equation}
Considering $a = b \neq c$ requires that $\zeta=0$. Next we set
$a = c \neq b$, which leads us to $\nu=0$ and $\zeta'=-\nu'$.
Similarly, $b = c \neq a$ imposes $\xi=0$ and $\zeta'=\xi'$ and leaves
us, for now, with $t=\zeta'=\xi'=-\nu'$. We need to check if the
remaining components of the Nomizu map are also invariant, e.g.,
\begin{equation}
  (\lambda_{B_c}\alpha)(H,H) = \mu P_c
\end{equation}
vanishes if and only if $\mu = 0$, while $\alpha(P_{a},H)$ and
$\alpha(H, P_{a})$ are invariant without further ado.

In summary, we get a one-parameter family of Nomizu maps, parametrised
by $t \in \RR$,
\begin{align}
  \alpha(P_{a},P_{b}) & = t \epsilon_{ab} H & \alpha(H,P_{a}) = - t \epsilon_{ab}P_{b} &&  \alpha(P_{a},H) &=  t \epsilon_{ab}P_{b}  \, .
\end{align}
It can be written in a more compact way using lorentzian $2+1$
dimensional tensors,
$\alpha(P_{\mu},P_{\nu})= \tilde t \epsilon_{\mu\nu\rho}
\eta^{\rho\sigma} P_{\sigma}$.

\subsubsection{$D=1$}
\label{sec:deq-1-lor}

Here we notice that $\lambda_B$ is the identity on $\m$, hence minus
the identity on $\m^*$.  Therefore, by parity, there are no zero
eigenvalues in $\m^*\otimes \m^* \otimes \m$ and hence no invariants
but the zero Nomizu map.

In summary, lorentzian homogeneous spacetimes have, with the exception
$D=2$, of a unique invariant connection given by the canonical
connection. As we will see in the next sections, there is more freedom
for galilean and carrollian spacetimes. However since we start with an
unique (vanishing) Nomizu map, only this vanishing case arises also as
a limit. The additional invariant non-relativistic and
ultra-relativistic connections can be seen as an intrinsic property
that does not originate from the relativistic spacetimes.

\subsection{Nomizu maps for riemannian spacetimes}
\label{sec:nomizu-maps-riemann}

The situation here is very similar to the lorentzian case.  Now the
boosts act as
\begin{equation}
  \lambda_{B_a} H = - P_a \qquad\text{and}\qquad \lambda_{B_a} P_b =
  \delta_{ab} H.
\end{equation}

The results are as in the lorentzian case: the only invariant
connection is the canonical connection, except in $D=2$ where there is
a one-parameter family.

\subsection{Nomizu maps for galilean spacetimes}
\label{sec:nomizu-maps-galileo}

On a galilean spacetime, the boosts act as
\begin{equation}
  \lambda_{B_a} H = P_a
\end{equation}
and the $P_a$ are invariant.  This results in the following invariant
Nomizu maps:
\begin{equation}\label{eq:nomizu-galilean}
  \begin{aligned}[m]
    \alpha(H,H) &=
    \begin{cases}
      (\nu + \xi) H & D>1\\
      (\nu + \xi) H + \mu' P & D=1
    \end{cases}\\
    \alpha(P_a,P_b) &= 0 
  \end{aligned}
  \qquad\qquad
  \begin{aligned}[m]
    \alpha(H,P_a) &=
    \begin{cases}
      \nu P_a & D\neq 2\\
      \nu P_a + \nu' \epsilon_{ab} P_b & D=2
    \end{cases}\\
    \alpha(P_a,H) &=
    \begin{cases}
      \xi P_a & D\neq 2\\
      \xi P_a - \nu' \epsilon_{ab} P_b & D=2.
    \end{cases}
  \end{aligned}
\end{equation}
We will now analyse the curvature and torsion for these Nomizu maps
for each galilean spacetime.

\subsubsection{Galilean spacetime ($\zG$)}

For $D\geq 3$, the torsion and curvature of the resulting connection
have the following non-zero components:
\begin{equation}
  \Theta(H,P_a) = (\nu - \xi) P_a \qquad\text{and}\qquad \Omega(H,P_a)H = -
  \xi^2 P_a.
\end{equation}
There is a unique torsion-free, flat invariant connection
corresponding to the canonical connection with $\nu = \xi = 0$.

For $D=2$, the torsion and curvature are given by the following non-zero
components:
\begin{equation}
  \Theta(H,P_a) = (\nu - \xi) P_a + 2 \nu' \epsilon_{ab} P_b
  \qquad\text{and}\qquad
  \Omega(H,P_a)H = (\nu'^2-\xi^2) P_a + 2 \nu' \xi \epsilon_{ab} P_b,
\end{equation}
so that again the canonical connection is the unique torsion-free,
flat invariant connection.

Finally, for $D=1$, torsion and curvature are given by
\begin{equation}
  \Theta(H,P) = (\nu - \xi) P \qquad\text{and}\qquad \Omega(H,P)H =
  -\xi^2 P.
\end{equation}
Since neither depend on $\mu'$, we now have a one-parameter family of
torsion-free, flat invariant connections, defined by the Nomizu map
\begin{equation}
  \alpha(H,H) = \mu' P.
\end{equation}

\subsubsection{Galilean de~Sitter spacetime ($\zdSG$)}
\label{sec:galilean-de}

Let $D\geq 3$.  The torsion and curvature, given by
equation~\eqref{eq:reductive_tor_and_curv}, have the following non-vanishing
components:
\begin{equation}
  \Theta(H, P_a) = (\nu-\xi)P_a \qquad \text{and} \qquad \Omega(H, P_{a})H = (1-\xi^2)P_a.
\end{equation}
Therefore, there are two torsion-free, flat invariant connections
corresponding to $\nu=\xi=\pm 1$.  The Nomizu maps for these two
connections are
\begin{equation}
  \begin{aligned}[m]
    \alpha (H, H) &= 2 H \\
    \alpha (H, P_{a}) &= P_{a} \\
    \alpha(P_{a}, H) &= P_a
  \end{aligned} \qquad \text{and} \qquad
  \begin{aligned}[m]
    \alpha (H, H) &= -2 H \\
    \alpha (H, P_a) &= - P_a \\
    \alpha (P_a, H) &= -P_a.
  \end{aligned}
\end{equation}

In $D=2$, the vector space of Nomizu maps is three-dimensional and the
non-vanishing curvature and torsion components in this dimension are
\begin{equation}
\Theta(H, P_a) = (\nu-\xi)P_a + 2\nu' \epsilon_{ab} P_b \qquad
\text{and} \qquad \Omega(H, P_{a})H = (\nu'^2 - \xi^2 + 1) P_a +
2\xi\nu'\epsilon_{ab}P_{b}.
\end{equation}
Again, there are two torsion-free, flat invariant connection
corresponding to $\nu=\xi= \pm 1$.

Finally, let $D=1$. The non-vanishing torsion and curvature components
are
\begin{equation}
\Theta(H, P) = (\nu - \xi)P \qquad \text{and} \qquad \Omega(H, P)H = (1 -\xi^{2})P.
\end{equation}
The torsion-free, flat connections are once again given by
$\nu=\xi=\pm 1$, but now there is a free parameter $\mu'$.

\subsubsection{Galilean anti de~Sitter spacetime ($\zAdSG$)}
\label{sec:galilean-anti-de}

The torsion and curvature have have the following non-zero components:
\begin{equation}
  \Theta(H,P_a) =
  \begin{cases}
    (\nu-\xi) P_a & D\neq 2\\
    (\nu-\xi) P_a + 2 \nu' \epsilon_{ab} P_b & D=2
  \end{cases}
\end{equation}
and
\begin{equation}
  \Omega(H, P_a) H = 
  \begin{cases}
    - (1+\xi^{2})P_{a} & D \neq 2\\
    -(1 + \xi^2 - \nu'^2) P_a + 2 \xi \nu' \epsilon_{ab} P_b & D=2.
  \end{cases}
\end{equation}
There are torsion-free connections, but none are flat.

\subsubsection{Torsional galilean de~Sitter spacetime ($\ztdSG_{\gamma=1}$)}
\label{sec:tdsg1}

Let $D\geq 3$. The torsion has the following non-zero components
\begin{equation}
  \Theta(H,P_a) = (\nu - \xi - 2 )P_a,
\end{equation}
whereas the only non-zero component of the curvature is
\begin{equation}
  \Omega(H,P_a)H = -(1 + \xi)^2 P_a.
\end{equation}
Therefore, there exists a unique invariant connection with zero torsion
and curvature corresponding to $\nu=1$ and $\xi = -1$:
\begin{equation}
  \alpha(H,P_a) = P_a \qquad\text{and}\qquad \alpha(P_a,H) = - P_a.
\end{equation}

If $D=2$, we have an additional parameter in our family of invariant
affine connections:
\begin{equation}
  \alpha(H,H) = (\nu + \xi) H, \qquad \alpha(H,P_a) = \nu P_a + \nu'
  \epsilon_{ab} P_b,  \qquad\text{and}\qquad
  \alpha(P_a,H) = \xi P_a - \nu' \epsilon_{ab} P_b.
\end{equation}
The only non-zero component of the torsion is
\begin{equation}
  \Theta(H,P_a) = (\nu- \xi - 2) P_a + 2 \nu' \epsilon_{ab} P_b,
\end{equation}
and the only non-zero component of the curvature is
\begin{equation}
  \Omega(H,P_a)H = ((\nu')^2 - (1+\xi)^2) P_a + 2 \nu' (1+\xi) \epsilon_{ab} P_b.
\end{equation}
We see that there is a unique torsion-free, flat invariant connection with Nomizu map
\begin{equation}
  \alpha(H,P_a) = P_a \qquad\text{and}\qquad \alpha(P_a,H) = - P_a.
\end{equation}

Finally, in $D=1$ we have a three-parameter family of Nomizu maps:
\begin{equation}
  \alpha(H,H) = (\nu + \xi) H + \mu' P, \qquad 
  \alpha(H,P) = \nu P, \qquad\text{and}\qquad \alpha(P,H) = \xi P.
\end{equation}
The torsion is given by
\begin{equation}
  \Theta(H,P) = (\nu - \xi - 2) P,
\end{equation}
and the curvature by
\begin{equation}
  \Omega(H,P)H = -(1+\xi)^2 P.
\end{equation}
Imposing zero torsion and zero curvature still leaves a one-parameter
family of invariant connections with Nomizu map
\begin{equation}
  \alpha(H,H) = \mu' P, \qquad \alpha(H,P) = P,
  \qquad\text{and}\qquad \alpha(P,H) = - P.
\end{equation}

\subsubsection{Torsional galilean de~Sitter spacetime ($\ztdSG_{\gamma\neq 1}$)}
\label{sec:other-invar-conn-tdsg}

For $D\geq 3$, the torsion is given by
\begin{equation}
  \Theta(H,P_a) = (\nu - \xi - (1 + \gamma)) P_a
\end{equation}
and the curvature by
\begin{equation}
  \Omega(H,P_a)H = -(\xi + 1) (\xi + \gamma) P_a.
\end{equation}
Therefore, there are precisely two torsion-free, flat invariant
connections, with Nomizu maps
\begin{equation}
  \label{eq:tff-2-gamma-hd}
  \begin{aligned}[m]
    \alpha(H,H) &= (\gamma - 1) H\\
    \alpha(H,P_a) &= \gamma P_a\\
    \alpha(P_a,H) &= -P_a
  \end{aligned}
  \qquad\qquad\text{and}\qquad\qquad
  \begin{aligned}[m]
    \alpha(H,H) &= (1-\gamma) H\\
    \alpha(H,P_a) &= P_a\\
    \alpha(P_a,H) &= -\gamma P_a.
  \end{aligned}
\end{equation}

If $D=2$, then there is a three-parameter family of invariant
connections with torsion and curvature that have the following non-zero
components:
\begin{equation}
  \begin{split}
    \Theta(H,P_a) &= (\nu - \xi - (1+\gamma))P_a + 2 \nu' \epsilon_{ab} P_b\\
    \Omega(H,P_a)H &= (\nu'^2 - (\xi + 1)(\xi + \gamma))P_a + \nu' (2\xi
    + 1 + \gamma) \epsilon_{ab} P_b.
  \end{split}
\end{equation}
There are precisely two torsion-free, flat invariant connections,
whose Nomizu maps are identical to those for $D\geq 3$ in
equation~\eqref{eq:tff-2-gamma-hd}.

In $D=1$, the torsion and curvature have the following non-zero components:
\begin{equation}
  \Theta(H,P) = (\nu - \xi - (1+\gamma)) P \qquad\text{and}\qquad \Omega(H,P)H =
  - (\xi + 1)(\xi + \gamma) P.
\end{equation}
There are two one-parameter families of torsion-free, flat invariant
connections.  They have Nomizu maps
\begin{equation}
  \begin{aligned}[m]
    \alpha(H,H) &= (\gamma - 1) H + \mu' P\\
    \alpha(H,P) &= \gamma P\\
    \alpha(P,H) &= -P
  \end{aligned}
  \qquad\qquad\text{and}\qquad\qquad
  \begin{aligned}[m]
    \alpha(H,H) &= (1-\gamma) H + \mu' P\\
    \alpha(H,P) &= P\\
    \alpha(P,H) &= -\gamma P.
  \end{aligned}
\end{equation}

\subsubsection{Torsional galilean anti de~Sitter spacetime ($\ztAdSG_\chi$)}
\label{sec:other-invar-conn-tadsg}

The torsion and curvature of the connection corresponding to this
Nomizu map in $D\geq 3$ are given by the following non-zero components:
\begin{equation}
  \Theta(H,P_a) = (\nu - \xi - 2\chi) P_a \qquad\text{and}\qquad \Omega(H,P_a)H
  = - (1 + (\xi + \chi)^2) P_a.
\end{equation}
Therefore, we see that there are no flat invariant connections;
although there is a one-parameter family of torsion-free invariant
connections.

For $D=2$, we have a three-parameter family of invariant connections
for which the torsion and curvature are given by the following non-zero
components:
\begin{equation}
  \Theta(H,P_a) = (\nu - \xi - 2\chi) P_a + 2 \nu' \epsilon_{ab} P_b \qquad\text{and}\qquad \Omega(H,P_a)H
  = \left((v')^2 - (\xi + \chi)^2 - 1\right) P_a + 2 \nu' (\xi +\chi) \epsilon_{ab} P_b.
\end{equation}
Again, there are no flat invariant connections, but there is a
two-parameter family of torsion-free invariant connections.

Let $D=1$.  We calculate the torsion and curvature to be
\begin{equation}
\Theta(H, P) = (\nu - \xi - 2\chi)P \qquad \text{and}\qquad \Omega(H, P)H = -(1+(\xi+\chi)^2) P, 
\end{equation}
respectively. As in higher dimensions, we thus find there to be no
flat invariant connections. There is, however, a two-parameter family
of torsion-free invariant connections.

\subsubsection{Spacetime $\text{\twodgal}_{\gamma,\chi}$}
\label{sec:other-invar-conn-twodgal}

Since this spacetime is particular to $D=2$ and reductive, we need only 
consider the $D=2$ case of \eqref{eq:nomizu-galilean} and we may use
equation~\eqref{eq:reductive_tor_and_curv} to obtain the following
torsion and curvature
\begin{equation}
  \begin{split}
    \Theta(H, P_a) &= (\nu - \xi - (1+\gamma))P_a + (2\nu' + \chi)\epsilon_{ab} P_b \\
    \Omega(H, P_a)H &= (\nu'(\nu' + \chi) - (\xi+1)(\xi+\gamma))P_a + (2\nu'\xi + (1+\gamma)\nu' + (1+\xi)\chi)\epsilon_{ab}P_b.
  \end{split}
\end{equation}
For the torsion to vanish we need $\nu' = -\chi/2$ and $\nu - \xi = 1 +
\gamma$.   If, in addition, the curvature were to vanish we would find
\begin{equation}
    0 = 2\nu'\xi + (1+\gamma)\nu' + (1+\xi)\chi = - \tfrac12 (\gamma -1)\chi.
\end{equation}
Hence torsion-free, flat invariant connections require either $\gamma=1$ or
$\chi = 0$.  Both of these values lie outside the range of their corresponding 
parameter.  From the vanishing of the $P_a$ term in the curvature, 
we see that $\chi= 0$ is necessary, which agrees with the previous 
results: torsional galilean de~Sitter spacetimes ($\hyperlink{S9}{\ztdSG_\gamma}$) 
admit torsion-free, flat invariant connections, but torsional galilean anti de~Sitter 
spacetimes ($\hyperlink{S11}{\ztAdSG_\chi}$) do not (unless $\chi = 0$).

\subsection{Nomizu maps for carrollian spacetimes}
\label{sec:nomizu-maps-carroll}

On a carrollian spacetime, the boosts act as
\begin{equation}
  \lambda_{B_a} P_b = \delta_{ab} H,
\end{equation}
and $H$ is invariant.  This results in the following invariant
Nomizu maps:
\begin{equation}\label{eq:nomizu-carrollian}
  \begin{aligned}[m]
    \alpha(H,H) &= 0\\
    \alpha(P_a,P_b) &=
    \begin{cases}
      \zeta \delta_{ab} H & D\geq 3\\
      \zeta \delta_{ab} H + \zeta' \epsilon_{ab} H & D=2\\
      \zeta H + (\nu'+ \xi') P & D=1
    \end{cases}
  \end{aligned}
  \qquad\qquad
  \begin{aligned}[m]
    \alpha(H,P_a) &=
    \begin{cases}
      0 & D\geq 2\\
      \nu' H & D=1
    \end{cases}\\
    \alpha(P_a,H) &=
    \begin{cases}
      0 & D\geq 2\\
      \xi' H & D=1.
    \end{cases}
  \end{aligned}
\end{equation}

\subsubsection{Carrollian spacetimes ($\zC$)}

For $D\geq 3$, the corresponding invariant connections are flat and torsion-free for
all values of $\zeta$.  

Letting $D=2$, we find the following non-vanishing torsion component
\begin{equation}
\Theta(P_a, P_b) = 2 \zeta' \epsilon_{ab} H.
\end{equation}
We, therefore, have the same torsion-free, flat invariant connections 
that were found in higher dimensions.

For $D=1$, the torsion and curvature are easily calculated to be
\begin{equation}
\Theta(H, P) = (\nu' - \xi') H \qquad \Omega(H, P)P = (\nu')^2 H.
\end{equation}
We thus find a one-parameter family of torsion-free, flat invariant
connections, as in higher dimensions:
\begin{equation}
  \alpha(P,P) = \zeta H.
\end{equation}

\subsubsection{(Anti) de~Sitter carrollian spacetimes ($\zdSC$ and $\zAdSC$)}

We will treat these two spacetimes together by introducing
$\varkappa=\pm 1$. Carrollian de~Sitter spacetime ($\hyperlink{S14}{\zdSC}$)
corresponds to $\varkappa=1$ and carrollian anti de~Sitter spacetime
($\hyperlink{S15}{\zAdSC}$) to $\varkappa =-1$.

If $D\geq 3$, the torsion vanishes and the curvature has the following
non-zero components:
\begin{equation}
  \Omega(H,P_a) P_b = \varkappa \delta_{ab} H \qquad\text{and}\qquad
  \Omega(P_a, P_b) P_c = \varkappa (\delta_{bc} P_a - \delta_{ac} P_b),
\end{equation}
which is never flat. Both of these results are independent of the
Nomizu map.

If $D=2$, the non-zero components of the torsion and curvature are given
by
\begin{equation}
  \begin{split}
    \Theta(P_a,P_b) &= 2 \zeta' \epsilon_{ab} H,\\
    \Omega(H,P_a) P_b &= \epsilon \delta_{ab} H, \text{and},\\
    \Omega(P_a, P_b) P_c &= \epsilon (\delta_{bc} P_a - \delta_{ac} P_b).
  \end{split}
\end{equation}
It is torsion-free if $\zeta' = 0$, but it is never flat.

Finally, if $D=1$, then the non-zero components of the torsion and
curvature are
\begin{equation}
  \begin{split}
    \Theta(H,P) &= (\nu' - \xi') H\\
    \Omega(H,P) P &= (\varkappa + \nu'^2) H,
  \end{split}
\end{equation}
which is never flat if $\varkappa = 1$ ($\zdSC\cong\zAdSG$), but if
$\varkappa=-1$ ($\zAdSC \cong \zdSG$) then we can take $\nu' = \xi'
= \pm 1$, to yield two one-parameter families of torsion-free, flat
connections with Nomizu maps:
\begin{equation}
  \begin{aligned}[m]
    \alpha(H,P) &= H\\
    \alpha(P,H) &= H\\
    \alpha(P,P) &= \zeta H + 2 P
  \end{aligned}
  \qquad\text{and}\qquad
  \begin{aligned}[m]
    \alpha(H,P) &= -H\\
    \alpha(P,H) &= -H\\
    \alpha(P,P) &= \zeta H - 2 P.
  \end{aligned}
\end{equation}

\subsubsection{Carrollian light cone ($\zLC$)}
\label{sec:other-invar-conn-flc}

As show in \cite{Figueroa-OFarrill:2018ilb}, this homogeneous
spacetime does not admit any invariant connections for $D\geq 2$. For
$D=1$, there is a three-parameter family of invariant connections and
a unique torsion-free, flat invariant connection.

\subsection{Nomizu maps for exotic two-dimensional spacetimes}
\label{sec:nomizu-maps-exotic}

In the bottom section of Table~\ref{tab:spacetimes} there are exotic
two-dimensional reductive spacetimes with no discernible structure,
and we must study their Nomizu maps separately. We can distinguish the
four types of spacetime by the action of $\lambda_B$ on the
two-dimensional space $\m$ spanned by $P$ and $H$.

In the case of spacetime \hyperlink{S17}{\xone}, $\lambda_B$ is not
diagonalisable.  Therefore, one needs to study the linear system
defined by $\lambda_B \alpha = 0$. Having done so, one deduces that
the only invariant Nomizu map is the zero map.

For all the remaining spacetimes, $\lambda_B$ acts semi-simply:
diagonally over $\RR$ for spacetimes \hyperlink{S18}{\xtwo} and
\hyperlink{S19}{\xthree$_\chi$} and diagonally over $\CC$ for
spacetime \hyperlink{S20}{\xfour$_\chi$}. In spacetime
\hyperlink{S18}{\xtwo}, $\lambda_B$ is minus the identity on $\m$,
hence the identity on $\m^*$. By parity, there are no zero eigenvalues
in $\m^* \otimes \m^* \otimes \m$, and hence the only invariant Nomizu
map is the zero map.

In spacetime \hyperlink{S19}{\xthree$_\chi$}, $\lambda_B$ acts
diagonally on $\m$ with eigenvalues $1-\chi$ and $-1-\chi$. Letting
$V_h$ denote the one-dimensional module of $B$ with weight $h$, we see
that as a $B$-module, $\m \cong V_{1-\chi} \oplus V_{-1-\chi}$, so
that $\m^* \cong V_{-1+\chi} \oplus V_{1+\chi}$. Therefore,
\begin{equation}
  \m^* \otimes \m^* \cong V_{2\chi-2} \oplus V_{2\chi+2} \oplus 2
  V_{2\chi} \qquad\text{and}\qquad
  \m^* \otimes \m^* \otimes \m \cong 3 V_{\chi+1} \oplus 3 V_{\chi-1}
  \oplus V_{\chi+3} \oplus V_{\chi-3}.
\end{equation}
Therefore, for generic $\chi>0$, there are no invariant Nomizu maps other
than the zero map.  But, for $\chi = 1$ there are three invariants:
\begin{equation}\label{eq:nomizu-x3-chi-1}
  \alpha(H,P) = \nu' H, \qquad \alpha(P,H) = \xi' H, \qquad\text{and}\qquad
  \alpha(P,P) = \zeta' P,
\end{equation}
and for $\chi = 3$ there is one invariant:
\begin{equation}\label{eq:nomizu-x3-chi-3}
  \alpha(P,P) = \zeta H.
\end{equation}
In the limit $\chi \to \infty$, spacetime
\hyperlink{S19}{\xthree$_\chi$} tends to spacetime
\hyperlink{S18}{\xtwo}. Since there are no non-zero invariant Nomizu
maps for generic $\chi$, we expect the same is true in the limit,
which agrees with our previous findings.

Finally, in spacetime \hyperlink{S20}{\xfour$_\chi$}, $\lambda_B$ is
semi-simple with complex eigenvalues, hence diagonalisable in the
complexification $\m_\CC$ of $\m$. If now $V_h$ denotes the
\emph{complex} one-dimensional $B$-module with weight $h$, we have
that as $B$-modules
\begin{equation}
  \m_\CC \cong V_{-\chi + i} \oplus V_{-\chi - i}\qquad\text{and
    hence}\qquad
    \m^*_\CC \cong V_{\chi - i} \oplus V_{\chi + i}.
\end{equation}
The imaginary parts of the weights of $\m_\CC$ and $\m^*_\CC$ are
$\pm i$, so (by parity) there cannot be any real weights in
$\m^*_\CC \otimes \m^*_\CC \otimes \m_\CC$ and, in particular, no zero
weights. Had there been a zero weight in
$\m^* \otimes \m^* \otimes \m$, this would have resulted in a zero
weight in $\m^*_\CC \otimes \m^*_\CC \otimes \m_\CC$ upon
complexification. Therefore there are no zero weights in
$\m^* \otimes \m^* \otimes \m$ and hence the only invariant Nomizu map
is the zero map.

\subsection{Nomizu maps for aristotelian spacetimes}
\label{sec:nomizu-maps-aristotle}

In this section, we study the geometrical properties of the
aristotelian spacetimes of Table~\ref{tab:aristotelian}. They are all
reductive, so there is a canonical invariant connection, and any other
invariant connection is determined uniquely by its Nomizu map. The
Nomizu maps $\alpha : \m \times \m \to \m$ are only subject to
equivariance under rotations and are given
by~\eqref{eq:nomizu-rotations}. They depend only on the dimension $D$
and not on the precise aristotelian spacetime; although, of course,
the precise expression for the torsion and curvature tensors does
depend on the spacetime. We will calculate the torsion and curvature
for each spacetime below.

\subsubsection{Static spacetime ($\zS$)}
\label{sec:staticinv}

For $D\geq 4$, the torsion and curvature of the most general invariant
connection has the following non-zero components:
\begin{equation}
  \begin{split}
    \Theta(H,P_a) &= (\nu - \xi) P_a,\\
    \Omega(H,P_a)H &= \xi (\nu-\mu) P_a,\\
    \Omega(H,P_a)P_b &= \zeta (\mu-\nu) \delta_{ab} H, \text{and}\\
    \Omega(P_a,P_b) P_c &= \zeta \xi (\delta_{bc} P_a - \delta_{ac} P_b).
  \end{split}
\end{equation}

There are three classes of torsion-free, flat invariant connections in addition
to the canonical connection:
\begin{enumerate}
\item $\zeta = 0$ and $\mu=\nu=\xi \neq 0$,
\item $\nu=\xi=\zeta = 0$ and $\mu \neq 0$, and
\item $\mu=\nu=\xi = 0$ and $\zeta \neq 0$.
\end{enumerate}

For $D=3$, the torsion and curvature have the following non-zero
components:
\begin{equation}
  \begin{split}
    \Theta(H,P_a) &= (\nu-\xi) P_a,\\
    \Theta(P_a,P_b) &= 2 \zeta' \epsilon_{abc} P_c,\\
    \Omega(H,P_a) H &= \xi(\nu-\mu) P_a,\\
    \Omega(H,P_a) P_b &= \zeta(\mu-\nu)\delta_{ab} H, \\
    \Omega(P_a,P_b) H &= 2 \xi \zeta' \epsilon_{abc} P_c, \text{and}\\
    \Omega(P_a,P_b) P_c &= (\zeta\xi - \zeta'^2)(\delta_{bc} P_a - \delta_{ac} P_b) + 2 \zeta \zeta' \epsilon_{abc} H.
  \end{split}
\end{equation}
The torsion-free condition implies that $\zeta' = 0$.  With this value
of $\zeta'$, the above components reduce
to those in the case $D\geq 4$. We, therefore, end up with the same  
torsion-free, flat invariant connections.

In $D=2$, the torsion and curvature have components
\begin{equation}
  \begin{split}
    \Theta(H,P_a) &= (\nu-\xi)P_a + (\nu'-\xi') \epsilon_{ab} P_b,\\
    \Theta(P_a,P_b) &= 2 \zeta' \epsilon_{ab} H,\\
    \Omega(H,P_a) H &= (\xi(\nu-\mu) - \xi'\nu') P_a + (\xi\nu' + (\nu-\mu)\xi') \epsilon_{ab} P_b,\\
    \Omega(H,P_a) P_b &= \left((\zeta(\mu-\nu) - \zeta'\nu') \delta_{ab} +
      (\zeta\nu' + (\mu-\nu)\zeta') \epsilon_{ab} \right) H,\\
    \Omega(P_a,P_b) H &= 2 (\xi\zeta' - \xi'\zeta) \epsilon_{ab} H, \text{and} \\
    \Omega(P_a,P_b) P_c &= (\zeta\xi+\zeta'\xi') (\delta_{bc} P_a -
    \delta_{ac} P_b) + (\zeta\xi' - \zeta'\xi) \epsilon_{ab} P_c.
  \end{split}
\end{equation}

Here we find a one-parameter family of torsion-free, flat invariant connections 
given by
\begin{equation}
\alpha (P_a, P_b) = \zeta \delta_{ab} H.
\end{equation}

Finally, in $D=1$, the torsion and curvature have the following
non-vanishing: components
\begin{equation}
\begin{split}
\Theta(H, P) &= (\nu'-\xi')H + (\nu - \xi)P, \\
\Omega(H, P)H &= (\xi\nu'-\zeta\mu')H + (\xi(\nu-\mu) + \mu'(\xi'-\zeta'))P, \text{and} \\
\Omega(H, P)P &= (\zeta\mu + \nu'(\zeta'-\xi'))H + (\zeta\mu'-\nu'\xi)P.
\end{split}
\end{equation}

Imposing torsion-free and flatness conditions, the following classes
of invariant connections are found
\begin{enumerate}
	\item $\mu=\nu=\xi=\mu'=0,$ and $\nu'=\xi'=\zeta'$, 
	\item $\nu=\xi=\zeta=0,$ and $\nu'=\xi'=\zeta'$,
	\item $\mu'=\nu'=\xi'=\zeta=0,$ and $\mu=\nu=\xi$,
	\item $\nu=\xi=\zeta=0,$ and $\nu'=\xi'$,
	\item $\zeta=\zeta'=0,$ and $\nu=\xi,\; \nu'=\xi'$,
	\item $\nu=\xi=\mu'=0,$ and $\nu'=\xi'$, and,
	\item $\mu=\nu=\xi=\mu'=\nu'=\xi'=0$.
\end{enumerate} 

Since the remaining aristotelian spacetimes, all have the same Nomizu
maps as this static case, all of them will have the above torsion and
curvature components as a base, with a few additional terms included
due to the additional non-vanishing brackets of the specific
spacetime.

\subsubsection{Torsional static spacetime ($\zTS$)}
\label{sec:other-invar-conn-tst}

For $D\geq 4$, the torsion and curvature are given by
\begin{equation}
\begin{split}
\Theta(H, P_a) &= (\nu - \xi - 1) P_a, \\
\Omega(H, P_a) H &= \xi(\nu - \mu - 1) P_a, \\
\Omega(H, P_a) P_b &= \zeta(\mu - \nu -1) \delta_{ab} H, \text{and} \\
\Omega(P_a, P_b) P_c &= \zeta\xi (\delta_{bc} P_a - \delta_{ac} P_b).
\end{split}
\end{equation}

As in the static case, we again find three classes of torsion-free, flat invariant connection:
\begin{enumerate}
	\item $\xi=\zeta=0,$ and $\nu=1$,
	\item $\mu=\xi=\nu -1,$ and $\zeta = 0,$ and,
	\item $\xi=0,\; \nu=1,$ and $\mu=2$.
\end{enumerate}

Letting $D=3$, we get the following non-vanishing torsion and curvature components:
\begin{equation}
\begin{split}
\Theta(H, P_a) &= (\nu - \xi - 1) P_a, \\
\Theta(P_a, P_b) &= 2 \zeta ' \epsilon_{abc} P_c, \\
\Omega(H, P_a) H &= \xi(\nu - \mu - 1) P_a, \\
\Omega(H, P_a) P_b &= \zeta (\mu - \nu -1) \delta_{ab} H - \zeta ' \epsilon_{abc} P_c, \\
\Omega(P_a, P_b) H &= 2\xi\zeta ' \epsilon_{abc}P_c, \text{and} \\
\Omega(P_a, P_b) P_c &= (\zeta\xi - \zeta '^2) (\delta_{bc} P_a - \delta_{ac} P_b) + 2\zeta\zeta ' \epsilon_{abc} H.
\end{split}
\end{equation}

Imposing the torsion-free condition makes $\zeta'$ vanish such that we
get the same three classes of torsion-free, flat invariant connections
as in the $D\geq 4$ case.

In $D=2$, the torsion and curvature are given by
\begin{equation}
\begin{split}
\Theta(H, P_a) &= (\nu - \xi - 1)P_a + (\nu'-\xi')\epsilon_{ab}P_b, \\
\Theta(P_a, P_b) &= 2\zeta'\epsilon_{ab} H, \\
\Omega(H, P_a)H &= (\xi(\nu-\mu-1) - \xi'\nu')P_a + (\xi'(\nu-\mu -1) + \xi\nu')\epsilon_{ab}P_b, \\
\Omega(H, P_a)P_b &= (\zeta(\mu-\nu-1) - \nu'\zeta')\delta_{ab}H + (\zeta'(\mu-\nu-1)+\nu'\zeta)\epsilon_{ab}H, \\
\Omega(P_a, P_b)H &= 2(\xi\zeta'-\xi'\zeta)\epsilon_{ab}H, \text{and} \\
\Omega(P_a, P_b) P_c &= (\zeta\xi+\zeta'\xi')(\delta_{bc}P_a-\delta_{ac}P_b) + (\zeta\xi'-\zeta'\xi)\epsilon_{ab}P_c.
\end{split}
\end{equation}

Here we find a unique torsion-free, flat invariant connection with 
\begin{equation}
\alpha (H, H) = 2 H \qquad \alpha (H, P_a) = 2P_a \qquad \text{and} \qquad \alpha (P_a, H) = P_a.
\end{equation}

Finally, let $D=1$.  The components of the torsion and curvature are 
\begin{equation}
\begin{split}
\Theta(H, P) &= (\nu'-\xi')H + (\nu - \xi - 1)P, \\
\Omega(H, P)H &= (\xi\nu'-\zeta\mu'-\xi')H + (\xi(\nu-\mu-1) + \mu'(\xi'-\zeta'))P, \text{and} \\
\Omega(H, P)P &= (\zeta(\mu-\nu) + \nu'(\zeta'-\xi'))H + (\zeta\mu'-\nu'\xi-\zeta')P.
\end{split}
\end{equation}

We find the following classes of torsion-free, flat invariant connections
\begin{enumerate}
	\item $\xi=\zeta=\nu'=\xi'=\zeta'=0,$ and $\nu=1$,
	\item $\xi=0,\; \mu=\nu=1,$ and $\nu'=\xi'=\zeta'=\mu'\zeta$,
	\item $\zeta = 0,\; \mu=\xi=\nu-1,$ and $\mu'=\nu'=\xi'=\zeta'=0$, and,
	\item $\mu'=0,\; \mu=\xi=1,\; \nu=2,$ and $\nu'=\xi'=-\zeta'=\sqrt{\frac{-\zeta}{2}},$ for when $\zeta \leq 0$.
\end{enumerate}

\subsubsection{Aristotelian spacetime \athree$_\varepsilon$}
\label{sec:other-invar-conn-a3}

In $D\geq 4$, the torsion and curvature are given by
\begin{equation}
\begin{split}
\Theta(H, P_a) &= (\nu - \xi) P_a, \\
\Omega(H, P_a) H &= \xi(\nu - \mu) P_a, \\
\Omega(H, P_a) P_b &= \zeta(\mu - \nu) \delta_{ab} H, \text{and} \\
\Omega(P_a, P_b) P_c &= (\zeta\xi + \varepsilon) (\delta_{bc} P_a - \delta_{ac} P_b).
\end{split}
\end{equation}

Imposing flatness, we find that this requires $\varepsilon$ to vanish;
therefore, since $\varepsilon = \pm 1$, we find no torsion-free, flat
invariant connections.

Let $D=3$.  The non-vanishing torsion and curvature components are 
\begin{equation}
\begin{split}
\Theta(H, P_a) &= (\nu - \xi) P_a, \\
\Theta(P_a, P_b) &= 2 \zeta ' \epsilon_{abc} P_c, \\
\Omega(H, P_a) H &= \xi(\nu - \mu) P_a, \\
\Omega(H, P_a) P_b &= \zeta (\mu - \nu) \delta_{ab} H, \\
\Omega(P_a, P_b) H &= 2\xi\zeta ' \epsilon_{abc}P_c, \text{and} \\
\Omega(P_a, P_b) P_c &= (\zeta\xi + \varepsilon - \zeta '^2) (\delta_{bc} P_a - \delta_{ac} P_b) + 2\zeta\zeta ' \epsilon_{abc} H.
\end{split}
\end{equation}

As in the static and torsional static cases, imposing the torsion-free
condition sets $\zeta'=0$. This means we get the same torsion-free,
flat invariant connections in this case as in $D\geq 4$. Therefore,
there are no torsion-free, flat invariant connections in this
dimension.

In $D=2$, the torsion and curvature become
\begin{equation}
\begin{split}
\Theta(H, P_a) &= (\nu - \xi)P_a + (\nu'-\xi')\epsilon_{ab}P_b, \\
\Theta(P_a, P_b) &= (2\zeta')\epsilon_{ab} H, \\
\Omega(H, P_a)H &= (\xi(\nu-\mu) - \xi'\nu')P_a + (\xi'(\nu-\mu) + \xi\nu')\epsilon_{ab}P_b, \\
\Omega(H, P_a)P_b &= (\zeta(\mu-\nu) - \nu'\zeta')\delta_{ab}H + (\zeta'(\mu-\nu)+\nu'\zeta)\epsilon_{ab}H, \\
\Omega(P_a, P_b)H &= 2(\xi\zeta'-\xi'\zeta)\epsilon_{ab}H, \text{and} \\
\Omega(P_a, P_b) P_c &= (\zeta\xi+\zeta'\xi'+\varepsilon)(\delta_{bc}P_a-\delta_{ac}P_b) + (\zeta\xi'-\zeta'\xi)\epsilon_{ab}P_c.
\end{split}
\end{equation}

Once again, we find no torsion-free, flat invariant connections for
this spacetime.

\subsubsection{Aristotelian spacetime \twoda}
\label{sec:other-invar-conn-2da}

The non-vanishing torsion and curvature components are
\begin{equation}
\begin{split}
\Theta(H, P_a) &= (\nu - \xi)P_a + (\nu'-\xi')\epsilon_{ab}P_b, \\
\Theta(P_a, P_b) &= (2\zeta' - 1)\epsilon_{ab} H, \\
\Omega(H, P_a)H &= (\xi(\nu-\mu) - \xi'\nu')P_a + (\xi'(\nu-\mu) + \xi\nu')\epsilon_{ab}P_b, \\
\Omega(H, P_a)P_b &= (\zeta(\mu-\nu) - \nu'\zeta')\delta_{ab}H + (\zeta'(\mu-\nu)+\nu'\zeta)\epsilon_{ab}H, \\
\Omega(P_a, P_b)H &= (2(\xi\zeta'-\xi'\zeta)-\mu)\epsilon_{ab}H, \text{and} \\
\Omega(P_a, P_b) P_c &= (\zeta\xi+\zeta'\xi'+ \nu')(\delta_{bc}P_a-\delta_{ac}P_b) + (\zeta\xi'-\zeta'\xi- \nu)\epsilon_{ab}P_c.
\end{split}
\end{equation}
We find a unique torsion-free, flat invariant connection.  The corresponding non-vanishing Nomizu maps are
\begin{equation}
\alpha (P_a, P_b) = \frac{9}{4} \delta_{ab} H + \frac{1}{2} \epsilon_{ab} H.
\end{equation}

\section{Pseudo-riemannian spacetimes and their limits}
\label{sec:metric}

Let us introduce parameters $\varkappa=0,\pm 1$, $\varsigma =0, \pm
1$, and $c$, and consider the following Lie brackets in addition to
\eqref{eq:kin}:
\begin{equation}
  \label{eq:Liewithlim}
  [H,\B] = \varsigma  \P, \qquad
  [H,\P] = - \varkappa  \B, \qquad 
  [\B,\P] = \frac1{c^2} H, \qquad 
  [\B,\B] = -\frac{\varsigma}{c^2} \J, \qquad 
  \text{and}\qquad
  [\P,\P] = -\frac{\varkappa}{c^2} \J.
\end{equation}
The parameter $\varsigma$ corresponds to the signature:
$\varsigma = 1$ for riemannian, $\varsigma=-1$ for lorentzian and
$\varsigma=0$ for carrollian. The parameter $\varkappa$ corresponds to
the curvature, so $\varkappa=1,0,-1$ for positive, zero and negative
curvature, respectively.\footnote{This has to be taken with a grain of
  salt. Indeed, it follows from Table~\ref{tab:symmetric} that the
  correspondence between $\varkappa$ and the sign of the curvature is
  a little fictitious in the galilean setting, at least: if we
  interpret them as limits of lorentzian spacetimes, then
  $\hyperlink{S8}{\zdSG}$ has ``positive'' curvature and
  $\hyperlink{S10}{\zAdSG}$ has ``negative'' curvature, but if we
  interpret them as limits of riemannian spaces, then it's the other
  way around. This means that these spacetimes are characterised by
  the product $\varsigma \varkappa$ (for $\hyperlink{S7}{\zG}$ the
  sign is irrelevant). Concerning the carrollian spacetimes it is
  useful to realise that subalgebra spanned by $\J$ and $\P$ is
  isomorphic to $\so(D+1)$ and $\so(D,1)$ for $\hyperlink{S14}{\zdSC}$
  and $\hyperlink{S15}{\zAdSC}$, respectively (see also Section
  \ref{sec:symm-carr-struct-1}). Compared to the limits of Section 5
  in \cite{Figueroa-OFarrill:2018ilb} we change
  $\tau^{2} \eta_{00} \to \varsigma$ and
  $\kappa^{2}\eta_{\natural \natural} \to \varkappa$.} The limit
$c \to \infty$ corresponds to the non-relativistic limit. In the
computations below we will work with unspecified values of
$\varsigma,\varkappa,c$ and only at the end will we set them to
appropriate values to recover the results for particular spacetimes.
Some of the expressions will have (removable) singularities whenever
$\varsigma$ or $\varkappa$ vanish, so will have to think of those
cases as limits: the ultra-relativistic limit $\varsigma\to 0$ and the
flat limit $\varkappa\to 0$. Table~\ref{tab:symmetric} shows the
spacetimes associated to different values of these parameters. They
can be characterised as those homogeneous kinematical spacetimes which
are symmetric, so the canonical invariant connection is torsion-free.
The table divides into four sections separated by horizontal rules
corresponding, from top to bottom, to lorentzian, euclidean, galilean
and carrollian symmetric spacetimes.

\begin{table}[h!]
  \centering
  \caption{Symmetric spacetimes}
  \label{tab:symmetric}
  \begin{tabular}{*{3}{>{$}c<{$}}|l}\toprule
    \varsigma & \varkappa & c^{-1} & Spacetime\\ \midrule
    -1 & 0 & 1 & Minkowski ($\hyperlink{S1}{\MM}$)\\
    -1 & 1 & 1 & de~Sitter ($\hyperlink{S2}{\zdS}$)\\
    -1 & -1 & 1 & anti de~Sitter ($\hyperlink{S3}{\zAdS}$)\\\midrule
    1 & 0 & 1 & euclidean ($\hyperlink{S4}{\EE}$)\\
    1 & 1 & 1 & sphere ($\hyperlink{S5}{\SS}$)\\
    1 & -1 & 1 & hyperbolic ($\hyperlink{S6}{\HH}$)\\\midrule
    \mp1 & 0 & 0 & galilean ($\hyperlink{S7}{\zG}$)\\
    \mp1 & \pm1 & 0 & galilean de~Sitter ($\hyperlink{S8}{\zdSG}$)\\
    \mp1 & \mp1 & 0 & galilean anti de~Sitter ($\hyperlink{S10}{\zAdSG}$)\\\midrule
    0 & 0 & 1 & carrollian ($\hyperlink{S13}{\zC}$)\\
    0 & 1 & 1 & carrollian de~Sitter ($\hyperlink{S14}{\zdSC}$)\\
    0 & -1 & 1 & carrollian anti de~Sitter ($\hyperlink{S15}{\zAdSC}$)\\\bottomrule
  \end{tabular}
\end{table}

\subsection{Invariant structures}
\label{sec:invariant-structures}

We will determine the form of the invariant tensors of small rank.
If $\k = \h \oplus \m$ is a reductive split then, as explained in
Section~\ref{sec:invariant-tensors}, invariant tensor
fields on a simply-connected homogeneous space $M=\Kgr/\Hgr$ are in
bijective correspondence with $\Hgr$-invariant tensors on $\m$, and
since $\Hgr$ is connected, these are in bijective correspondence with
$\h$-invariant tensors on $\m$.

The action of $\h$ on $\m$ is the linear isotropy representation,
which is the restriction to $\h$ of the adjoint action:
\begin{equation}\label{eq:LIR}
  \begin{aligned}[m]
    J_{ab} \cdot H &= 0\\
    J_{ab} \cdot P_c &= \delta_{bc} P_a - \delta_{ac} P_b
  \end{aligned}
  \qquad\text{and}\qquad
  \begin{aligned}[m]
    B_a \cdot H &= - \varsigma P_a\\
    B_a \cdot P_b &= \frac1{c^2} \delta_{ab} H.
  \end{aligned}
\end{equation}
With respect to the canonical dual basis $\eta$, $\pi_a$ for $\m^*$,
the dual linear isotropy representation is the restriction of the
coadjoint action:
\begin{equation}\label{eq:LIR-dual}
  \begin{aligned}[m]
    J_{ab} \cdot \eta &= 0\\
    J_{ab} \cdot \pi^c &= -\delta_a^c \pi_b + \delta_b^c \pi_a
  \end{aligned}
  \qquad\text{and}\qquad
  \begin{aligned}[m]
    B_a \cdot \eta &= - \frac1{c^2} \pi_a\\
    B_a \cdot \pi^b &= \varsigma \delta_a^b \eta.
  \end{aligned}
\end{equation}
It follows that $H$ is invariant in the $\sigma \to 0$ limit,
whereas $\eta$ is invariant in the $c\to\infty$ limit.

Concerning the rotationally invariant tensors of second rank, let us
observe that
\begin{equation}
  \alpha_1 H^2 + \beta_1 \P^2 \quad\text{is invariant}\quad \iff \quad
  \sigma \alpha_1  = \frac1{c^2} \beta_1
\end{equation}
and
\begin{equation}
  \alpha_2 \eta^2 + \beta_2 \bpi^2 \quad\text{is invariant}\quad \iff \quad
  \frac1{c^2} \alpha_2 = \sigma \beta_2.
\end{equation}
It is interesting to note that the sign $\varkappa$ of the curvature
has played no rôle thus far.

We shall now specialise to the different classes of spacetimes and
determine whether and how the structures are induced in the limit.

\subsubsection{Lorentzian and riemannian case}
\label{sec:lorentz-riem-case}

It is clear that for the (pseudo\nobreakdash-)riemannian case, where
$\varsigma\neq 0 \neq \frac{1}{c^{2}}$, only the metric and its
co-metric are invariant.  Keeping in mind that we wish the limit in
which the parameters $\varsigma$ and $c$ tend to zero to exist, we set
$\alpha_{1}=\frac{1}{c^{2}}$ and $\beta_{1}=\varsigma$ and similarly
for the co-metric, which leads to the invariants
\begin{equation}
  \label{eq:metrics}
  \frac{1}{c^{2}} H^{2} + \varsigma \P^{2} \qquad \text{and} \qquad \varsigma \eta^{2} + \frac{1}{c^{2}} \mathbold{\pi}^{2}  \,.
\end{equation}
For negative (positive) $\varsigma$ this is the invariant lorentzian
(riemannian) structure. The metric and the co-metric are not per se
the inverse of each other, although using definite values for the
limiting parameters they can be made to be.

\subsubsection{Non- and ultra-relativistic limits}
\label{sec:non-ultra-relat}

Let us now investigate the limits. Taking the non-relativistic limit
($c \to \infty$) of the metrics leads to the invariants
\begin{equation}
  \label{eq:gallim}
  \varsigma \P^{2} \qquad \text{and} \qquad \varsigma \eta^{2},
\end{equation}
which can be interpreted as the invariants that properly arise from
the lorentzian structure. However, as \eqref{eq:LIR} shows also
$\eta$ itself is an invariant in this limit. This does not follow from the
contractions, but can be anticipated from the metrics. We could now
take the ultra-relativistic limit ($\varsigma \to 0$) of
\eqref{eq:gallim} leading to no invariant tensor. Of course, this
spacetime has the invariants $H, \P^{2}, \eta, \mathbold{\pi}^{2}$,
but none of these arise from the limit of the original lorentzian and
riemannian metrics. For the ultra-relativistic limit, we may apply 
the same logic.

Concluding, we have the galilean structure $\eta, \varsigma \P^2$ and the
carrollian structure $H, \frac{1}{c^{2}} \mathbold{\pi}^{2}$, where we
have left the contraction parameters for the invariants that arise
from a limit.

\subsection{Action of the boosts}
\label{sec:action-boosts}

The actions of the boosts for all the lorentzian, riemannian, galilean,
and reductive carrollian spacetimes were determined in
Section~\ref{sec:boosts}, where we arrived at
equation~\eqref{eq:boost-orbit} for the orbit of $(t_0,\x_0)$ under
the one-parameter family of boosts generated by $\w \cdot \B$, which
we rewrite here as follows:
\begin{equation}
    \begin{split}
      t(s) &= t_0 \cosh(s z) + \frac1{c^2} \frac{\sinh(s z)}{z}\x_0 \cdot \w\\
      \x(s) &= \x_0^\perp - \varsigma t_0 \frac{\sinh(s z)}{z} \w +
      \cosh(s z) \frac{(\x_0 \cdot \w)}{w^2} \w ,
  \end{split}
\end{equation}
where $\x_0^\perp := \x_0 - \frac{\x_0 \cdot \w}{w^2}\w$ and $z^2 :=
-\frac1{c^2} \varsigma w^2$.  Notice that the orbits of $(0,\x_0)$
with $\x_0 \cdot \w = 0$ are point-like. To understand the nature of
the other (generic) orbits, we choose values for the
parameters. Notice that in our coset parametrisation the boosts do
not depend on $\varkappa$, but only on $\varsigma$ and $c$.  Therefore,
we shall be able to treat each class of spacetime uniformly.

\subsubsection{Lorentzian boosts}
\label{sec:Lorentzian-boosts}
Here we take $\varsigma = -1$ and keep $c^{-1}$ non-zero. Then
$z^2 = \frac{w^2}{c^2}$, so $z = \left|\frac{\w}{c}\right|$, and the
orbits of the boosts are
\begin{equation}\label{eq:lor-boost-orbit}
  \begin{split}
    t(s) &= t_0 \cosh(s \left|\tfrac{\w}{c}\right|) + \frac1{c^2} \frac{\sinh(s \left|\tfrac{\w}{c}\right|)}{\left|\tfrac{\w}{c}\right|}\x_0 \cdot \w\\
    \x(s) &= \x^\perp_0 + t_0 \frac{\sinh(s \left|\tfrac{\w}{c}\right|)}{\left|\tfrac{\w}{c}\right|} \w +
    \cosh(s \left|\tfrac{\w}{c}\right|) \frac{(\x_0 \cdot \w)}{w^2} \w.
  \end{split}
\end{equation}
Let $\x = \x^\perp + y \w$, where $\x^\perp \cdot \w = 0$.  Then
$\x^\perp(s) = \x^\perp_0$ for all $s$ and the orbit takes place in
the $(t,y)$ plane.  Letting $|\w| = 1$ and $c=1$, we find
\begin{equation}\label{eq:lor-boost-orbit-too}
  t(s) = t_0 \cosh(s) + \sinh(s) y_0 \qquad\text{and}\qquad y(s) = t_0 \sinh(s) + \cosh(s) y_0,
\end{equation}
which is either a point (if $t_0 = y_0 = 0$), a straight line (if
$t_0 = \pm y_0 \neq 0$), or a hyperbola (otherwise). The nature of the
orbits in the exponential coordinates is clear, but only in the case
of Minkowski spacetime do the exponential coordinates provide a global
chart and hence only in that case can we deduce from this calculation
that the generic orbits are not compact. For (anti) de~Sitter
spacetime, we must argue in a different way.

Let $\overline{\zdS}$ denote the quotient of $\hyperlink{S2}{\zdS}$
which embeds as a quadric hypersurface in Minkowski spacetime. The
covering map $\zdS \to \overline{\zdS}$ relates the orbits of the
boosts on $\hyperlink{S2}{\zdS}$ and in the quotient $\overline{\zdS}$
and since continuous maps send compact sets to compact sets, it is
enough to show the non-compactness of the orbits in $\overline{\zdS}$.
The embedding $\overline{\zdS} \subset \RR^{D+1,1}$ is given by the
quadric
\begin{equation}
  x_1^2 + \cdots + x_D^2 + x_{D+1}^2 - x_{D+2}^2 = R^2,
\end{equation}
which is acted on transitively by $\SO(D+1,1)$. The stabiliser Lie
algebra of the point $(0,\cdots,0,R,0)$ is spanned by the $\so(D+1,1)$
generators $J_{ab}$ and $J_{a,D+2}$, so that $B_a = J_{a,D+2}$, which
is a boost in $\RR^{D+1,1}$. We have just shown that boosts in
Minkowski spacetime have non-compact orbits; therefore, this is the
case in $\overline{\zdS}$ and hence also in $\hyperlink{S2}{\zdS}$.

Similarly, let $\overline{\zAdS}$ denote the quotient of
$\hyperlink{S3}{\zAdS}$ which embeds in $\RR^{D,2}$ as the quadric
\begin{equation}
  x_1^2 + \cdots + x_D^2 - x_{D+1}^2 - x_{D+2}^2 = -R^2.
\end{equation}
The Lie algebra $\so(D,2)$ acts transitively on this quadric and the
stabiliser Lie algebra at the point $(0,\cdots,0,0,R)$ is spanned by
the $\so(D,2)$ generators $J_{ab}$ and $J_{a,D+1}$, so that
$B_a = J_{a,D+1}$ which is a ``boost'' in $\RR^{D,2}$. The calculation
of the orbit, in this case, is formally identical to the one for
Minkowski spacetime (in fact, it takes place in the lorentzian plane
with coordinates $(x_a, x_{D+2})$) and we see that they are
non-compact, so the same holds in $\overline{\zAdS}$ and thus also in
$\hyperlink{S3}{\zAdS}$.

\subsubsection{Euclidean ``boosts''}
\label{sec:euclidean-boosts}

Here we take $\varsigma = 1$ and keep $c^{-1}$ non-zero.  Then $z^2 =
-\frac{w^2}{c^2}$, so $z = i \left|\frac{\w}{c}\right|$, and the orbits
of the boosts are
\begin{equation}\label{eq:riem-boost-orbit}
  \begin{split}
    t(s) &= t_0 \cos(s \left|\tfrac{\w}{c}\right|) + \frac1{c^2}
    \frac{\sin(s
      \left|\tfrac{\w}{c}\right|)}{\left|\tfrac{\w}{c}\right|}\x_0
    \cdot \w\\
    \x(s) &= \x^\perp_0 - t_0 \frac{\sin(s
      \left|\tfrac{\w}{c}\right|)}{\left|\tfrac{\w}{c}\right|} \w +
    \cos(s \left|\tfrac{\w}{c}\right|) \frac{(\x_0 \cdot \w)}{w^2}
    \w.
  \end{split}
\end{equation}

As before, letting $\x = \x^\perp + y \w$, and choosing $|\w|=1$ and
$c=1$, we find that the orbit is such that $\x^\perp$ is constant and
$(t,y)$ evolve as
\begin{equation}\label{eq:riem-boost-orbit-too}
  t(s) = t_0 \cos(s) + \sin(s) y_0 \qquad\text{and}\qquad y(s) = -t_0 \sin(s) + \cos(s) y_0,
\end{equation}
which is either a point (if $t_0 = y_0 = 0$) or a circle (otherwise)
and in any case compact. This suffices for $\hyperlink{S4}{\EE}$ and
$\hyperlink{S6}{\HH}$ since the exponential coordinates give a global
chart. For $\hyperlink{S5}{\SS}$ it is clear that the boosts act with
compact orbits because the kinematical Lie group $\SO(D+2)$ is itself
compact, therefore, so are the one-parameter subgroups.

\subsubsection{Galilean boosts}
\label{sec:galilean-boosts}

Here we take the limit $c \to \infty$ and, for definiteness,
$\varsigma = -1$.  The orbits of the boosts are then the limit $c \to
\infty$ of equation~\eqref{eq:lor-boost-orbit}:
\begin{equation}\label{eq:gal-boost-orbit}
  \begin{split}
    t(s) &= t_0 \\
    \x(s) &= \x_0 + s t_0 \w.
  \end{split}
\end{equation}
Here the orbits of $(0,\x_0)$ are point-like. The generic orbit
($t_0\neq 0$) is not periodic and hence not compact. This suffices for
$\hyperlink{S7}{\zG}$ and $\hyperlink{S8}{\zdSG}$, since the
exponential coordinates define a global chart. For
$\hyperlink{S10}{\zAdSG}$ we need to argue differently and this is
done in Section~\ref{sec:adsg}.

\subsubsection{Carrollian boosts}
\label{sec:carrollian-boosts}

Here we keep $c^{-1}$ non-zero, but take the limit $\varsigma \to 0$
in equation~\eqref{eq:boost-orbit}:
\begin{equation}\label{eq:car-boost-orbit}
  \begin{split}
    t(s) &= t_0 + s \frac1{c^2} \x_0 \cdot \w\\
    \x(s) &= \x_0.
  \end{split}
\end{equation}
Here the orbits $(t_0,\x_0)$ with $\x_0 \cdot \w =0$ are point-like,
but the other orbits are not periodic, hence not compact. This settles
it for $\hyperlink{S15}{\zAdSC}$, since the exponential coordinates
give a global chart. For the other carrollian spacetimes we can argue
in a different way.

As shown in \cite{Duval:2014uoa}, a carrollian spacetime admits an
embedding as a null hypersurface in a lorentzian spacetime. For the
homogeneous examples in this paper, this was done in
\cite{Figueroa-OFarrill:2018ilb} following the embeddings of the
carrollian spacetimes $\hyperlink{S13}{\zC}$ and
$\hyperlink{S16}{\zLC}$ as null hypersurfaces of Minkowski spacetime
described already in \cite{Duval:2014uoa}.

As explained in Section~\ref{sec:boosts}, for $\hyperlink{S14}{\zdSC}$
it is enough to work with the discrete quotient $\overline{\zdSC}$,
which embeds as a null hypersurface in the hyperboloid model
$\overline{\zdS}$ of de~Sitter spacetime, which itself is a quadric
hypersurface in Minkowski spacetime. In
\cite{Figueroa-OFarrill:2018ilb} we showed that the boosts in
$\overline{\zdSC}$ can be interpreted as null rotations in the
(higher-dimensional) pseudo-orthogonal Lie group and the orbits of
null rotations are never compact. This is done in detail in
Section~\ref{sec:action-boosts-1} for $\hyperlink{S16}{\zLC}$.

\subsection{Fundamental vector fields}
\label{sec:fund-vect-fields}

The fundamental vector fields for rotations and boosts are linear in
exponential coordinates and given by equations~\eqref{eq:fvf-rot} and
\eqref{eq:fvf-boost}, respectively. To determine the fundamental
vector fields for the translations we must work harder.

Now let $A = t H + \x \cdot \P$.  Then we have that
\begin{equation}\label{eq:adA-on-gens}
  \begin{aligned}[m]
    \ad_A H &= \varkappa \x \cdot \B\\
    \ad_A B_a &= \varsigma t P_a - \frac{1}{c^2} x_a H\\
    \ad_A P_a &= \frac{\varkappa}{c^2} J_{ab} x^b - \varkappa t B_a\\
    \ad_A J_{ab} &= x_a P_b - x_b P_a
  \end{aligned}
  \qquad\text{and}\qquad
  \begin{aligned}[m]
    \ad^2_A H &= \varkappa\varsigma t \x \cdot \P - \frac{\varkappa}{c^2} x^2 H\\
    \ad^2_A B_a &= \frac{\varkappa}{c^2} \varsigma t x^b J_{ab} -
    \varkappa \varsigma t^2 B_a - \frac{\varkappa}{c^2} x_a \x \cdot \B\\
    \ad^2_A P_a &= -\varkappa (\frac1{c^2} x^2 + \varsigma t^2) P_a + \frac{\varkappa}{c^2}
    x_a \x \cdot P + \frac{\varkappa}{c^2} t x_a H\\
    \ad^2_A J_{ab} &= -\varkappa t(x_a B_b - x_b B_a) + \frac{\varkappa}{c^2}
    x^c(x_a J_{bc} - x_b J_{ac}),
  \end{aligned}
\end{equation}
so that in general we have
\begin{equation}
  \ad_A^3 = -\varkappa (\frac1{c^2} x^2 + \varsigma t^2)\ad_A.
\end{equation}
Letting $x_\pm$ denote the two complex square roots of
$-\varkappa (\frac1{c^2} x^2 + \varsigma t^2)$, with $x_- = - x_+$, we
can rewrite this equation as $\ad_A^3 = x_+^2 \ad_A$.

Now, if $f(z)$ is analytic in $z$ and admits a power series expansion
$f(z) = \sum_{n=0}^\infty c_n z^n$, then
\begin{equation}
  f(\ad_A) = f(0) + \frac1{x_+}\sum_{k=0}^\infty c_{2k+1} x_+^{2k+1}
  \ad_A + \frac1{x_+^2} \sum_{k=1}^\infty c_{2k} x_+^{2k} \ad_A^2.
\end{equation}
Observing that
\begin{equation}
  \sum_{k=0}^\infty c_{2k+1} x_+^{2k+1} = \tfrac12 ( f(x_+) - f(x_-) )
  \qquad\text{and}\qquad
  \sum_{k=1}^\infty c_{2k} x_+^{2k} = \tfrac12 ( f(x_+) + f(x_-) -
  2 f(0) ),
\end{equation}
we arrive finally at
\begin{equation}
  \label{eq:fad}
  f(\ad_A) =f(0) + \frac1{2x_+} (f(x_+)-f(x_-)) \ad_A +
  \frac1{2x_+^2} (f(x_+) + f(x_-) - 2 f(0)) \ad_A^2.
\end{equation}
Introducing the shorthand notation:
\begin{equation}\label{eq:shorthand}
  f^+ := \tfrac12 (f(x_+) + f(x_-))\qquad\text{and}\qquad
  f^- := \frac1{2x_+}(f(x_+) - f(x_-)),
\end{equation}
equation~\eqref{eq:fad} becomes
\begin{equation}
  f(\ad_A) =f(0) + f^- \ad_A + \frac1{x_+^2} (f^+ - f(0)) \ad_A^2.
\end{equation}

It follows from the above equation and
equation~\eqref{eq:adA-on-gens}, that for $f(z)$ analytic in $z$,
\begin{equation}\label{eq:fadA-on-gens}
  \begin{split}
    f(\ad_A) H &= f(0) H + f^- \varkappa \x \cdot \B +
    \tfrac1{x_+^2}(f^+-f(0))\left(\varkappa\varsigma t \x \cdot \P -
      \tfrac{\varkappa}{c^2} x^2 H\right)\\
    f(\ad_A) B_a &= f(0) B_a + f^-(\varsigma t P_a - \tfrac1{c^2} x_a
    H) + \tfrac1{x_+^2} (f^+-f(0))\left(- \varkappa\varsigma t^2 B_a -
    \tfrac{\varkappa}{c^2} x_a \x \cdot \B + \tfrac{\varkappa}{c^2}
    \varsigma t J_{ab} x^b\right)\\
  f(\ad_A) P_a &= f^+ P_a + f^- (-\varkappa t B_a +
  \tfrac{\varkappa}{c^2} J_{ab} x^b) + \tfrac1{x_+^2}(f^+-f(0))
  \tfrac{\varkappa}{c^2} x_a ( t H + \x \cdot \P)\\
  f(\ad_A) J_{ab} &= f(0) J_{ab} + f^-(x_a P_b - x_b P_a) +
  \tfrac1{x_+^2}(f^+ - f(0)) \varkappa\left(- t (x_a B_b - x_b B_a) +
    \tfrac1{c^2} x^c(x_a J_{bc} - x_b J_{ac})\right).
  \end{split}
\end{equation}

Let us calculate $\xi_H = \tau \frac{\d}{\d t} + y^a \frac{\d}{\d
  x^a}$, where by equation~\eqref{eq:master}
\begin{equation}
  \tau H + \y \cdot \P = G(\ad_A) H - F(\ad_A)\bbeta \cdot \B,
\end{equation}
for some $\bbeta$.  From equation~\eqref{eq:fadA-on-gens}, we have
\begin{multline}
  \tau H + \y \cdot \P = H + G^- \varkappa \x \cdot \B + \frac1{x_+^2}
  (G^+-1) \left(\varkappa\varsigma t \x \cdot \P -
    \tfrac{\varkappa}{c^2} x^2 H\right)\\
  -\left(\bbeta\cdot \B + F^-(\varsigma t\bbeta \cdot \P - \tfrac1{c^2} \x
    \cdot\bbeta H ) + \frac1{x_+^2}(F^+-1)\left( -\varkappa\varsigma t^2\bbeta \cdot\B - \tfrac{\varkappa}{c^2} \x \cdot\bbeta \x \cdot \B +
      \tfrac{\varkappa}{c^2} \varsigma t J_{ab} \beta^a x^b\right) \right).
\end{multline}
By $\so(D)$-covariance, $\bbeta$ has to be proportional to $\x$, since
that is the only other vector appearing in the $\B$ terms, which means
that the $J_{ab}$ term above vanishes. This leaves terms in $\B$, $H$,
and $\P$, which allow us to solve for $\bbeta$, $\tau$, and $\y$,
respectively. The $\B$ terms cancel if and only if
\begin{equation}
  \bbeta = \frac{G^-}{F^+} \varkappa \x,
\end{equation}
which we can reinsert into the equation to solve for $\tau$ and $\y$.
Doing so we find
\begin{equation}
    \tau = 1 - \left(\frac{x_+ \coth x_+ - 1}{x_+^2}\right)
    \frac{\varkappa}{c^2} x^2 \qquad\text{and}\qquad
    y^a =\left(\frac{x_+ \coth x_+ - 1}{x_+^2}\right) \varkappa
    \varsigma t x^a,
\end{equation}
so that
\begin{equation}
\xi_H = \frac{\d}{\d t} + \left(\frac{x_+ \coth x_+ -
      1}{x_+^2}\right) \varkappa \left( \varsigma t x^a \frac{\d}{\d
      x^a} - \frac{1}{c^2} x^2 \frac{\d}{\d t} \right).
\end{equation}

To calculate $\xi_{\v \cdot \P} = \tau \frac{\d}{\d t} + y^a
\frac{\d}{\d x^a}$, equation \eqref{eq:master} says we must solve
\begin{equation}
  \tau H + \y \cdot \P = G(\ad_A) \v \cdot \P - F(\ad_A) \left(\bbeta \cdot \B
  + \tfrac12 \lambda^{ab} J_{ab}\right),
\end{equation}
for $\lambda^{ab}$, $\bbeta$, $\tau$, and $\y$ from the components along
$J_{ab}$, $\B$, $H$, and $\P$, respectively.  The details of the
calculation are not particularly illuminating.  Let us simply remark
that we find
\begin{equation}
  \lambda^{ab} = h_1 (v^a x^b - v^b x^a) + h_2 (\beta^a x^b - \beta^b x^a)
\end{equation}
for
\begin{equation}
  h_1 = \frac{G^-\frac{\varkappa}{c^2}}{1 -
    \frac1{x_+^2}(F^+-1)\frac{\varkappa}{c^2} x^2}
  \qquad\text{and}\qquad
  h_2 = \frac{-\frac1{x_+^2} (F^+-1)\frac{\varkappa}{c^2}\varsigma t}{1 -
    \frac1{x_+^2}(F^+-1)\frac{\varkappa}{c^2} x^2},
\end{equation}
and
\begin{equation}
  \bbeta = -\frac{G^-}{F^+} \varkappa t \v,
\end{equation}
so that
\begin{equation}
  \lambda^{ab} = -\frac{\varkappa}{c^2} \frac{\tanh(x_+/2)}{x_+} (v^a
  x^b - v^b x^a).
\end{equation}
Re-inserting these expressions into the equation we solve for $\tau$
and $\y$, resulting in
\begin{equation}
  \tau = \frac{x_+\coth x_+ - 1}{x_+^2} \frac{\varkappa}{c^2} t \x
  \cdot \v
\end{equation}
and
\begin{equation}
  y^a = x_+ \coth(x_+) v^a + \frac{x_+\coth x_+ - 1}{x_+^2}
  \frac{\varkappa}{c^2}  \x \cdot \v x^a.
\end{equation}
Finally, we have that
\begin{equation}
  \xi_{P_a} = \frac{x_+\coth x_+ - 1}{x_+^2}
    \frac{\varkappa}{c^2} x_a \left(t \frac{\d}{\d t} + x^b
      \frac{\d}{\d x^b}\right) + x_+ \coth x_+ \frac{\d}{\d x^a}.
\end{equation}

Let us summarise all the fundamental vector fields and remember that
$x_{+}= \sqrt{-\varkappa(\frac1{c^2} x^2 + \varsigma t^2)}$
\begin{equation}
  \label{eq:fundvec}
  \begin{split}
    \xi_{J_{ab}} &= x^b \frac{\d}{\d x^a} - x^a \frac{\d}{\d x^b}
    \\
    \xi_{B_a} &= \frac{1}{c^2} x^a \frac{\d}{\d t} - \varsigma t \frac{\d}{\d x^a}
    \\
    \xi_H &= \frac{\d}{\d t}
    + \left(
      \frac{x_+ \coth x_+ - 1}{x_+^2}
    \right)
    \varkappa \left(
      \varsigma t x^a \frac{\d}{\d x^a} - \frac{1}{c^2} x^2 \frac{\d}{\d t}
    \right)
    \\
    \xi_{P_a} &= \frac{x_+\coth x_+ - 1}{x_+^2}
    \frac{\varkappa}{c^2} x_a
    \left(
      t \frac{\d}{\d t} + x^b \frac{\d}{\d x^b}
    \right)
    + x_+ \coth x_+ \frac{\d}{\d x^a}.
\end{split}
\end{equation}
We can now calculate the Lie brackets of the vector fields which indeed
shows the anti-homomorphism with respect to \eqref{eq:Liewithlim}
\begin{equation}
  [\xi_{H},\xi_{\B}] = - \varsigma  \xi_{\P}, \quad
  [\xi_{H},\xi_{\P}] =  \varkappa  \xi_{\B}, \quad 
  [\xi_{\B},\xi_{\P}] =- \frac1{c^2} \xi_{H}, \quad 
  [\xi_{\B},\xi_{\B}] = \frac{\varsigma}{c^2} \xi_{\J}, \quad 
  \text{and}\quad
  [\xi_{\P},\xi_{\P}] = \frac{\varkappa}{c^2} \xi_{\J}.
\end{equation}
Let us emphasise that taking the limit of the vector fields and then
calculating their Lie bracket leads to the same result as just taking
just the limit of the Lie brackets, i.e., these operations commute.

\subsection{Soldering form and connection one-form}
\label{sec:sold-form-conn}

The soldering form and the connection one-form are the two components
of the pull-back of the left-invariant Maurer--Cartan form on $\Kgr$.
We will calculate it first for all the (pseudo\nobreakdash-)riemannian
cases and then take the flat, non-relativistic and ultra-relativistic
limit. As we will see, the exponential coordinates are well adapted
for that purpose, and the limits can then be systematically studied.
That the limits are well defined follows from our construction since
the quantities we calculate are a power series of the contraction
parameters, $\epsilon=c^{-1}, \varkappa, \tau$ in the $\epsilon \to 0$
limit and not of their inverse. Let us however stress that for some
quantities like, e.g., the galilean structure, modified exponential
coordinates are more economical (see, e.g., Appendix
\ref{app:modexp}).

For the non-flat (pseudo\nobreakdash-)riemannian geometries
our exponential coordinates are, except for the hyperbolic case,
neither globally valid nor are quantities like the curvature very
compact. Since coordinate systems for these cases are well studied, we
will focus in the following mainly on the remaining cases. It is
useful to derive the soldering form, the invariant connection and the
vielbein in full generality since we take the limit and use them to
calculate the remaining quantities of interest.

We start by calculating the Maurer--Cartan form via
equation~\eqref{eq:MC-pullback} for which we again use
equation~\eqref{eq:fadA-on-gens}. We find that
\begin{multline}
  \theta + \omega
  =
  dt H
  + D^- \varkappa  dt \x \cdot \B
  + \frac{1}{x_+^2}(D^+-1)
  \left(
    \varkappa \varsigma t  dt \x \cdot \P
    - \tfrac{\varkappa}{c^2} x^2 dt H
  \right)
  \\
  + D^+ d\x \cdot \P + D^-
  (
  -\varkappa t d\x \cdot \B
  + \tfrac{\varkappa}{c^2} dx^a x^b J_{ab}
  )
  + \frac1{x_+^2}(D^+-1) \tfrac{\varkappa}{c^2} \x \cdot d\x (t H + \x \cdot \P),
\end{multline}
which, using that
\begin{equation}
  D^- = \frac{1 - \cosh x_+}{x_+^2}, \qquad D^+ = \frac{\sinh
    x_+}{x_+} \qquad\text{and hence}\qquad \frac1{x_+^2}(D^+-1) =
  \frac{\sinh x_+ - x_+}{x_+^3},
\end{equation}
gives the following expressions:
\begin{equation}
  \begin{split}
    \theta &= dt H
    + \frac{\sinh x_+}{x_+} d\x \cdot \P
    + \frac{\sinh x_+ - x_+}{x_+^3} \varkappa
    \left(
      \varsigma t dt \x\cdot\P
      + \tfrac{1}{c^2}
      (t \, \x \cdot d\x  H - x^2 dt H + \x\cdot d\x\, \x \cdot \P)
     \right)\\
    \omega &= \frac{1-\cosh x_+}{x_+^2} \varkappa
    \left(
      dt \x \cdot\B
      - t d\x \cdot \B
      - \tfrac1{c^2}  x^a dx^b J_{ab}
    \right).
  \end{split}
\end{equation}

We can also evaluate the vielbein $E=E_{H} \eta + E_{\P} \cdot \bpi$
which leads us to
\begin{align}
  E_{H} &=
          \frac{\varkappa}{x_{+}^{2}}
          \left[
          \left(
          - \sigma t^{2} - \frac{x^{2}}{c^{2}} x_{+} \csch x_{+}
          \right)
          \frac{\pd}{\pd t}
          +
          \sigma 
          \left(
          -1+ x_{+} \csch x_{+}
          \right)
          t x^{a}
          \frac{\pd}{\pd x^{a}}
          \right]
  \\
  E_{P_{a}}
        &=
          \frac{\varkappa x^{a}}{c^{2} x_{+}^{2}}
          \left(
          -1 + x_{+} \csch x_{+}
          \right)
          \left(
          t \frac{\pd}{\pd t} + x^{b}\frac{\pd}{\pd x^{b}}
          \right)
          +
           x_{+} \csch x_{+} \frac{\pd}{\pd x^{a}} .
\end{align}

\subsection{Flat limit, Minkowski ($\MM$) and euclidean spacetime ($\EE$)}
\label{sec:flat-limit}

In the flat limit $\varkappa \to 0$ the soldering form and connection
one-form are given by
\begin{equation}
    \theta = dt H + d\x \cdot \P \qquad \text{and} \qquad \omega = 0 ,
\end{equation}
respectively, where $(t,\x)$ are global coordinates. The vielbein is
given by
\begin{align}
      E&= \frac{\partial}{\partial t} \eta +  \frac{\partial}{\partial \x} \cdot \mathbold{\pi}
\end{align}
and the fundamental vector fields, taking the limit of
\eqref{eq:fundvec}, by
\begin{equation}
    \xi_{B_a} = \frac{1}{c^2} x^a \frac{\d}{\d t} - \varsigma t \frac{\d}{\d x^a}, \qquad
    \xi_H = \frac{\d}{\d t}, \qquad \text{and} \qquad
    \xi_{P_a} =  \frac{\d}{\d x^a}.
\end{equation}
Using the soldering form and the vielbein we can now write the metric
and co-metric, given in equation \eqref{eq:metrics}, in coordinates
\begin{align}
 g &= \sigma dt^{2} + \frac{1}{c^{2}} d \x \cdot d \x & \tilde g &=  \frac{1}{c^{2}}  \frac{\pd}{\pd t}\otimes \frac{\pd}{\pd t} + \sigma \delta^{ij} \frac{1}{\pd x^{i}} \otimes  \frac{1}{\pd x^{j}}.
\end{align}
Since the connection one-form vanishes the torsion and curvature evaluate to
\begin{align}
  \Omega&=0 & \Theta&=0 .
\end{align}

We can now set $\sigma$ and $c$ to definite values to obtain the
Minkowski spacetime ($\sigma=-1$, $c=1$), Euclidean space
($\sigma=-1$, $c=1$), galilean spacetime ($\sigma=1$, $c^{-1}=0$), and
carrollian spacetime ($\sigma=0$, $c=1$). This is obvious enough for
the first two cases so that we go straight to the galilean spacetime.

\subsection{Galilean spacetime ($\zG$)}
\label{sec:galilean-spacetime}

For galilean spacetimes we have the fundamental vector fields
\begin{align}
    \xi_{B_a} &=  t \frac{\d}{\d x^a} &
    \xi_H &= \frac{\d}{\d t} &
    \xi_{P_a} &=  \frac{\d}{\d x^a},
\end{align}
and the invariant galilean structure  which is characterised by the clock one-form
$\tau = dt$ and the spatial metric on one-forms
$h = \delta^{ab} \frac{\d}{\d x^a} \otimes \frac{\d}{\d x^b}$.

\subsection{Carrollian spacetime ($\zC$)}
\label{sec:carrollian-spacetime}

The fundamental vector fields for the carrollian spacetime are
\begin{align}
  \label{eq:carrvec}
    \xi_{B_a} &=  x^a \frac{\d}{\d t}  &
    \xi_H &= \frac{\d}{\d t} &
    \xi_{P_a} &=  \frac{\d}{\d x^a},
\end{align}
and the invariant carrollian structure is given by
$\kappa = \frac{\d}{\d t}$ and $b = \delta_{ab} dx^a dx^b$. 

\subsection{Non-relativistic limit}
\label{sec:non-relat-limit}
In the non-relativistic limit $c \to \infty$ we get
$x_{+}= \sqrt{-\varkappa \varsigma t^2}$ and the soldering form and
connection one-form are given by
\begin{equation}
  \begin{split}
    \theta &= dt H + \frac{\sinh x_+}{x_+} d\x \cdot \P
    + \frac{\sinh  x_+ - x_+}{x_+^3} \varkappa \varsigma t dt \x\cdot\P \\
    \omega &= \frac{1-\cosh x_+}{x_+^2} \varkappa
    \left(
      dt \x \cdot \B- t d\x \cdot \B 
    \right)
  \end{split}.
\end{equation}
We take the non-relativistic limit of the vielbein and obtain
\begin{equation}
  \begin{split}
  E_{H} &=
          \frac{\pd}{\pd t}
          +
          \left(
          1 - x_{+} \csch x_{+}
          \right)
          \frac{x^{a}}{t}
          \frac{\pd}{\pd x^{a}}
  \\
  E_{P_{a}}
        &=
           x_{+} \csch x_{+} \frac{\pd}{\pd x^{a}} .
 \end{split}
\end{equation}
We can now calculate the invariant galilean structure which is given
by the clock one-form and the spatial co-metric ($h=\varsigma \P^2$):
\begin{align}
  \tau &= \eta(\theta)= \sigma dt
  &
    h &=  x_{+}^{2} \csch^{2} x_{+}  \delta^{ab}\frac{\pd}{\pd x^{a}} \otimes \frac{\pd}{\pd x^{b}} .
\end{align}
The fundamental vector fields are given by
\begin{equation}
  \begin{split}
    \xi_{B_a} &= - \varsigma t \frac{\d}{\d x^a}
    \\
    \xi_H &= \frac{\d}{\d t}
    + \left(
      \frac{x_+ \coth x_+ - 1}{x_+^2}
    \right)
    \varkappa
      \varsigma t x^a \frac{\d}{\d x^a}
    \\
    \xi_{P_a} &= 
    x_+ \coth x_+ \frac{\d}{\d x^a} .
\end{split}
\end{equation}

\subsection{Galilean de~Sitter spacetime ($\zdSG$)}
\label{sec:dsg}

We start be setting $\sigma=-1$ and $\varkappa=1$ so that $x_{+}=t$ and see that
\begin{equation}
  \begin{split}
    \theta &= dt \left(H + \frac{t - \sinh (t)}{t^2}\x\cdot\P\right) + \frac{\sinh (t)}{t} d\x\cdot\P \\
    \omega &= \frac{1 - \cosh (t)}{t^2} \left( dt \x\cdot\B - t
      d\x\cdot\B \right).
  \end{split}
\end{equation}
The soldering form is invertible for all $(t,\x)$, since
$\sinh(t)/t \neq 0$ for all $t \in \RR$. From the above soldering
form, it is easily seen that the torsion two-form vanishes and the
curvature two-form is given by
\begin{equation}
  \Omega = \frac{1}{t}\sinh (t) B_{a} (dt \wedge dx^{a}).
\end{equation}

The vielbein is given by
\begin{equation}
  E_H = \frac{\d}{\d t} + \left(1 - t \csch t\right)\frac{x^a}{t} \frac{\d}{\d
    x^a}\qquad\text{and}\qquad
  E_{P_a} = t\csch t \frac{\d}{\d x^a}.
\end{equation}
We can thus find the invariant galilean structure: the clock
one-form is given by $\tau = \eta(\theta) = dt$ and the spatial metric
is given by
\begin{equation}
  h = t^2 \csch^2t \delta^{ab} \frac{\d}{\d x^{a}} \tensor
  \frac{\d}{\d x^{b}}.
\end{equation}
Finally, the fundamental vector fields are
\begin{align}
  \xi_{B_a} &= t \frac{\d}{\d x^{a}}
              \\
\xi_{H} &= \frac{\d}{\d t} + \left(\frac{1}{t} -
          \coth(t)\right)x^{a}\frac{\d}{\d x^{a}}
          \\
  \xi_{P_a} &= t \coth(t) \frac{\d}{dx^{a}} .
\end{align}

\subsection{Galilean anti de~Sitter spacetime ($\zAdSG$)}
\label{sec:adsg}

For $\sigma=-1$ and $\varkappa=1$ the soldering form and connection
one-form for the canonical invariant connection are
\begin{equation}\label{eq:sold-conn-adsg}
\begin{split}
  \theta &= dt \left(H + \frac{t - \sin t}{t^2}\x\cdot\P\right) + \frac{\sin
    t}{t} d\x\cdot\P \\
  \omega &= \frac{1 - \cos t}{t^2}\left( t \x\cdot\B - t d\x\cdot\B
  \right).
\end{split}
\end{equation}
Because of the zero of $\sin(t)/t$ at $t = \pm \pi$, the soldering
form is an isomorphism for all $\x$ and for $t \in (-\pi,\pi)$, so
that the exponential coordinates are invalid outside of that region.
Let $t_0 \in (-\pi,\pi)$ and $\x_0 \in \RR^D$.  The orbit of the point
$(t_0,\x_0)$ under the one-parameter subgroup of boosts generated by
$\w \cdot \B$ is
\begin{equation}
  t(s) = t_0 \qquad\text{and}\qquad \x(s) = \x_0 + s t_0 \w.
\end{equation}
The orbits are point-like for $t_0 = 0$ and straight lines for
$t_0 \neq 0$. These orbits remain inside the domain of validity of the
exponential coordinates. The generic orbits are, therefore,
non-compact.

The torsion two-form again vanishes and the curvature form is
\begin{equation}\label{eq:curv-adsg}
  \Omega = \frac{1}{t}\sin t B_{a} (dt \wedge dx^{a}).
\end{equation}
The vielbein is given by
\begin{equation}\label{eq:viel-adsg}
  E_H = \frac{\d}{\d t} + \left(1 - \frac{t}{\sin t} \right) \frac{x^a}{t} \frac{\d}{\d
    x^a}\qquad\text{and}\qquad
  E_{P_a} = \frac{t}{\sin t} \frac{\d}{\d x^a},
\end{equation}
so that the invariant galilean structure has a clock one-form
$\tau = \eta(\theta) = dt$ and a spatial metric
\begin{equation}\label{eq:nc-adsg}
  h = \left(\frac{t}{\sin t}\right)^2 \delta^{ab} \frac{\d}{\d x^{a}} \tensor
  \frac{\d}{\d x^{b}}.
\end{equation}
The fundamental vector fields for galilean AdS are
\begin{equation}\label{eq:fvf-adsg}
\begin{split}
      \xi_{B_{a}} &= t \frac{\d}{\d x^{a}} \\
      \xi_{P_{a}} &= t \cot t \frac{\d}{\d x^{a}} \\
      \xi_{H} &= \frac{\d}{\d t} + \left(\tfrac{1}{t} - \cot t \right)
      x^{a}\frac{\d}{\d x^{a}} .
    \end{split}
\end{equation}

\subsection{Ultra-relativistic limit}
\label{sec:ultra-relat-limit}

In the ultra-relativistic limit $\sigma \to 0$ to the carrollian (anti)
de Sitter spacetimes we get
$x_{+}= \sqrt{-\frac{\varkappa}{c^2} x^2 }$ and the soldering form and
invariant connection are
\begin{equation}
  \begin{split}
    \theta &= 
    \frac{\sinh x_+}{x_+}
    \left(
      dt H +  d\x \cdot \P
    \right)
    + 
    \left(
      1 - \frac{\sinh x_+}{x_+}
    \right) \frac{\x \cdot d\x}{x^{2}}
    \left(
      t H  +  \x \cdot \P
     \right)\\
    \omega &= \frac{\cosh x_+-1}{x^2} c^{2}
    \left(
      dt \x \cdot\B
      - t d\x \cdot \B
      - \tfrac1{c^2} J_{ab} x^a dx^b
    \right) .
  \end{split}
\end{equation}
The vielbein in the ultra-relativistic limit has the following form
\begin{equation}
  \begin{split}
  E_{H} &=
          x_{+} \csch x_{+}
          \frac{\pd}{\pd t}
  \\
  E_{P_{a}}
        &=
          \frac{x^{a}}{x^{2}}
          \left(
          1 - x_{+} \csch x_{+}
          \right)
          \left(
          t \frac{\pd}{\pd t} + x^{b}\frac{\pd}{\pd x^{b}}
          \right)
          +
           x_{+} \csch x_{+} \frac{\pd}{\pd x^{a}} .
 \end{split}
\end{equation}

The ultra-relativistic limit leads to carrollian structure consisting
of $\kappa=E_H$ and the spatial metric
$b=\frac{1}{c^{2}} \mathbold{\pi}^{2}$ given by
\begin{align}
  b=
  \frac{1}{c^{2}}
  \left(
  \frac{\sinh x_+}{x_+}
  \right)^{2} d \x \cdot d \x
  +
  \frac{1}{c^{2}}
  \left(
  1 - 
  \left(
  \frac{\sinh x_+}{x_+}
  \right)^{2}
  \right)
  \frac{(\x \cdot d\x)^{2}}{x^{2}}
  .
\end{align}
The fundamental vector fields are
\begin{equation}
  \begin{split}
    \xi_{B_a} &= \frac{1}{c^2} x^a \frac{\d}{\d t} 
    \\
    \xi_H &=
      x_+ \coth x_+ 
       \frac{\d}{\d t}
    \\
    \xi_{P_a} &=
     \frac{x^{a}}{x^{2}}
    \left(
      1 - x_+\coth x_+ 
    \right)
    \left(
      t \frac{\d}{\d t} + x^b \frac{\d}{\d x^b}
    \right)
    + x_+ \coth x_+ \frac{\d}{\d x^a} .
\end{split}
\end{equation}

\subsection{(Anti) de~Sitter carrollian spacetimes ($\zdSC$ and $\zAdSC$)}
\label{sec:de-sitter-carroll}

We will treat these two spacetimes together, such that $\varkappa=1$
corresponds to carrollian de~Sitter ($\hyperlink{S14}{\zdSC}$) and
$\varkappa =-1$ to carrollian anti de~Sitter
($\hyperlink{S15}{\zAdSC}$) spacetimes. Furthermore we set $c=1$.

We find that the soldering form is given by
\begin{equation}
  \begin{split}
    \theta^{(\varkappa=1)} &= \frac{\sin|\x|}{|\x|} (dt H + d\x \cdot
    \P) + \left( 1 - \frac{\sin|\x|}{|\x|}\right) \frac{\x \cdot
      d\x}{x^2} (t H + \x \cdot \P)\\
    \theta^{(\varkappa=-1)} &= \frac{\sinh|\x|}{|\x|}(dt H + d\x \cdot
    \P) + \left( 1 - \frac{\sinh|\x|}{|\x|}\right) \frac{\x \cdot
      d\x}{x^2} (t H + \x \cdot \P).
  \end{split}
\end{equation}
These soldering forms are invertible whenever the functions
$\frac{\sin|\x|}{|\x|}$ (for $\varkappa=1$) or
$\frac{\sinh|\x|}{|\x|}$ (for $\varkappa=-1$) are invertible. The
latter function is invertible for all $\x$, whereas the former
function is invertible in the open ball $|\x|<\pi$.

The connection one-form is given by
\begin{equation}
  \begin{split}
    \omega^{(\varkappa=1)} &= \frac{\cos|\x| -1}{x^2}( dt \x \cdot
    \B - t d\x \cdot \B + dx^a x^b J_{ab})\\
    \omega^{(\varkappa=-1)} &= \frac{\cosh|\x| -1}{x^2}( dt \x \cdot
    \B - t d\x \cdot \B + dx^a x^b J_{ab}).
  \end{split}
\end{equation}

The canonical connection is torsion-free, since $\mathsf{(A)dSC}$ is
symmetric, but it is not flat. The curvature is given by
\begin{equation}
\begin{split}
  \Omega^{(\varkappa=1)} = & \left(\frac{\sin\,|\x|}{|\x|}\right)^2 \, dt\wedge d\x\cdot\B - \frac{\sin\,|\x|}{|\x|}\left( \frac{\sin\,|\x|}{|\x|} - 1\right) \frac{\x\cdot\B}{\x\cdot\x} dt \wedge d\x\cdot\x + \\
  & \left(\frac{\sin\,|\x|}{|\x|}\right)^2 J_{ab} dx^{a}\wedge dx^{b} + \frac{2 \sin\,|\x|}{|\x|}\left( \frac{\sin\,|\x|}{|\x|} - 1\right) (x^{c}x^{b} J_{ac} - t x^{b}B_{a})dx^{a} \wedge dx^{b}, \\
  \Omega^{(\varkappa=-1)} = & -\left(\frac{\sinh\,|\x|}{|\x|}\right)^2  dt\wedge d\x\cdot\B + \frac{\sinh\,|\x|}{|\x|}\left( \frac{\sinh\,|\x|}{|\x|} - 1\right) \frac{\x\cdot\B}{\x\cdot\x} dt \wedge d\x\cdot\x - \\
  & \left(\frac{\sinh\,|\x|}{|\x|}\right)^2 J_{ab} dx^{a}\wedge dx^{b} -
  \frac{2 \sinh\,|\x|}{|\x|}\left( \frac{\sinh\,|\x|}{|\x|} - 1\right)
  (x^{c}x^{b} J_{ac} - t x^{b}B_{a})dx^{a} \wedge dx^{b}.
\end{split}
\end{equation}
Using the soldering form, we find the vielbein E to have components
\begin{equation}
\begin{split}
E_{H}^{(\varkappa=1)} = |\x|\csc|\x| \frac{\d}{\d t} \qquad &\text{and} \qquad E_{P_a}^{(\varkappa=1)} = \frac{x^a}{x^2} \left(1-|\x|\csc|\x|\right) \left( t \frac{\d}{\d t} + x^b \frac{\d}{\d x^b}\right) + |\x|\csc|\x|\frac{\d}{\d x^a}, \\
E_{H}^{(\varkappa=-1)} = |\x|\csch|\x| \frac{\d}{\d t} \qquad &\text{and} \qquad E_{P_a}^{(\varkappa=-1)} = \frac{x^a}{x^2} \left(1-|\x|\csch|\x|\right) \left( t \frac{\d}{\d t} + x^b \frac{\d}{\d x^b}\right) + |\x|\csch|\x|\frac{\d}{\d x^a}.
\end{split}
\end{equation}
The invariant carrollian structure is given by $\kappa=E_H$ and the spatial metric
\begin{equation}
\begin{split}
b^{(\varkappa=1)} &= \left( \frac{\sin\, |\x|}{|\x|}\right)^2 d\x\cdot d\x + \left( 1 - \left( \frac{\sin\, |\x|}{|\x|}\right)^2\right)   \frac{(\x \cdot d\x)^{2}}{x^{2}}
 \\
b^{(\varkappa=-1)} &= \left( \frac{\sinh\, |\x|}{|\x|}\right)^2 d\x\cdot d\x + \left( 1 - \left( \frac{\sinh\, |\x|}{|\x|}\right)^2\right)   \frac{(\x \cdot d\x)^{2}}{x^{2}}
.
\end{split}
\end{equation}
Finally, the fundamental vector field of our ultra-relativistic algebras are
\begin{equation}
\begin{split}
  \xi_{B_a} &= x^a \frac{\d}{\d t} \\
  \xi_H^{(\varkappa=1)} &= |\x|\cot|\x| \frac{\d}{\d t}\\
  \xi_H^{(\varkappa=-1)} &= |\x|\coth|\x| \frac{\d}{\d t} \\
  \xi^{(\varkappa=1)}_{P_a} &= \frac{x^a}{x^2}(1-|\x|\cot|\x|)\left(t
    \frac{\d}{\d t} + x^b \frac{\d}{\d x^b}\right) + |\x|\cot|\x|
  \frac{\d}{\d x^a}\\
  \xi^{(\varkappa=-1)}_{P_a} &= \frac{x^a}{x^2}(1-|\x|\coth|\x|)\left(t
    \frac{\d}{\d t} + x^b \frac{\d}{\d x^b}\right) + |\x|\coth|\x|
  \frac{\d}{\d x^a}.
  \end{split}
\end{equation}

\section{Torsional galilean spacetimes}
\label{sec:galilean}

Unlike the galilean symmetric spacetimes discussed in Section
\ref{sec:metric}, some galilean spacetimes do not arise as limits from
the (pseudo\nobreakdash-)riemannian spacetimes: namely, the torsional
galilean de~Sitter ($\hyperlink{S9}{\ztdSG_\gamma}$) and anti
de~Sitter ($\hyperlink{S11}{\ztAdSG_\chi}$) spacetimes and spacetime
$\hyperlink{S12}{\text{\twodgal}_{\gamma,\chi}}$, which are the
subject of this section. Galilean spacetimes can be seen as null
reductions of lorentzian spacetimes one dimension higher and it would
be interesting to exhibit these galilean spacetimes as null
reductions. We hope to return to this question in the future.

\subsection{Torsional galilean de~Sitter spacetime ($\ztdSG_{\gamma\neq 1}$)}
\label{sec:tdsg}

The additional brackets not involving $\J$ for $\hyperlink{S9}{\ztdSG_\gamma}$ are
$[H,\B] = - \P$ and $[H,\P] = \gamma \B + (1+\gamma) \P$, where
$\gamma \in (-1,1)$.

\subsubsection{Fundamental vector fields}
\label{sec:fund-vect-fields-tdsg}

We start by determining the expressions for the fundamental vector
fields $\xi_{B_a}$, $\xi_{P_a}$, and $\xi_H$ relative to the
exponential coordinates.  The boosts are galilean and hence act in the
usual way, with fundamental vector field
\begin{equation}
  \xi_{B_a} = t \frac{\d}{\d x^a}.
\end{equation}
To determine the other fundamental vector fields we must work harder.
The matrix $\ad_A$ in this basis is given by
\begin{equation}
  \ad_A  = t
  \begin{pmatrix}
    \zero  &  \gamma \\
    -1 & 1+\gamma
  \end{pmatrix},
\end{equation}
which is diagonalisable (since $\gamma \neq 1$) with eigenvalues $1$
and $\gamma$, so that $\ad_A = S \Delta S^{-1}$, with
\begin{equation}
  \Delta =
  \begin{pmatrix}
    t & \zero \\
    \zero & t \gamma
  \end{pmatrix} \qquad\text{and}\qquad S
  =
  \begin{pmatrix} \gamma & 1 \\
    1 & 1
  \end{pmatrix}.
\end{equation}
Therefore if $f(z)$ is analytic,
\begin{equation}
  f(\ad_A)
  = S
  \begin{pmatrix}
    f(t) & 0 \\
    0 & f(\gamma t)
  \end{pmatrix}
  S^{-1},
\end{equation}
so that
\begin{equation}
  \begin{split}
    f(\ad_A) \B &= \frac{f(\gamma t) -\gamma f(t)}{1-\gamma} \B +
    \frac{f(\gamma t) - f(t)}{1-\gamma} \P\\
    f(\ad_A) \P &= \frac{\gamma(f(\gamma t) - f(t))}{\gamma-1} \B +
    \frac{\gamma f(\gamma t) - f(t)}{\gamma-1} \P.
  \end{split}
\end{equation}
On the other hand, $\ad_A H = -\gamma \x \cdot \B - (1+\gamma) \x
\cdot\P$, so if $f(z) = 1 + z \widetilde{f}(z)$, then
\begin{equation}
  \begin{split}
      f(\ad_A) H &= H - \gamma \widetilde{f}(\ad_A) \x \cdot \B - (1+\gamma)
      \widetilde{f}(\ad_A) \x \cdot \P\\
      &= H + \frac{\gamma}{1-\gamma} \left(\gamma \widetilde{f}(\gamma t) - \widetilde{f}(t) \right)
      \x \cdot \B + \frac{1}{1-\gamma}\left( \gamma^2 \widetilde{f}(\gamma t) - \widetilde{f}(t) \right) \x \cdot \P,
  \end{split}
\end{equation}
where $\widetilde{f}(t) = (f(t)-1)/t$. With these expressions we can
now use equation \eqref{eq:master} to solve for the fundamental vector
fields.

Put $X = \v\cdot \P$ and $Y'(0) = \bbeta \cdot \B$ in
equation~\eqref{eq:master} to obtain that $\tau = 0$ and
\begin{equation}
  \begin{split}
    \y \cdot \P &= \tfrac1{\gamma-1}
    \left[
      \gamma
      \left(G(\gamma t) -
        \gamma G(t)
      \right) \v \cdot \B
      + \left(
        \gamma G(\gamma t) -  G(t)
      \right) \v \cdot \P
    \right] \\
    & \quad
    - \tfrac1{1-\gamma}
    \left[
      \left(
        F(\gamma t) - \gamma F(t)
      \right)
      \bbeta \cdot \B
      + \left(
        F(\gamma t) - F(t)
      \right) \bbeta \cdot \P
    \right].
\end{split}
\end{equation}
This requires
\begin{equation}
  \bbeta = - \gamma \frac{G(\gamma t) - G(t)}{F(\gamma t) - \gamma F(t)} \v,
\end{equation}
and hence, substituting back into the equation for $\y$ and
simplifying, we obtain
\begin{equation}
  \y = t \left( -1 + \frac{(\gamma-1)e^t}{e^{\gamma t} - e^t} \right) \v,
\end{equation}
so that
\begin{equation}\label{eq:xi-P-tdsg}
  \xi_{P_a} = t \left( -1 + \frac{(\gamma-1)e^t}{e^{\gamma t} - e^t} \right)  \frac{\d}{\d x^a}.
\end{equation}

Finally, let $X = H$ and $Y'(0) = \bbeta \cdot \B$ in
equation~\eqref{eq:master} to obtain that $\tau = 1$ and
\begin{equation}
  \begin{split}
  \y \cdot \P &= \tfrac{\gamma}{1-\gamma} \left(\gamma h(\gamma t) - h(t) \right) 
      \x \cdot \B + \tfrac{1}{1-\gamma}\left( \gamma^2 h(\gamma t) -
        h(t) \right) \x \cdot \P\\
      & \quad {} - \tfrac1{1-\gamma} \left( F(\gamma t) - \gamma F(t) \right) \bbeta \cdot \B  - \tfrac1{1-\gamma} \left(F(\gamma t) -
      F(t) \right) \bbeta \cdot \P,
  \end{split}
\end{equation}
where $h(t) = (G(t)-1)/t$.  This requires
\begin{equation}
  \bbeta = \gamma \frac{\gamma h(\gamma t) - h(t)}{F(\gamma t) - \gamma,
    F(t)} \x
\end{equation}
so that
\begin{equation}
  \y = \left( 1 + \frac1t + \frac{(1-\gamma) e^t}{e^{\gamma t} - e^t}\right)\x.
\end{equation}
This means that
\begin{equation}\label{eq:xi-H-tdsg}
  \xi_H = \frac{\d}{\d t} + \left( 1 + \frac1t + \frac{(1-\gamma) e^t}{e^{\gamma t} - e^t}\right) x^a \frac{\d}{\d x^a}.
\end{equation}
We can easily check that $[\xi_H, \xi_{B_a}] = \xi_{P_a}$ and
$[\xi_H,\xi_{P_a}] = -\gamma \xi_{B_a} - (1 + \gamma) \xi_{P_a}$.

\subsubsection{Soldering form and canonical connection}
\label{sec:sold-form-canon-tdsg}

This homogeneous spacetime is reductive, so we have not just a
soldering form, but also a canonical invariant connection, which can
be determined via equation~\eqref{eq:MC-pullback}:
\begin{equation}
  \begin{split}
    \theta + \omega &= D(\ad_A) (dt H + d\x \cdot \P)\\
    &= dt (H + \tfrac\gamma{1-\gamma}(\gamma \widetilde{D}(\gamma t) - \widetilde{D}(t)) \x
    \cdot \B + \tfrac1{1-\gamma}(\gamma^2 \widetilde{D}(\gamma t) - \widetilde{D}(t)) \x \cdot\P\\
    & \quad {} + \tfrac\gamma{\gamma-1}(D(\gamma t) - D(t)) d\x\cdot\B
    + \tfrac1{\gamma-1}(\gamma D(\gamma t) - D(t)) d\x \cdot \P,
  \end{split}
\end{equation}
where now $\widetilde{D}(z) = (D(z)-1)/z$.  Substituting $D(z) = (1-e^{-z})/z$, we
find that the soldering form is given by
\begin{equation}\label{eq:theta-tdsg}
  \theta = dt \left( H + \frac1t \x \cdot \P \right) + \frac{e^{-t}-e^{-\gamma
        t}}{t^2(1-\gamma)} \left( dt \x - t d\x \right) \cdot \P,
\end{equation}
from where it follows that $\theta$ is invertible for all $(t,\x)$.
The canonical invariant connection is given by
\begin{equation}\label{eq:omega-tsdg}
  \omega = \left( \frac1{t^2} + \frac{\gamma e^{-t} - e^{-\gamma
        t}}{t^2(1-\gamma)} \right) (dt \x  - t d \x) \cdot \B.
\end{equation}
The torsion and curvature of the canonical invariant connection are easily
determined from equations~\eqref{eq:torsion} and \eqref{eq:curvature},
respectively:
\begin{equation}
  \Theta =\left(\frac{1+\gamma}{1-\gamma}\right) \frac{e^{-t} - e^{-\gamma t}}{t}
  dt \wedge d\x \cdot \P \qquad\text{and}\qquad
  \Omega = \left(\frac{\gamma}{1-\gamma}\right) \frac{e^{-t} - e^{-\gamma t}}{t}
  dt \wedge d\x \cdot \B.
\end{equation}

This spacetime admits an invariant galilean structure with clock form
$\tau = \eta(\theta) = dt$ and spatial metric on one-forms
$h = \delta^{ab} E_{P_a} \otimes E_{P_b}$, where $E$ is the vielbein
obtained by inverting the soldering form:
\begin{equation}
  E_H = \frac{\d}{\d t} + \left(\frac1t - \frac{\gamma-1}{e^{-t} -
      e^{-t\gamma}}\right) x^a \frac{\d}{\d x^a}\qquad\text{and}\qquad
  E_{P_a} = \frac{t(\gamma-1)}{e^{-t}-e^{-\gamma t}} \frac{\d}{\d x^a}.
\end{equation}
Therefore, the spatial metric of the galilean structure is given by
\begin{equation}
  h = \frac{t^2(\gamma-1)^2}{(e^{-t} - e^{-\gamma t})^2}
  \delta^{ab}\frac{\d}{\d x^a} \otimes \frac{\d}{\d x^b}.
\end{equation}

\subsection{Torsional galilean de~Sitter spacetime ($\ztdSG_{\gamma=1}$)}
\label{sec:tdsg-1}

This is $\hyperlink{S9}{\ztdSG_1}$, which is the $\gamma\to1$ limit of
the previous example. Some of the expressions in the previous section
have removable singularities at $\gamma = 1$, so it seems that
treating that case in a separate section leads to a more transparent
exposition.

The additional brackets not involving $\J$ are now $[H,\B] = - \P$ and
$[H,\P] = 2 \P + \B$. We start by determining the expressions for the
fundamental vector fields $\xi_{B_a}$, $\xi_{P_a}$, and $\xi_H$
relative to the exponential coordinates $(t,\x)$, where
$\sigma(t,\x) = \exp(t H + \x \cdot \P)$.

\subsubsection{Fundamental vector fields}
\label{sec:fund-vect-fields-}

The bracket $[H,\B] = -\P$ shows that $\B$ acts as a galilean
boost.  We can, therefore, immediately write down
\begin{equation}
  \xi_{B_a} = t \frac{\d}{\d x^a}.
\end{equation}
To find the other fundamental vector fields requires solving
equation~\eqref{eq:master} with $A = tH + \x \cdot \P$ and
$Y'(0)= \bbeta \cdot \B$ (for this Lie algebra) for $X = P_a$ and $X = H$.
To apply equation \eqref{eq:master} we must first determine how to act
with $f(\ad_A)$ on the generators, where $f(z)$ is analytic in $z$.

We start from
\begin{equation}
  \begin{split}
    \ad_A H &= - \x \cdot \B - 2 \x \cdot \P\\
    \ad_A \P &= 2 t \P + t \B\\
    \ad_A \B &= -t \P.
  \end{split}
\end{equation}
It follows from the last two expressions that
\begin{equation}
  \ad_A \begin{pmatrix} \B & \P \end{pmatrix} = 
  \begin{pmatrix} \B & \P \end{pmatrix}
  \begin{pmatrix}
    \zero & t \\ -t & 2 t
  \end{pmatrix},
\end{equation}
where the matrix
\begin{equation}
  M =
  \begin{pmatrix}
    \zero & 1 \\ -1 & 2 
  \end{pmatrix}
\end{equation}
is not diagonalisable, but may be brought to Jordan normal form
$M = S J S^{-1}$, where
\begin{equation}
  J = 
  \begin{pmatrix}
    1 & \zero \\ 1 & 1
  \end{pmatrix}
  \qquad\text{and}\qquad
  S = S^{-1} =
  \begin{pmatrix}
    1 & -1 \\ \zero & -1
  \end{pmatrix}.
\end{equation}
It follows that for $f(z)$ analytic in $z$,
\begin{equation}
  f(\ad_Z) \begin{pmatrix} \B & \P \end{pmatrix} = 
  \begin{pmatrix} \B & \P \end{pmatrix} S f(tJ) S.
\end{equation}
If $f(z) = \sum_{n=0}^\infty c_n z^n$,
\begin{equation}
  f(t J) = \sum_{n=0}^\infty c_n t^n
  \begin{pmatrix}
    1 & \zero \\ n & 1
  \end{pmatrix} =
  \begin{pmatrix}
    f(t) & \zero \\ t f'(t) & f(t)
  \end{pmatrix}.
\end{equation}
Performing the matrix multiplication, we arrive at
\begin{equation}
  \begin{split}
    f(\ad_A) \B & = (f(t) - t f'(t)) \B - t f'(t)) \P\\
    f(\ad_A) \P & = t f'(t)\B + (f(t) + t f'(t)) \P.
  \end{split}
\end{equation}
Similarly,
\begin{equation}
  f(\ad_A) H = f(0) H - 2 \x \cdot \widetilde{f}(\ad_A) \P - \x \cdot \widetilde{f}(\ad_A) \B,
\end{equation}
where $\widetilde{f}(z) = (f(z)-f(0))/z$.

We are now ready to apply equation~\eqref{eq:master}. Let
$X = \v \cdot \P$. Then equation~\eqref{eq:master} becomes
\begin{equation}
  \begin{split}
    \tau H + \y \cdot \P &= G(\ad_A) \v \cdot \P - F(\ad_A) \bbeta \cdot \B\\
    &= (G(t) + t G'(t)) \v \cdot \P + t G'(t) \v \cdot \B - (F(t) - t
    F'(t))\bbeta \cdot \B + t F'(t) \bbeta \cdot \P,
  \end{split}
\end{equation}
from where we find that $\tau = 0$,
\begin{equation}
  \bbeta = \frac{t G'(t)}{F(t) - tF'(t)} \v \qquad\text{and hence}\qquad
  \y = \frac{F(t)G(t) + t (F(t) G'(t)-F'(t)G(t))}{F(t) - t F'(t)} \v =
  (1-t) \v,
\end{equation}
so that
\begin{equation}
  \xi_{P_a} = (1-t) \frac{\d}{\d x^a},
\end{equation}
which is indeed the limit $\gamma \to 1$ of
equation~\eqref{eq:xi-P-tdsg}.

Now let $X = H$, so that equation~\eqref{eq:master} becomes
\begin{equation}
  \begin{split}
    \tau H + \y \cdot \P &= G(\ad_A) H - \bbeta \cdot F(\ad_A) \B \\
    &= H - 2 \x \cdot \widetilde{G}(\ad_A) \P - \x \cdot
    \widetilde{G}  (\ad_A) \B -  \bbeta \cdot  F(\ad_A)\B\\
    &= H - (\widetilde{G}(t) + t \widetilde{G}'(t)) \x \cdot \B -
    (F(t) -t F'(t)) \bbeta \cdot \B - (2 \widetilde{G}(t) + t
    \widetilde{G}'(t))\x \cdot \P + t F'(t) \bbeta\cdot \P, 
  \end{split}
\end{equation}
from where $\tau = 1$,
\begin{equation}
  \bbeta = \frac{\widetilde{G}(t) + t \widetilde{G}'(t)}{t F'(t) - F(t)}
  \x \qquad\text{and hence}\qquad
  \y = \frac{t(F'(t) \widetilde{G}(t) - F(t) \widetilde{G}'(t)) - 2
    F(t) \widetilde{G}(t)}{F(t) - t F'(t)} \x = \x.
\end{equation}
In summary,
\begin{equation}
  \xi_H = \frac{\d}{\d t} + x^a \frac{\d}{\d x^a},
\end{equation}
which is indeed the $\gamma\to 1$ limit of
equation~\eqref{eq:xi-H-tdsg}.

\subsubsection{Soldering form and canonical connection}
\label{sec:sold-form-canon-tdsg1}

To calculate the soldering form and the connection one-form for the
canonical invariant connection, we apply equation
\eqref{eq:MC-pullback}:
\begin{equation}
  \begin{split}
    \sigma^*\vartheta &= D(\ad_A) (dt H + d\x \cdot \P)\\
    &= dt \left( H - 2 \x \cdot \widetilde{D}(\ad_A) \P - \x \cdot
      \widetilde{D}(\ad_A)\B\right) + d\x \cdot D(\ad_A) \P\\
    &= dt \left( H - ( \widetilde{D}(t) + t  \widetilde{D}'(t))
      \x\cdot\B - (2  \widetilde{D}(t) + t  \widetilde{D}'(t)) \x
      \cdot \P \right) + (D(t) + t D'(t)) d\x \cdot \P + t D'(t) d\x
    \cdot \B.
  \end{split}
\end{equation}
Performing the calculation,
\begin{equation}
  \begin{split}
    \theta &= dt \left( H + \frac{1-e^{-t}}{t} \x \cdot \P \right) + e^{-t} d\x
    \cdot \P\\
    \omega &= \frac1t \left(\frac{1-e^{-t}}{t} - e^{-t}\right) (\x
    \cdot \B dt - t d\x \cdot \B),
  \end{split}
\end{equation}
which are equations~\eqref{eq:theta-tdsg} and \eqref{eq:omega-tsdg} in
the limit $\gamma\to 1$. Notice that $\theta$ is an isomorphism for
all $(t,\x)$.

The torsion and curvature two-forms for the canonical
invariant connection are given by
\begin{equation}
  \Theta = -2 e^{-t} dt \wedge d \x \cdot \P \qquad\text{and}\qquad
  \Omega = - e^{-t} dt \wedge d \x \cdot \B.
\end{equation}
The vielbein $E$ has components
\begin{equation}
  E_H = \frac{\d}{\d t} + \frac{1-e^{t}}{t} x^a \frac{\d}{\d x^a}
  \qquad\text{and}\qquad
  E_{P_a} = e^t \frac{\d}{\d x^a}.
\end{equation}

The invariant galilean structure has clock form $\tau
=\eta(\theta) = dt$ and inverse spatial metric
\begin{equation}
  h = \delta^{ab} E_{P_a}\otimes E_{P_b} = e^{2t}  \delta^{ab}
  \frac{\d}{\d x^a} \otimes \frac{\d}{\d x^b}.
\end{equation}

\subsection{Torsional galilean anti de~Sitter spacetime ($\ztAdSG_\chi$)}
\label{sec:spacetime-tadsg}

Here $[H,\B] = - \P$ and $[H,\P] = (1+\chi^2)\B + 2\chi \P$.

\subsubsection{Fundamental vector fields}
\label{sec:fund-vect-fields-tadsg}

Since $\B$ acts via galilean boosts we can immediately write down
\begin{equation}
  \xi_{B_a} = t \frac{\d}{\d x^a}.
\end{equation}
To calculate the other fundamental vector fields we employ
equation~\eqref{eq:master}. The adjoint action of
$A = t H + \x \cdot \P$ is given by
\begin{equation}
  \begin{split}
    \ad_A H &= -(1+\chi^2) \x \cdot \B - 2 \chi \x \cdot \P\\
    \ad_A \B &= - t \P\\
    \ad_A \P &= t (1+\chi^2) \B + 2 t \chi \P.
  \end{split}
\end{equation}
In matrix form,
\begin{equation}
  \ad_A \begin{pmatrix} \B & \P \end{pmatrix}= \begin{pmatrix} \B & \P \end{pmatrix}
  \begin{pmatrix}
    \zero & (1+\chi^2)t \\ -t & 2 t \chi
  \end{pmatrix}.
\end{equation}
We notice that this matrix is diagonalisable:
\begin{equation}
  \begin{pmatrix}
    \zero & (1+\chi^2) \\ -1 & 2 \chi
  \end{pmatrix} = S \Delta S^{-1}, \qquad\text{where}\qquad S :=
  \begin{pmatrix}
    \chi+i & \chi - i\\ 1 & 1
  \end{pmatrix}
  \qquad\text{and}\qquad
  \Delta :=
  \begin{pmatrix}
    \chi -i & \zero \\ \zero & \chi + i
  \end{pmatrix}.
\end{equation}
So if $f(z)$ is analytic in $z$,
\begin{equation}
  f(\ad_A) \begin{pmatrix} \B & \P \end{pmatrix} = \begin{pmatrix} \B
    & \P \end{pmatrix} S f(t\Delta) S^{-1}~,
\end{equation}
or letting $t_\pm := t(\chi \pm i)$,
\begin{equation}
  \begin{split}
    f(\ad_A) \B &= \tfrac{i}2 (f(t_+) - f(t_-)) (\P + \chi \B) + \tfrac12 (f(t_+) + f(t_-))\B\\
    f(\ad_A) \P &=-\tfrac{i}2 (f(t_+) - f(t_-))(\chi \P +
    (1+\chi^2) \B)  + \tfrac12 (f(t_+) + f(t_-))\P.
  \end{split}
\end{equation}
Similarly,
\begin{equation}
  \begin{split}
    f(\ad_A) H &= f(0) H + \frac1{\ad_A}(f(\ad_A)-f(0)) \ad_A H\\
    &= f(0) H - (1+\chi^2) \x \cdot \widetilde{f}(\ad_A) \B - 2 \chi \x \cdot \widetilde{f}(\ad_A) \P,
  \end{split}
\end{equation}
where $\widetilde{f}(z) := (f(z)-f(0))/z$. With these formulae we can
now use equation~\eqref{eq:master} to find out the expressions for the
fundamental vector fields $\xi_H$ and $\xi_{P_a}$. Putting
$X = \v \cdot \P$ and $Y'(0) = \bbeta \cdot \B$ in
equation~\eqref{eq:master} we arrive at
\begin{equation}
  \bbeta = \frac{-i(1+\chi^2)(G(t_+)-G(t_-))}{F(t_+)+F(t_-) + i
    \chi(F(t_+)-F(t_-))} \v
\end{equation}
and hence
\begin{equation}\label{eq:xi-P-tadsg}
  \xi_{P_a} = t (\cot t - \chi) \frac{\d}{\d x^a}.
\end{equation}

Similarly, putting $X = H$ and $Y'(0) = \bbeta \cdot \B$ in 
equation~\eqref{eq:master} we find
\begin{equation}
  \bbeta = \frac{i\chi(\widetilde{G}(t_+) - \widetilde{G}(t_-)) -
    (\widetilde{G}(t_+)+\widetilde{G}(t_-))}{F(t_+) + F(t_-) + i
    \chi (F(t_+) - F(t_-))} \x
\end{equation}
and hence
\begin{equation}\label{eq:xi-H-tadsg}
  \xi_H = \frac{\d}{\d t} + \left(\tfrac1t + \chi - \cot
      t\right) x^a \frac{\d}{\d x^a}.
\end{equation}
We check that $[\xi_H, \xi_{B_a}] = \xi_{P_a}$ and
$[\xi_H, \xi_{P_a}] = -(1+\chi^2) \xi_{B_a} -2 \chi \xi_{P_a}$, as
expected. Another check is that taking $\chi \to 0$, we recover the
fundamental vector fields for galilean anti de~Sitter spacetime given
by equation~\eqref{eq:fvf-adsg}.

\subsubsection{Soldering form and canonical connection}
\label{sec:sold-form-canon-tadsg}

Let us now use equation~\eqref{eq:MC-pullback} to calculate the
soldering form $\theta$ and the connection one-form $\omega$ for the
canonical invariant connection:
\begin{equation}
  \begin{split}
    \theta + \omega &= D(\ad_A) (dt H + d\x \cdot \P)\\
    &= dt \left(H - (1+\chi^2) \x \cdot \widetilde{D}(\ad_A) \B - 2
      \chi \x \cdot \widetilde{D}(\ad_A) \P\right) + d\x \cdot
    D(\ad_A) \P, \\
  \end{split}
\end{equation}
where $\widetilde{D}(z) = (D(z) -1)/z$.  Evaluating these expressions,
we find
\begin{equation}
  \theta = dt \left(H + \frac{(t-e^{\chi t}\sin t)}{t^2} \x \cdot
    \P\right) + \frac1{t} e^{-\chi t}\sin t d\x \cdot \P
\end{equation}
and
\begin{equation}
  \omega = \frac{1 - e^{-\chi t}(\cos t + \chi \sin t)}{t^2} (dt
  \x \cdot \B - t d\x \cdot \B).
\end{equation}
Again, the zeros of $\frac{e^{-\chi t}\sin t}{t}$ at $t = \pm \pi$
invalidate the exponential coordinates for $t \not\in
(-\pi,\pi)$.

The torsion and curvature of the canonical invariant connection are
easily calculated to be
\begin{equation}
  \begin{split}
    \Theta &=  -\frac{2\chi}{t} e^{-\chi t} \sin t dt \wedge d\x
    \cdot \P\\
    \Omega &= - \frac{(1+\chi^2)}{t}  e^{-\chi t} \sin t dt \wedge d\x
    \cdot \B.
  \end{split}
\end{equation}
As $\chi \to 0$, the torsion vanishes and the curvature agrees with
that of the galilean anti de~Sitter spacetime (\adsg) in
equation~\eqref{eq:curv-adsg}.

The vielbein $E$ has components
\begin{equation}
  \begin{split}
    E_H &= \frac{\d}{\d t} + \left(\frac1t - e^{\chi t}\csc t\right)x^a\frac{\d}{\d x^a}\\
    E_{P_a} &= t e^{\chi t}\csc t \frac{\d}{\d x^a},
  \end{split}
\end{equation}
whose $\chi \to 0$ limit agrees with equation~\eqref{eq:viel-adsg}.
The invariant galilean structure has clock form $\tau =
\eta(\theta) = dt$ and inverse spatial metric
\begin{equation}
  h = t^2 e^{2 \chi t} \csc^2t \delta^{ab}  \frac{\d}{\d x^a}
  \otimes  \frac{\d}{\d x^b},
\end{equation}
which again agrees with equation~\eqref{eq:nc-adsg} in the limit $\chi
\to 0$.

\subsection{Spacetime $\text{\twodgal}_{\gamma,\chi}$}
\label{sec:three-twodgal}

There is a two-parameter family of spacetimes which is unique to
$D=2$. Here the additional brackets are $[H, \B] = - \P$, and
$[H, \P] = (1+\gamma) \P - \chi \Pt + \gamma \B - \chi \tilde{\B}$. To
make the following calculations easier we may complexify the algebra
by defining $\PP = \P_{1} + i \P_{2}$ and $\BB = \B_{1} + i \B_{2}$
such that the brackets become
$[H, \BB] = - \PP, [H, \PP] = (1+ z) \PP + z \BB$, where
$z = \gamma + i \chi$ . We start by determining the expressions for
the fundamental vector fields $\xi_{B_a}$, $\xi_{P_a}$, and $\xi_H$.

\subsubsection{Fundamental vector fields}
\label{sec:fund-vect-fields-twodgal}

Since $\B$ acts via galilean boosts we can immediately write down
\begin{equation}
  \xi_{B_a} = t \frac{\d}{\d x^a}.
\end{equation}
To calculate the other fundamental vector fields we employ
equation~\eqref{eq:master}. The adjoint action of $A = tH + \x\cdot\P$
on a basis $(\BB, \PP)$ is given by
\begin{equation}
  \ad_A  =  t
  \begin{pmatrix}
    \zero  &  z \\
    -1 & 1+z
  \end{pmatrix}.
\end{equation}
Notice that this matrix is diagonalisable:
\begin{equation}
  \begin{pmatrix}
    \zero & z \\
    -1 & 1 + z
  \end{pmatrix}
  = S \Delta S^{-1}, \qquad\text{where}\qquad S :=
  \begin{pmatrix}
    z& 1\\ 1 & 1
  \end{pmatrix}
  \qquad\text{and}\qquad
  \Delta :=
  \begin{pmatrix}
    1 & \zero \\
    \zero & z
  \end{pmatrix}.
\end{equation}
So if $f(\zeta)$ is an analytic function of $\zeta$,
\begin{equation}
  f(\ad_A) \begin{pmatrix} \BB & \PP \end{pmatrix} = \begin{pmatrix} \BB
    & \PP \end{pmatrix} S f(t\Delta) S^{-1}~,
\end{equation}
such that
\begin{equation}
  \begin{split}
    f(\ad_A) \BB &= \frac{ f(z t) - z f(t)}{1-z} \BB +
    \frac{f(zt) - f(t)}{1-z} \PP\\
    f(\ad_A) \PP &= \frac{z(f(t) - f(z t))}{1-z} \BB +
    \frac{f(t) - z f(z t)}{1-z} \PP.
  \end{split}
\end{equation}
Let $\widetilde f(\zeta) := (f(\zeta) -
f(0))/\zeta$.  Then we may write, using the notation $\xx = x^1 + i
x^2$,
\begin{equation}
\begin{split}
  \ad_{A} H &= -\tfrac1t \Re\left(\xbar \ad_{A} \PP\right) \\
  f(\ad_{A}) H &= H - \Re\left( \xbar \widetilde f(\ad_{A})(z \BB + (1+z)\PP)\right).
\end{split}
\end{equation}
Similarly, let $\vv$, $\ww$, and $\yy$ now be complex numbers.  Setting
$X = \Re(\vbar\PP)$ and $Y'(0)=\Re(\wbar\BB)$ we obtain $\tau = 0$ and
\begin{equation}
  \begin{split}
    \Re(\ybar \PP) = \Re
    \left(
      \vbar \frac{1}{1-z} \left(z(G(t) - G(zt)) \BB + (G(t) - z G(z t)) \PP \right)\right. \\
     \left. - \wbar
      \frac{1}{z-1}\left( (z F(t) - F(z t) ) \BB + (F(t) - F(z t))
        \PP\right)
  \right).
\end{split}
\end{equation}
This requires
\begin{equation}
  \wbar = \frac{z(G(z t) - G(t))}{z F(t) - F(z t)} \vbar.
\end{equation}
Substituting back into the equation we find
\begin{equation}
  \yy = t \left( -1 + \frac{e^t(-1 + \gamma - i \chi)}{-e^t+
      e^{t(\gamma - i \chi)}}\right) \vv =: (a + i b) \vv,
\end{equation}
where we have introduced $a$ and $b$ as the real and imaginary parts
of the expression multiplying $\vv$.  In full
glory,
\begin{equation}
  \begin{split}
    a &= \frac{t((\gamma-1)\cos(t\gamma) +
      (1+\gamma)\cosh(t(\gamma-1)) - \chi \sin(t\chi) +
      (\gamma-1)\sinh(t(1-\gamma)))}{2(\cos(t\chi) -
      \cosh(t(\gamma-1)))}\\
    b &= \frac{t(\chi \cos(t \chi) + (1-\gamma) \sin(t\chi) -
      e^{t(1-\gamma)}\chi)}{2(\cos(t\chi) - \cosh(t(\gamma-1)))},
  \end{split}
\end{equation}
so that $y^a = a v^a - b \epsilon_{ab} v^b$ and hence
\begin{equation}
  \xi_{P_a} = a \frac{\d}{\d x^a} + b \epsilon_{ab} \frac{\d}{\d x^b}.
\end{equation}

Now letting $X = H$ and $Y'(0) = \Re(\wbar\BB)$, we obtain $\tau = 1$ and 
\begin{equation}
  \begin{split}
    \Re(\ybar\PP) &= - \Re(\xbar \tG(\ad_A) ((1+z)\PP + z \BB) -
    \Re(\wbar F(\ad_A)\BB)\\
    &= -\Re\left( \frac{\xbar (1+z)}{1-z}\left( z (\tG(t) - \tG(z
        t))\BB + (\tG(t) - z \tG(z t))\PP\right)\right)\\
    &\quad {} - \Re\left( \frac{\xbar z}{1-z}\left( (\tG(z t) - z
        \tG(t))\BB + (\tG(z t) - \tG(t))\PP \right) \right)\\
    &\quad {} - \Re\left( \frac{\wbar}{1-z} \left( (F(z t) - z F(t))\BB
        + (F(z t) - F(t)) \PP\right) \right).
  \end{split}
\end{equation}
We solve for $\ww$ to find
\begin{equation}
  \wbar = \frac{z(z \tG(z t) - \tG(t))}{F(z t) - z F(t) } \xbar.
\end{equation}
Substituting this back in to the equation, we find
\begin{equation}
  \yy = \left(1 + \tfrac{1}{t} + \frac{(-1+\gamma-i \chi)e^{t}}{e^t -
      e^{(\gamma - i \chi)t})}\right)\xx =: (c + i d) \xx,
\end{equation}
where $c, d$ are the real and imaginary parts of the expression
multiplying $\xx$.  Expanding we find
\begin{equation}
  \begin{split}
    c &= \frac{e^{2\gamma t}(1+t) + e^{2t}(1+t\gamma) -
      e^{t(1+\gamma)} \left(2 + t(1+\gamma)\cos(t\chi) +
        t\chi\sin(t\chi)\right)}{t\left(e^{2t} + e^{2 t\gamma} - 2
        e^{t(1+\gamma)} \cos(t\chi)\right)}\\
    d &= \frac{-e^{2t}\chi + e^{t(1+\gamma)} \left(\chi \cos(t\chi) +
        (1-\gamma)\sin(t\chi)\right)}{t\left(e^{2t} + e^{2 t\gamma} -
        2 e^{t(1+\gamma)} \cos(t\chi)\right)},
  \end{split}
\end{equation}
so that $y^a = c x^a - d \epsilon_{ab} x^b$ and hence
\begin{equation}
  \xi_{H} = \frac{\d}{\d t} + c x^{a} \frac{\d}{\d x^{a}} + d
    \epsilon_{ab} x^a \frac{\d}{\d x^{b}}.
\end{equation}
One can check that $[\xi_{H}, \xi_{P_a}] = \xi_{[P_a, H]}$ and
$[\xi_{H}, \xi_{B_a}] = \xi_{[B_a, H]}$.

\subsubsection{Soldering form and canonical connection}
\label{sec:sold-form-conn-twodgal}

We can now use equation~\eqref{eq:MC-pullback} in order to calculate
the soldering form $\theta$ and the connection one-form $\omega$ for
the canonical invariant connection:
\begin{equation}
  \begin{split}
    \theta + \omega &= D(\ad_{A}) (dt H + d\x\cdot\P) \\
    &= dt H - dt \Re \left( \xbar \tD(\ad_{A})(z \BB + (1+z)\PP)\right) + \Re\left(d\xbar D(\ad_{A})\PP\right),
  \end{split}
\end{equation}
where $\tD(\zeta) = (D(\zeta) - 1)/\zeta$.  Evaluating these expressions we find
\begin{equation}
  \theta = dt H + \frac{dt}{t} \Re (\xbar \PP) + \Re\left(\frac{t
      d\xbar - \xbar dt}{t^2(z-1)}\left(e^{-t} - e^{-t z}\right) \PP \right)
\end{equation}
and
\begin{equation}
  \omega = \Re\left(\frac{\xbar dt - t d\xbar}{t^2} \left( 1 +
      \frac{e^{-t z} - z e^{-t}}{z-1} \right) \BB\right).
\end{equation}
It is not immediately obvious from the expression for $\theta$ whether
it fails to be an isomorphism. Because $\theta^H_t = 1$, the soldering
form is invertible provided that the determinant of $\theta^{P_a}_b$
does not vanish. Unpacking the complex notation, we find that the
determinant is given by
\begin{equation}
  \frac{e^{-2 t (\gamma+1)} \left(e^{2 t \gamma}+e^{2 t}- 2 e^{t(\gamma+1)} \cos (t
      \chi) \right)}{t^2 \left((\gamma-1)^2+\chi^2\right)}.
\end{equation}
This is nowhere zero for $\gamma \in [-1,1)$.  But if $\gamma = 1$,
then it becomes
\begin{equation}
 \frac{2 e^{-2 t} \left(1 - \cos (t\chi) \right)}{t^2 \chi^2},
\end{equation}
which vanishes whenever $t\chi = 2 \pi k$, $k = \pm1, \pm2,\cdots$.
Therefore, for $\chi >0$ and $\gamma \in [-1,1)$, the soldering form
is invertible everywhere, whereas if $\gamma = 1$ then it is
invertible for $t \in (-\frac{2\pi}{\chi}, \frac{2\pi}{\chi})$ and for
all $\x \in \RR^2$. For $\chi = 0$, the soldering form is invertible
everywhere. This agrees with $\hyperlink{S9}{\ztdSG_\gamma}$ and
$\ztAdSG_{2/\chi}$, which are the $\chi\to 0$ and $\gamma\to 1$ limits
of $\hyperlink{S12}{\text{\twodgal}_{\gamma,\chi}}$,
respectively.\footnote{One might ask why in $\ztAdSG_{2/\chi}$ the
  range of $t$ does not involve $\chi$ but here it does. It has to do
  with the complex change of basis which gives the isomorphism
  $\text{\twodgal}_{1,\chi} \cong \ztAdSG_{2/\chi}$.}

The torsion and curvature of the canonical invariant connection are
calculated to be
\begin{equation}
  \begin{split}
    \Theta &= - \Re\left( \frac{1+z}{t(z-1)} \left(e^{-t} - e^{-t z}
      \right) dt \wedge d\xbar \PP\right)\\
    \Omega &= - \Re\left( \frac{z}{t(z-1)} \left(e^{-t} - e^{-t z}
      \right) dt \wedge d\xbar \BB\right).
  \end{split}
\end{equation}

Using the soldering form we can read-off the vielbein and deduce the
invariant galilean structure. The clock one-form is
$\tau = \eta(\theta) = dt$ and the inverse spatial metric

\begin{equation}
  h = \left(\Re\left(\frac{(z-1)t}{e^{-t} - e^{-zt}}\right)\right)^2 \delta^{a b}
  \frac{\d}{\d x^{a}} \tensor \frac{\d}{\d x^{b}}.
\end{equation}

\subsection{The action of the boosts}
\label{sec:action-boosts-2}

In this section we show that the generic orbits of boosts are not
compact in the torsional galilean spacetimes discussed above. This
requires a different argument to the ones we used for the symmetric
spaces.

Let $M$ be one of the torsional galilean spacetimes discussed in this
section; that is, $\hyperlink{S9}{\ztdSG_\gamma}$,
$\hyperlink{S11}{\ztAdSG_\chi}$ or
\hyperlink{S12}{$\text{\twodgal}_{\gamma,\chi}$}, for the relevant
ranges of their parameters. The following discussion applies verbatim
to the torsional galilean (anti) de Sitter, whereas for
$\hyperlink{S12}{\text{\twodgal}_{\gamma,\chi}}$ the exposition is
more cumbersome; although, as we will see, the result still holds.

Our default description of $M$ is as a simply-connected kinematical
homogeneous spacetime $\Kgr/\Hgr$, where $\Kgr$ is a simply-connected
kinematical Lie group and $\Hgr$ is the connected subgroup generated
by the boots and rotations. Our first observation is that we may
dispense with the rotations and also describe $M$ as $\Sgr/\Bgr$,
where $\Sgr$ is the simply-connected solvable Lie group generated by
the boosts and spatio-temporal translations and $\Bgr$ is the
connected abelian subgroup generated by the boosts. The Lie algebra
$\s$ of $\Sgr$ is spanned by $H,B_a,P_a$ and the Lie algebra $\b$ of
$\Bgr$ is spanned by $B_a$ with non-zero brackets
\begin{equation}
  [H,B_a] = -P_a \qquad\text{and}\qquad [H,P_a] = \alpha B_a + \beta P_a~,
\end{equation}
for some real numbers $\alpha,\beta$ depending on the parameters
$\gamma$, $\chi$.  We may identify $\s$ with the Lie subalgebra of
$\gl(2D+1,\RR)$ given by
\begin{equation}
  \s = \left\{
    \begin{pmatrix}
      \zero & t \alpha \1 & \y \\
      -t \1 & t \beta \1 & \x \\
      \zero & \zero & \zero      
    \end{pmatrix}
\middle | (t,\x,\y) \in \RR^{2D+1}\right\},
\end{equation}
where $\1$ is the $D\times D$ identity matrix and $\b$ with the Lie
subalgebra
\begin{equation}
  \b = \left\{
    \begin{pmatrix}
      \zero & \zero & \y \\
      \zero & \zero & \zero \\
      \zero & \zero & \zero      
    \end{pmatrix}
\middle | \y \in \RR^D\right\}.
\end{equation}
The Lie algebras $\b \subset \s \subset \gl(2D+1,\RR)$ are the Lie
algebras of the subgroups
$\Bgrbar \subset \Sgrbar \subset \GL(2D+1,\RR)$ given by
\begin{equation}
  \Sgrbar = \left\{
    \begin{pmatrix}
      a(t) \1 & b(t) \1 & \y \\
      c(t) \1 & d(t) \1 & \x \\
      \zero & \zero & 1      
    \end{pmatrix}
    \middle | (t,\x,\y) \in \RR^{2D+1}\right\} \qquad\text{and}\qquad
  \Bgrbar = \left\{
    \begin{pmatrix}
      \1 & \zero & \y \\
      \zero & \1 & \zero \\
      \zero & \zero & 1      
    \end{pmatrix}
    \middle | \y \in \RR^D\right\},
\end{equation}
for some functions $a(t),b(t),c(t),d(t)$ which are given explicitly by
\begin{equation}\label{eq:abcd-tdsg}
  \begin{pmatrix}
    a(t) & b(t) \\ c(t) & d(t) 
  \end{pmatrix} = \frac{1}{\gamma -1}
  \begin{pmatrix}
    \gamma e^t - e^{\gamma t} & \gamma \left( e^{\gamma t} - e^t\right)\\
    e^t - e^{\gamma t} & \gamma e^{t\gamma} - e^t
  \end{pmatrix}
\end{equation}
for $\hyperlink{S9}{\ztdSG_\gamma}$ with $\gamma \in (-1,1)$,
\begin{equation}\label{eq:abcd-tdsg1}
  \begin{pmatrix}
    a(t) & b(t) \\ c(t) & d(t) 
  \end{pmatrix} =
  \begin{pmatrix}
    e^t (1-t) & e^t t \\
    -e^t t & e^t (1+t)
  \end{pmatrix}
\end{equation}
for $\hyperlink{S9}{\ztdSG_1}$, and
\begin{equation}\label{eq:abcd-tadsg}
  \begin{pmatrix}
    a(t) & b(t) \\ c(t) & d(t) 
  \end{pmatrix} = 
  \begin{pmatrix}
    e^{t\chi} (\cos t - \chi \sin t) & e^{t\chi} (1+\chi^2) \sin t\\
    -e^{t\chi} \sin t & e^{t\chi} (\cos t + \chi \sin t)
  \end{pmatrix}
\end{equation}
for $\hyperlink{S11}{\ztAdSG_\chi}$ with $\chi > 0$. The homogeneous
space $\Mbar = \Sgrbar/\Bgrbar$, if not simply connected, is
nevertheless a discrete quotient of the simply-connected $M$ and, as
argued at the end of Section~\ref{sec:boosts}, it is enough to show
that the orbits of boosts in $\Mbar$ are generically non-compact to
deduce that the same holds for $M$.

Let us denote by $g(t,\x,\y) \in \Sgrbar$ the generic group element
\begin{equation}
  g(t,\x,\y) = \begin{pmatrix}
    a(t) \1 & b(t) \1 & \y \\
    c(t) \1 & d(t) \1 & \x \\
    \zero & \zero & 1      
  \end{pmatrix}\in \Sgrbar,
\end{equation}
so that the generic boost is given by
\begin{equation}
  g(0,0,\y) = \begin{pmatrix}
    \1 & \zero & \y \\
    \zero & \1 & \zero \\
    \zero & \zero & 1      
  \end{pmatrix}\in \Bgrbar.
\end{equation}

Parenthetically, let us remark that while it might be tempting to
identify $\Mbar$ with the submanifold of $\Sgrbar$ consisting of
matrices of the form $g(t,\x,0)$, this would not be correct. For this
to hold true, it would have to be the case that given $g(t,\x,\y)$,
there is some $g(0,0,\w)$ such that
$g(t,\x,\y) g(0,0,\w) = g(t',\x',0)$ for some $t',\x'$. As we now
show, this is only ever the case provided that $a(t) \neq 0$. Indeed,
\begin{equation}
  g(t,\x,\y)g(0,0,\w) = g(t, c(t) \w + \x, a(t)\w + \y),
\end{equation}
and hence this is of the form $g(t',\x',0)$ if and only if we can
solve $a(t) \w + \y = 0$ for $\w$. Clearly this cannot be done if
$a(t) = 0$, which may happen for
$\hyperlink{S9}{\ztdSG_{\gamma\in(0,1)}}$ at
$t = \frac{\log\gamma}{\gamma -1}$ and for
$\hyperlink{S11}{\ztAdSG_{\chi>0}}$ at
$\cos t = \pm \frac{\chi}{\sqrt{1+\chi^2}}$.

The action of the boosts on $\Mbar$ is induced by left multiplication
on $\Sgrbar$:
\begin{equation}
  g(0,0,\v)  g(t,\x,\y)= g(t,\x,\y + \v)
\end{equation}
which simply becomes a translation $\y \mapsto \y + \v$ in $\RR^D$.
This is non-compact in $\Sgrbar$, but we need to show that it is
non-compact in $\Mbar$.

The right action of $\Bgrbar$ is given by
\begin{equation}
  g(t,\x,\y) g(0,0,\w) = g(t,\x + c(t)\w, \y + a(t) \w),
\end{equation}
which is again a translation
$(\x,\y) \mapsto (\x + c(t)\w, \y + a(t)\w)$ in $\RR^{2D}$. The
quotient $\RR^{2D}/\Bgrbar$ is the quotient vector space
$\RR^{2D}/\mathbb{B}$, where $\mathbb{B}\subset \RR^{2D}$ is the image
of the linear map $\RR^D \to \RR^{2D}$ sending
$\w \to (c(t) \w, a(t) \w)$. Notice that $(a(t),c(t)) \neq (0,0)$ for
all $t$, since the matrices in $\Sgrbar$ are invertible, hence
$\BB \cong \RR^D$ and hence the quotient vector space
$\RR^{2D}/\mathbb{B} \cong \RR^{D}$. By the Heine--Borel theorem, it
suffices to show that the orbit is unbounded to conclude that it is
not compact. Let $[(\x,\y)] \in \RR^{2D}/\mathbb{B}$ denote the
equivalence class modulo $\mathbb{B}$ of $(\x,\y) \in \RR^{2D}$. The
distance $d$ between $[(\x,\y)]$ and the boosted $[(\x, \y + \v)]$ is
the minimum of the distance between $(\x,\y)$ and any point on the
coset $[(\x, \y + \v)]$; that is,
\begin{equation}
  d = \min_{\w} \|(\x + c(t)\w, \y + \v + a(t) \w) - (\x,\y)\| =
  \min_{\w} \|(c(t)\w, \v + a(t)\w)\|.
\end{equation}
Completing the square, we find
\begin{equation}
    \|(c\w, \v + a\w)\|^2 = (a^2 + c^2) \left\| \w  +
      \frac{a}{a^2+c^2} \v\right\|^2 + \frac{c^2}{a^2+c^2} \|\v\|^2,
\end{equation}
whose minimum occurs when $\w = - \frac{a}{a^2+c^2} \v$, resulting in
\begin{equation}
  d = \frac{|c(t)|}{\sqrt{a(t)^2+c(t)^2}} \|\v\|.
\end{equation}
As we rescale $\v \mapsto s \v$, this is unbounded provided that
$c(t) \neq 0$. From equations~\eqref{eq:abcd-tdsg},
\eqref{eq:abcd-tdsg1} and \eqref{eq:abcd-tadsg}, we see that for
$\hyperlink{S9}{\ztdSG_{\gamma\in(-1,1]}}$, $c(t)=0$ if and only if
$t=0$, whereas for $\hyperlink{S11}{\ztAdSG_{\chi>0}}$, $c(t) = 0$ if
and only if $t = n \pi$ for $n \in \ZZ$, and hence, in summary, the
generic orbits are non compact.

Let us remark that for $\hyperlink{S11}{\ztAdSG_{\chi>0}}$, if
$t = n \pi$ for $n\neq 0$ then the exponential coordinate system
breaks down, so that we should restrict to $t \in (-\pi,\pi)$. Indeed,
using the explicit matrix representation, one can determine when the
exponential coordinates on $\Mbar$ stop being injective; that is, when
there are $(t,\x)$ and $(t',\x')$ such that
$\exp(t H + \x\cdot \P) = \exp(t'H + \x'\cdot \P) B$ for some
$B \in \Bgrbar$. In $\hyperlink{S9}{\ztdSG_{\gamma\in(-1,1]}}$ this
only happens when $t=t'$ and $\x = \x'$, but in
$\hyperlink{S11}{\ztAdSG_{\chi>0}}$ it happens whenever $t=t'= n \pi$
($n\neq 0$) and, if so, for all $\x$, $\x'$.

It now remains to look at the case of spacetime
$\hyperlink{S12}{\text{\twodgal}_{\gamma,\chi}}$. This case is very
similar to $\ztdSG_{\gamma}$ in $D=1$ except for two important
changes: we work over the complex numbers and $\gamma$ is replaced by
$z = \gamma + i \chi$. This means that the (real) subalgebras
$\b \subset \s \subset \gl(3,\CC)$ are given by
\begin{equation}
  \b = \left\{
    \begin{pmatrix}
      \zero & \zero & \yy \\ \zero & \zero & \zero \\ \zero & \zero & \zero
    \end{pmatrix}
    ~\middle |~ \yy \in \CC \right\}
  \qquad\text{and}\qquad
  \s = \left\{
    \begin{pmatrix}
      \zero & t z & \yy \\ -t & t (1+z) & \xx \\ \zero & \zero & \zero
    \end{pmatrix}
~\middle |~ t \in \RR, \xx,\yy \in \CC \right\},
\end{equation}
whereas the (real) subgroups $\Bgrbar \subset \Sgrbar
\subset \GL(3,\CC)$ are given by
\begin{equation}
  \Bgrbar = \left\{
    \begin{pmatrix}
      1 & \zero & \yy \\ \zero & 1 & \zero \\ \zero & \zero & 1
    \end{pmatrix}
    ~\middle |~ \yy \in \CC \right\}
  \qquad\text{and}\qquad
  \Sgrbar = \left\{
    \begin{pmatrix}
      \frac{z e^t-e^{zt}}{z-1} & \frac{z(e^{zt} - e^t)}{z-1} & \yy \\
      \frac{e^t - e^{zt}}{z-1} & \frac{z e^{zt} - e^t}{z-1} & \xx \\ \zero & \zero & 1
    \end{pmatrix}
    ~\middle |~ t \in \RR, \xx,\yy \in \CC \right\}.
\end{equation}
Let $g(t,\xx,\yy)$ denote the typical element (shown above) in
$\Sgrbar$ and let $g(0,0,\yy)$ denote the typical element of
$\Bgrbar$. Then we have
\begin{equation}
  g(0,0,\vv) g(t,\xx,\yy) = g(t,\xx,\yy+\vv) \qquad\text{and}\qquad
  g(t,\xx,\yy) g(0,0,\ww) = g(t , \xx + a(t) \ww, \yy + c(t) \ww),
\end{equation}
where
\begin{equation}
  \label{eq:a-and-c}
  a(t) = \frac{z e^t-e^{zt}}{z-1} \qquad\text{and}\qquad c(t) =
  \frac{e^t - e^{zt}}{z-1}.
\end{equation}
Hence the left and right action of the boosts takes place in $\CC^2$:
under the left action $(\xx,\yy) \mapsto (\xx, \yy + \vv)$, whereas
under the right action $(\xx,\yy) \mapsto (\xx + c(t) \ww, \yy + a(t)
\ww)$.

Now $\CC^2$ is equivalent to $\RR^4$ as a metric space and hence the
Heine--Borel theorem applies and all we need to show is that the
generic orbits are not bounded.  The squared distance (in the quotient
$\Sgrbar/\Bgrbar$) between a point $[g(t,\xx,\yy)] $ and its boost
$[g(t,\xx,\yy+\vv)]$ with parameter $\vv$ is
\begin{equation}
  \min_{\ww} \left\| ( \xx + c(t) \ww, \yy + \vv + a(t) \ww) -
    (\xx,\yy)\right\|^2 = \min_{\ww} \left\| (c(t) \ww, \vv + a(t)\ww)
  \right\|^2 = \min_{\ww} \left( |c(t)|^2 |\ww|^2 + | \vv + a(t)\ww|^2  \right).
\end{equation}
We complete the square and write this as
\begin{equation}
  \min_{\ww} \left( (|a|^2+|c|^2) \left| \ww + \frac{\bar a
        \vv}{|a|^2+|c|^2} \right|^2 + |\vv|^2 \left( 1 -
      \frac{|a|^2}{|a|^2+|c|^2}\right)  \right) =
  \frac{|c|^2|\vv|^2}{|a|^2+|c|^2},
\end{equation}
where we have used that $a(t)$ and $c(t)$ cannot both be zero because
$g(t,\xx,\yy)$ is invertible for all $t$. This grows without bound
with $\vv$ provided that $c(t) \neq 0$. Since $z \neq 1$,
equation~\eqref{eq:a-and-c} says that $c(t) = 0$ for those $t$
satisfying
\begin{equation}
  e^{zt} = e^t \iff e^{(z-1)t} = 1 \iff (z-1)t = 2 \pi i n \quad
  \exists n \in \ZZ.
\end{equation}
But $z -1 = (\gamma-1) + i \xi$ and $\gamma \neq 1$, so that this can
only be true for $n =0$ and hence $t=0$. Hence the generic orbit
($t \neq 0$) is unbounded and hence not compact. Here too one can show
that the exponential coordinate system is everywhere valid, by working
explicitly with the matrices and checking that the equation
$\exp(t H + \x \cdot \P) = \exp(t' H + \x'\cdot \P) B$ for some
$B \in \Bgrbar$ has the unique solution $t = t'$ and $\x = \x'$ (and
hence $B = \1$).

\section{Carrollian light cone ($\zLC$)}
\label{sec:spacetime-flc}

The carrollian light cone $\hyperlink{S16}{\zLC}$ is a hypersurface in
Minkowski spacetime, identifiable with the future light cone. It does
not arise as a limit and has additional brackets $[H,\B] = \B$,
$[H,\P] = -\P$ and $[\B,\P] = H + \J$, which shows that it is a
non-reductive homogeneous spacetime.

\subsection{Action of the boosts}
\label{sec:action-boosts-1}

Although it might be tempting to use that the boosts in Minkowski
spacetime act with generic non-compact orbits to deduce the same about
the boosts in $\hyperlink{S16}{\zLC}$, one has to be careful because
what we call boosts in $\hyperlink{S16}{\zLC}$ might not be
interpretable as boosts in the ambient Minkowski spacetime. Indeed, as
we will now see, boosts in $\hyperlink{S16}{\zLC}$ are actually null
rotations in the ambient Minkowski spacetime.

We first exhibit the isomorphism between the $\hyperlink{S16}{\zLC}$ Lie algebra and
$\so(D+1,1)$.  In the $\hyperlink{S16}{\zLC}$ Lie algebra, the boosts
and translations obey the following brackets:
\begin{equation}
  [H,\B] = \B, \qquad [H,\P] = -\P, \qquad\text{and}\qquad [\B,\P] = H
  + \J.
\end{equation}
If we let $L_{\mu\nu}$ be the standard generators of $\so(D+1,1)$ with
$\mu = (0,a,\natural)$, $a=1,\dots,D$, and with Lie brackets
\begin{equation}
  [L_{\mu\nu}, L_{\rho\sigma}] = \eta_{\nu\rho} L_{\mu\sigma} - \eta_{\mu\rho} L_{\nu\sigma} -  \eta_{\nu\sigma} L_{\mu\rho} + \eta_{\mu\sigma} L_{\nu\rho},
\end{equation}
where $\eta_{ab} = \delta_{ab}$, $\eta_{00} = -1$, and
$\eta_{\natural\natural} = 1$, then the correspondence is:
\begin{equation}
  J_{ab} = L_{ab}, \qquad B_a = \tfrac1{\sqrt2}
  (L_{0a} + L_{a\natural}), \qquad P_a = \tfrac1{\sqrt2} (L_{0a} -
  L_{a\natural}), \qquad\text{and}\qquad H = - L_{0\natural}.
\end{equation}
We see that, as advertised, the boosts $B_a$ are indeed null
rotations.

The boosts act linearly on the ambient coordinates $X^\mu$ in
Minkowski spacetime, with fundamental vector fields
\begin{equation}
  \zeta_{B_a} = \frac1{\sqrt2}\left( -X^0\frac{\d}{\d X^a} - X^a
    \frac{\d}{\d X^0} + X^a \frac{\d}{\d X^\natural} - X^\natural
    \frac{\d}{\d X^a}\right).
\end{equation}
Consider a linear combination $B = w^a B_a$ and let $T := X^0$, $X :=
w^a X^a$, and $Y := X^\natural$, so that in terms of these coordinates
and dropping the factor of $\frac1{\sqrt2}$,
\begin{equation}
  \zeta_B =  -T \frac{\d}{\d X} - X \frac{\d}{\d T} + X \frac{\d}{\d Y} - Y \frac{\d}{\d X}.
\end{equation}
This allows us to examine the orbit of this vector field while
focussing on the three-dimensional space with coordinates $T,X,Y$.  The vector field is
linear, so there is a matrix $A$ such that
\begin{equation}
  \zeta_B = \begin{pmatrix} T & X & Y \end{pmatrix} A
  \begin{pmatrix}
    \frac{\d}{\d T} \\ \frac{\d}{\d X} \\  \frac{\d}{\d Y}
  \end{pmatrix}
  \implies
  A =
  \begin{pmatrix}
    \zero & -1 & \zero\\
    -1 & \zero & 1 \\
    \zero & -1 & \zero
  \end{pmatrix}.
\end{equation}
The matrix $A$ obeys $A^3 =0$, so its exponential is
\begin{equation}
  \exp(s A) =
  \begin{pmatrix}
    1 + \frac12 s^2 & -s & -\frac12 s^2\\
    -s & 1 & s \\
    \frac12 s^2 & -s & 1 - \frac12 s^2
  \end{pmatrix}
\end{equation}
and hence the orbit of $(T_0,X_0,Y_0,\dots)$ is given by
\begin{equation}
  \begin{split}
    T(s) &= (1 + \tfrac12 s^2) T_0 - s X_0 - \tfrac{1}{2} s^2 Y_0\\
    X(s) &= -s T_0 + X_0 +s Y_0\\
    Y(s) &= \tfrac{1}{2} s^2 T_0 - s X_0 + (1-\tfrac12 s^2) Y_0,
  \end{split}
\end{equation}
with all other coordinates inert, which is clearly non-compact in the
Minkowski spacetime. But of course, this orbit lies on the future
light cone (indeed, notice that
$-T(s)^2 + X(s)^2 + Y(s)^2 = - T_0^2 + X_0^2 + Y_0^2$), which is a
submanifold, and hence the orbit is also non-compact on
$\hyperlink{S16}{\zLC}$, provided with the subspace topology.

\subsection{Fundamental vector fields}
\label{sec:fund-vect-fields-flc}

Let $A = t H + \x \cdot \P$ and let us calculate the action of $\ad_A$
on the generators, this time with the indices written explicitly:
\begin{equation}
  \begin{split}
    \ad_A B_a &= t B_a - x^a H - x^b J_{ab}\\
    \ad_A P_a &= -t P_a \\
    \ad_A H &= x^a P_a\\
    \ad_A J_{ab} &= x^a P_b - x^b P_a.
  \end{split}
\end{equation}
In order to compute the fundamental vector fields using
equation~\eqref{eq:master} and the soldering form using equation
\eqref{eq:MC-pullback}, we need to calculate the action of certain
universal power series on $\ad_A$ on the generators. To this end, let
us derive formulae for the action of $f(\ad_A)$, for $f(z)$ an
analytic function of $z$, on the generators. We will do this by first
calculating powers of $\ad_A$ on generators. It is clear, first of
all, that on $\P$,
\begin{equation}
  f(\ad_A) \P = f(-t) \P.
\end{equation}
On $H$ and $\J$ we just need to treat the constant term
separately:
\begin{equation}
  \begin{split}
    f(\ad_A) H &= f(0) H - \frac1t \left(f(-t) - f(0)\right) \x \cdot\P\\
    f(\ad_A) J_{ab} &= f(0) J_{ab} - \frac1t \left(f(-t) - f(0)\right)
    (x^a P_b - x^b P_a).
  \end{split}
\end{equation}
On $\B$ it is a little bit more complicated.  Notice first of all that whereas
\begin{equation}
  \ad_A^2 B_a = t \ad_A B_a - 2 x^a x^b P_b + x^2 P_a,
\end{equation}
$\ad_A^3 B_a = t^2 \ad_A B_a$.  Therefore, by induction, for all $n
\geq 1$,
\begin{equation}
  \ad_A^n B_a =
  \begin{cases}
    t^{n-1}\ad_A B_a & n~\text{odd}\\
    t^{n-1}\ad_A B_a + t^{n-2} (x^2 P_a - 2 x^a \x \cdot \P) & n~\text{even},
  \end{cases}
\end{equation}
and therefore
\begin{equation}
  f(\ad_A) B_a = f(t) B_a - \tfrac1t (f(t)-f(0))(x^a H + x^b J_{ab}) +
  \tfrac1{t^2}(\tfrac12(f(t)+f(-t)) - f(0))(x^2 P_a - 2 x^a\x \cdot \P).
\end{equation}
Using these formulae, we can now apply equation \eqref{eq:master} in
order to determine the expression of the fundamental vector fields in
terms of exponential coordinates.

Let us take $X = \v \cdot \P$ in equation~\eqref{eq:master}.  We must
take $Y'(0)=0$ here and find that
\begin{equation}
  \y \cdot \P = G(\ad_A) \v \cdot \P = G(-t) \v \cdot \P  \implies \y
  = \frac{t}{1-e^{-t}}\v,
\end{equation}
resulting in
\begin{equation}
  \xi_{P_a} = \frac{t}{1-e^{-t}} \frac{\d}{\d x^a}.
\end{equation}

Taking $X=H$ in equation~\eqref{eq:master}, we again must take
$Y'(0)=0$.  Doing so, we arrive at
\begin{equation}
  \tau H + \y \cdot \P = G(\ad_A) H = H - \frac1t \left(G(-t) - 1\right) \x
  \cdot \P \implies \tau = 1 \qquad\text{and}\qquad \y =
 \left(\frac{1}{t}-1-\frac{1}{e^t-1}\right)\x,
\end{equation}
resulting in
\begin{equation}
  \xi_H = \frac{\d}{\d t} +
     \left(\frac{1}{t}-1-\frac{1}{e^t-1}\right) x^a \frac{\d}{\d x^a}.
\end{equation}
One checks already that $[\xi_H,\xi_{P_a}] = \xi_{P_a}$, as expected.

Finally, put $X=\v \cdot \B$ in equation~\eqref{eq:master} and hence
now $Y'(0)= \bbeta \cdot \B + \tfrac12 \lambda^{ab} J_{ab}$.  Substituting
this in equation~\eqref{eq:master} and requiring that the $\h$-terms
vanish, we find
\begin{equation}
  \bbeta = \frac{G(t)}{F(t)} \v = e^{-t} \v \qquad\text{and}\qquad
  \lambda^{ab} = \frac{1-e^{-t}}{t} (v^a x^b - v^b x^a).
\end{equation}
Comparing the $H$ terms, we see that
\begin{equation}
  \tau =\frac{1-e^{-t}}{t} \x \cdot \v,
\end{equation}
whereas the $\P$ terms give
\begin{equation}
  \y =\frac{1-e^{-t}}{2t} x^2 \v + \frac{1-t-e^{-t}}{t^2} \x\cdot\v \x,
\end{equation}
resulting in
\begin{equation}
  \xi_{B_a} = \frac{1-e^{-t}}{t} x^a \frac{\d}{\d t} +
    \frac{1-e^{-t}}{2t} x^2 \frac{\d}{\d x^a} + \frac{1-t-e^{-t}}{t^2}
    x^a x^b \frac{\d}{\d x^b}.
\end{equation}
One checks that, as expected, $[\xi_H, \xi_{B_a}] = - \xi_{B_a}$ and
that $[\xi_{B_a},\xi_{P_b}] = - \delta_{ab} \xi_H - \xi_{J_{ab}}$,
where $\xi_{J_{ab}} = x^b \frac{\d}{\d x^a} - x^a \frac{\d}{\d x^b}$.

\subsection{Soldering form and canonical connection}
\label{sec:sold-form-canon-flc}

The soldering form can be calculated from
equation~\eqref{eq:MC-pullback} and projecting the result to $\k/\h$:
\begin{equation}
  \begin{split}
    \theta &= D(\ad_A)(dt \Hbar + d\x\cdot \Pbar) = dt\left(\Hbar -
    \frac{D(-t)-1}{t} \x \cdot \Pbar\right) + D(-t) d\x \cdot \Pbar \\
    &= dt \Hbar + \frac{1+t-e^t}{t^2} \x \cdot \Pbar dt + \frac{e^t-1}{t} d
    \x \cdot \Pbar.
  \end{split}
\end{equation}
It follows from the expression of $\theta$ that it is invertible for
all $(t,\x)$, since $\frac{e^t-1}{t} \neq 0$ for all $t \in \RR$.  Its
inverse, the vielbein $E$, has components
\begin{equation}
  E_{\Hbar} = \frac{\d}{\d t} + \left( \frac1t - \frac1{e^t-1}\right) x^a
  \frac{\d}{\d x^a} \qquad\text{and}\qquad
  E_{\overline{P}_a} = \frac{t}{e^t-1} \frac{\d}{\d x^a}.
\end{equation}
The invariant carrollian structure is given by $\kappa = E_{\Hbar}$ and spatial
metric $b = \pi^2(\theta,\theta)$, given by
\begin{equation}
  b = \frac{(1+t-e^t)^2}{t^4} x^2 dt^2 + \frac{(e^t-1)^2}{t^2} d\x
  \cdot d\x + 2 \frac{(e^t-1)(1+t-e^t)}{t^3} \x \cdot d\x dt.
\end{equation}

\section{Exotic two-dimensional spacetimes}
\label{sec:exotic}

In this section, we discuss the two-dimensional homogeneous spacetimes
in Table~\ref{tab:spacetimes}. These spacetimes can be treated
together. They are reductive, symmetric and even affine, but have no
invariant metrics, galilean or carrollian structures. Relative to
exponential coordinates $(t,x)$, where
$\sigma(t,x) = \exp(t H + x P)$, the soldering form is
\begin{equation}
  \theta = dt H + dx P,
\end{equation}
and the invariant connection $\omega = 0$.  The vielbein are
\begin{equation}
  E_H = \frac{\d}{\d t} \qquad\text{and}\qquad E_P = \frac{\d}{\d x}.
\end{equation}
The exponential coordinates are affine, so that
\begin{equation}
  \xi_H = \frac{\d}{\d t} \qquad\text{and}\qquad \xi_P = \frac{\d}{\d x}.
\end{equation}
The only distinguishing feature is the action of the boosts.  We will
see that in all cases the fundamental vector field $\xi_B$ is linear
in the affine coordinates, so we will be able to determine the
orbits simply by exponentiating the corresponding matrix.  Indeed, we
will see that
\begin{equation}\label{eq:linear_vf}
  \xi_B =
  \begin{pmatrix}
    t & x 
  \end{pmatrix}
  \begin{pmatrix}
    a & b \\ c & d
  \end{pmatrix}
  \begin{pmatrix}
    \frac{\d}{\d t} \\
    \frac{\d}{\d x}
  \end{pmatrix} =
  (a t + c x) \frac{\d}{\d t} + (b t + d x) \frac{\d}{\d x},
\end{equation}
and hence the orbit of the boost through $(t_0,x_0)$ is given by
\begin{equation}\label{eq:boost_orbit}
  \begin{pmatrix}
    t(s) \\ x(s)
  \end{pmatrix} = \exp(s A) \begin{pmatrix} t_0 \\ x_0 \end{pmatrix}
  \qquad\text{for}\qquad A = \begin{pmatrix} a & b \\ c & d \end{pmatrix}. 
\end{equation}

As we saw in Section~\ref{sec:nomizu-maps-exotic}, in all cases but
\xthree$_\chi$, the only invariant connection is the canonical
connection.

\subsection{Spacetime \xone}
\label{sec:spacetime-x1}

Here $[B,H] = P$ and $[B,P]=-H-2P$, so that
\begin{equation}
  \xi_B = - x\frac{\d}{\d t} - (2x-t) \frac{\d}{\d x}.
\end{equation}

From equation~\eqref{eq:linear_vf}, we see that the matrix $A$ in
equation~\eqref{eq:boost_orbit} is given by
\begin{equation}
  A =
  \begin{pmatrix}
    0 & 1 \\ -1 & -2 
  \end{pmatrix}
  \implies
  \exp(s A) = e^{-s}
  \begin{pmatrix}
     1+s & s \\
     - s & 1-s \\
  \end{pmatrix}.
\end{equation}
The vector field is complete, and the orbits are homeomorphic to the
real line, except for the critical point at the origin which is its
own orbit.

\subsection{Spacetime \xtwo}
\label{sec:spacetime-x2}

Here $[B,H] = -H$ and $[B,P]=-P$, so that
\begin{equation}
  \xi_B = - t\frac{\d}{\d t} - x \frac{\d}{\d x}.
\end{equation}

From equation~\eqref{eq:linear_vf}, we see that the matrix $A$ in
equation~\eqref{eq:boost_orbit} is given by
\begin{equation}
  A =
  \begin{pmatrix}
    -1 & 0 \\ 0 & -1
  \end{pmatrix}
  \implies
  \exp(s A) = e^{-s}
  \begin{pmatrix}
    1 & 0 \\ 0 & 1
  \end{pmatrix}.
\end{equation}
Again, the vector field is complete, and the orbits are homeomorphic
to the real line, except for the critical point at the origin which is
its own orbit.

\subsection{Spacetime \xthree$_\chi$}
\label{sec:spacetime-x3-chi}

Here $[B,H] = -(1+\chi) H$ and $[B,P] = (1-\chi) P$, so that
\begin{equation}
  \xi_B = - (1+\chi) t \frac{\d}{\d t} + (1-\chi) x \frac{\d}{\d x}.
\end{equation}

From equation~\eqref{eq:linear_vf}, we see that the matrix $A$ in
equation~\eqref{eq:boost_orbit} is given by
\begin{equation}
  A =
  \begin{pmatrix}
    -(1+\chi) & 0 \\ 0 & 1-\chi
  \end{pmatrix}
  \implies
  \exp(s A) = 
  \begin{pmatrix}
    e^{-s(1+\chi)} & 0 \\ 0 & e^{s(1-\chi)}
  \end{pmatrix}.
\end{equation}
Here $\chi > 0$.  The vector field is complete, and for $\chi\neq 1$
the orbits are homeomorphic to the real line, except for the critical
point at the origin which is its own orbit.  For $\chi = 1$, every
point on the $x$-axis ($t=0$) is its own orbit, but the other orbits
are non-compact.

If $\chi = 1$, we have a three-parameter family of invariant
connections characterised by the Nomizu map in
equation~\eqref{eq:nomizu-x3-chi-1}.  The torsion and curvature have
components
\begin{equation}
  \Theta(H,P) = (\nu' - \xi') H \qquad\text{and}\qquad
  \Omega(H,P) P = \nu' (\zeta' - \xi') H.
\end{equation}
Therefore, there is a two-parameter family of torsion-free invariant
connections and two one-parameter families of torsion-free, flat connections:
\begin{equation}
  \begin{aligned}[m]
    \alpha(P,P) &= \zeta' P
  \end{aligned}
  \qquad\text{and}\qquad
  \begin{aligned}[m]
    \alpha(H,P) &= \nu' H\\
    \alpha(P,H) &= \nu' H\\
    \alpha(P,P) &= \nu' P.
  \end{aligned}
\end{equation}

If $\chi=3$, we have a one-parameter family of invariant connections,
which are flat and torsion-free, with Nomizu map given by
equation~\eqref{eq:nomizu-x3-chi-3}.

\subsection{Spacetime \xfour$_\chi$}
\label{sec:spacetime-x4-chi}

Here $[B,H] = P$ and $[B,P] = -(1+\chi^2) H - 2\chi P$, so that
\begin{equation}
  \xi_B = -(1+\chi^2) x \frac{\d}{\d t} + (t - 2 \chi x) \frac{\d}{\d x}.
\end{equation}

From equation~\eqref{eq:linear_vf}, we see that the matrix $A$ in
equation~\eqref{eq:boost_orbit} is given by
\begin{equation}
  A =
  \begin{pmatrix}
    0 & 1 \\ -(1+\chi^2) & -2\chi
  \end{pmatrix}
  \implies
  \exp(s A) = e^{-\chi s}
  \begin{pmatrix}
    \cos s + \chi  \sin s & \sin s \\
    -\left(1+ \chi^2\right) \sin s & \cos s - \chi  \sin s
  \end{pmatrix}.
\end{equation}
The vector field is complete, and for $\chi>0$ the orbits are
homeomorphic to the real line, except for the critical point at the
origin which is its own orbit. For $\chi = 0$, the orbits are circles,
as expected since, as seen in Figure~\ref{fig:d=2-graph},
$\text{\xfour}_{\chi=0} = \EE$, the euclidean space.

\section{Aristotelian spacetimes}
\label{sec:aristotelian}

In this section we introduce coordinates for the aristotelian
spacetimes of Table~\ref{tab:aristotelian} and study their geometric
properties.

\subsection{Static spacetime ($\zS$)}
\label{sec:static}

This is an affine space and the exponential coordinates $(t,\x)$ are
affine, so that
\begin{equation}
  \xi_H= \frac{\d}{\d t} \qquad\text{and}\qquad \xi_{P_a} =
  \frac{\d}{\d x^a}.
\end{equation}
Similarly, the soldering form is $\theta = dt H + d\x \cdot \P$,
the canonical invariant connection vanishes, and so does the torsion.
The vielbein is
\begin{equation}
  E_H =   \xi_H \qquad\text{and}\qquad E_{P_a} = \xi_{P_a}.
\end{equation}

\subsection{Torsional static spacetime ($\zTS$)}
\label{sec:tst}

Here $[H,\P] = \P$.

\subsubsection{Fundamental vector fields}
\label{sec:fund-vect-fields-tst}

Letting $A= tH + \x \cdot \P$, we find $\ad_A H =
- \x \cdot \P$ and $\ad_A \P = t \P$.  Therefore, for any analytic
function $f$, we conclude that
\begin{equation}
  f(\ad_A) \P = f(t) \P \qquad\text{and}\qquad f(\ad_A) H = f(0) H -
  \frac1t (f(t) - f(0)) \x \cdot \P.
\end{equation}
Applying this to equation~\eqref{eq:master}, we find
\begin{equation}
  \begin{split}
    \xi_H &= \frac{\d}{\d t} + \left(\frac1t - \frac1{e^t-1} -1\right)
    x^a \frac{\d}{\d x^a}\\
    \xi_{P_a} &= \frac{t}{1-e^{-t}} \frac{\d}{\d x^a},
  \end{split}
\end{equation}
which one can check obey $[\xi_H, \xi_{P_a}] = - \xi_{P_a}$, as
expected.

\subsubsection{Soldering form and canonical connection}
\label{sec:sold-form-canon-tst}

Applying the same formula to equation~\eqref{eq:MC-pullback}, we find
that the canonical invariant connection one-form vanishes in this
basis and that the soldering form is given by
\begin{equation}
  \theta = dt \left(H + \frac1t \left(1-\frac{1-e^{-t}}{t}\right)\x
    \cdot \P\right) + \frac{1-e^{-t}}{t} d\x \cdot \P,
\end{equation}
so that the corresponding vielbein is
\begin{equation}
  E_H = \frac{\d}{\d t} + \left(\frac1t - \frac1{1-e^{-t}} \right) x^a
  \frac{\d}{\d x^a} \qquad\text{and}\qquad E_{P_a} =
  \frac{t}{1-e^{-t}} \frac{\d}{\d x^a}.
\end{equation}
It is clear from the fact that the function $\frac{1-e^{-t}}{t}$ is
never zero that $\theta$ is invertible for all $(t,\x)$.

Although the canonical connection is flat, its torsion 2-form does not
vanish:
\begin{equation}
  \Theta = \frac{e^{-t}-1}{t} dt \wedge d\x \cdot \P.
\end{equation}

\subsection{Aristotelian spacetime \athree$_\varepsilon$}
\label{sec:spacetime-a3}

Here $[P_a, P_b] = -\varepsilon J_{ab}$, where $D\geq 2$.

\subsubsection{Fundamental vector fields}
\label{sec:fund-vect-fields-a3}

Let $A = t H + \x \cdot \P$.  Then $\ad_A H = 0$ and
$\ad_A P_b = -\varepsilon x^a J_{ab}$.  Continuing, we find
\begin{equation}
  \ad_A^2 P_b = \varepsilon x^b \x \cdot \P - \varepsilon x^2 P_b
  \qquad\text{and}\qquad \ad_A^3 P_b = (-\varepsilon x^2) \ad_A P_b.
\end{equation}
Therefore, an induction argument shows that
\begin{equation}
  \ad_A^n P_b = (-\varepsilon x^2) \ad_A^{n-2} P_b \qquad \forall n
  \geq 3.
\end{equation}
If $f(z)$ is analytic in $z$, then $f(\ad_A) H = f(0) H$ and
\begin{equation}
  f(\ad_A) P_b = \tfrac12 \left( f(x_+) + f(x_-) \right) P_b -
    \tfrac12 \left( f(x_+) + f(x_-) - 2 f(0) \right) \frac{x^b
      \x\cdot\P}{x^2} - \frac{\varepsilon}{2 x_+} \left(f(x_+) -
      f(x_-)\right) x^a J_{ab},
\end{equation} 
where
\begin{equation}
  x_\pm = \pm \sqrt{-\varepsilon x^2} =
  \begin{cases}
    \pm |\x| & \varepsilon = -1\\
    \pm i |\x| & \varepsilon = 1.
  \end{cases}
\end{equation}
Similarly, $\ad_A J_{ab} = x^a P_b - x^b P_a$, so that
\begin{equation}
  f(\ad_A) J_{ab} = f(0) J_{ab} + \tfrac12
  \left(\widetilde{f}(x_+)+\widetilde{f}(x_-)\right)(x^a P_b - x^b P_a) - \frac{\varepsilon}{2
    x_+} \left(\widetilde{f}(x_+) - \widetilde{f}(x_-)\right) x^c (x^a J_{cb} - x^b J_{ca}),
\end{equation}
where $\widetilde{f}(z) = (f(z) - f(0))/z$.

Inserting these formulae in equation~\eqref{eq:master} with $X=H$ and
$Y'(0)=0$, we see that
\begin{equation}
  \xi_H = \frac{\d}{\d t}.
\end{equation}
If instead $X = \v \cdot \P$ and $Y'(0) = \tfrac12
\lambda^{ab}J_{ab}$, we see first of all that $\tau = 0$ and that
demanding that the $J_{ab}$ terms cancel,
\begin{equation}
  \lambda^{ab} = \frac{-\varepsilon \left(G(x_+)-G(x_-)\right)}{x_+
    \left(F(x_+) + F(x_-)\right)}  (x^a v^b - x^b v^a),
\end{equation}
and reinserting into equation~\eqref{eq:master}, we find that
\begin{multline}
  y^a =\tfrac12 \left(G(x_+) + G(x_-) -
    \frac{\left(G(x_+)-G(x_-)\right) \left( F(x_+)-F(x_-) \right)}{F(x_+) +
        F(x_-)}\right) v^a \\
    - \tfrac12 \left(G(x_+) + G(x_-) - 2 - \frac{\left(G(x_+)-G(x_-)\right) \left( F(x_+)-F(x_-) \right)}{F(x_+) +
        F(x_-)}\right) \frac{\v \cdot \x}{x^2} x^a.
\end{multline}
From this we read off the expression for $\xi_{P_a}$:
\begin{equation}
  \xi_{P_a} = \frac{F(x_+) G(x_-) + F(x_-)G(x_+)}{F(x_+) + F(x_-)}\frac{\d}{\d x^a} 
  + \left(1 - \frac{F(x_+) G(x_-) + F(x_-)G(x_+)}{F(x_+) + F(x_-)}\right) \frac{x^a x^b}{x^2} \frac{\d}{\d x^b},
\end{equation}
which simplifies to
\begin{equation}
  \begin{split}
    \xi_{P_a}^{(\varepsilon=1)} &= |\x| \cot|\x| \frac{\d}{\d x^a} + (1- |\x| \cot|\x|) \frac{x^a x^b}{x^2} \frac{\d}{\d x^b}\\
    \xi_{P_a}^{(\varepsilon=-1)} &= |\x| \coth|\x| \frac{\d}{\d x^a} + (1- |\x| \coth|\x|) \frac{x^a x^b}{x^2} \frac{\d}{\d x^b}.
  \end{split}
\end{equation}

\subsubsection{Soldering form and canonical connection}
\label{sec:sold-form-canon-a3}

The soldering form and connection one-form for the canonical
connection are obtained from equation~\eqref{eq:MC-pullback}, which
says that
\begin{equation}
  \begin{split}
    \theta + \omega &= dt H + dx^b D(\ad_A) P_b\\
    &=  dt H + \tfrac12 (D(x_+)+D(x_-)) d\x \cdot \P  \\
    & \quad  - \tfrac12 (D(x_+)D(x_-) - 2) \frac{\x \cdot d\x}{x^2} \x\cdot \P-
    \frac{\varepsilon}{2 x_+}(D(x_+)-D(x_-)) x^a dx^b J_{ab},
  \end{split}
\end{equation}
whence
\begin{equation}
  \begin{split}
    \theta^{(\varepsilon=1)} &= dt H + \frac{\sin|\x|}{|\x|} d\x \cdot \P + \left(1 - \frac{\sin|\x|}{|\x|}\right) \frac{\x \cdot d\x}{x^2} \x\cdot \P\\
    \theta^{(\varepsilon=-1)} &= dt H + \frac{\sinh|\x|}{|\x|} d\x \cdot \P + \left(1 - \frac{\sinh|\x|}{|\x|}\right) \frac{\x \cdot d\x}{x^2} \x\cdot \P
  \end{split}
\end{equation}
and
\begin{equation}
  \begin{split}
    \omega^{(\varepsilon=1)} &=   \frac{1 - \cos|\x|}{x^2} x^a dx^b J_{ab}\\
    \omega^{(\varepsilon=-1)} &=   \frac{1 - \cosh|\x|}{x^2} x^a dx^b J_{ab}.
  \end{split}
\end{equation}
It follows that if $\varepsilon=-1$ the soldering form is invertible
for all $(t,\x)$, whereas if $\varepsilon=1$ then it is invertible for
all $t$ but inside the open ball $|\x|<\pi$.

The torsion of the canonical connection vanishes, since
$[\theta,\theta]_\m = 0$.  The curvature is given by
\begin{equation}
  \begin{split}
    \Omega^{(\varepsilon=1)} &= \tfrac12 \frac{\sin^2|\x|}{x^2} dx^a
    \wedge dx^b J_{ab} + \frac{\sin|\x|}{|x|}\left(1 -
      \frac{\sin|\x|}{|x|}\right) \frac{x^bx^c}{x^2} dx^a \wedge dx^c J_{ab}\\
    \Omega^{(\varepsilon=-1)} &= -\tfrac12 \frac{\sinh^2|\x|}{x^2} dx^a
    \wedge dx^b J_{ab} - \frac{\sinh|\x|}{|x|}\left(1 -
      \frac{\sinh|\x|}{|x|}\right) \frac{x^bx^c}{x^2} dx^a \wedge dx^c J_{ab}.
  \end{split}
\end{equation}

\subsection{Aristotelian spacetime \twoda}
\label{sec:spacetime-2da}

Here $D=2$ and $[P_a, P_b] = \epsilon_{ab} H$.

\subsubsection{Fundamental vector fields}
\label{sec:fund-vect-fields-2da}

Letting $A = t H + \x \cdot \P$, we have that $\ad_A H = 0$ and
$\ad_A P_a = - \epsilon_{ab} x^b H$, whence $\ad_A^2 P_a = 0$.  So if
$f(z)$ is analytic in $z$,
\begin{equation}
  f(\ad_A) H = f(0) H \qquad\text{and}\qquad f(\ad_A) P_a = f(0) P_a -
  f'(0) \epsilon_{ab} x^b H.
\end{equation}
Since $G(z) = 1 - \frac12 z + O(z^2)$, from
equation~\eqref{eq:master} we see that 
\begin{equation}
 \xi_H = \frac{\d}{\d t} \qquad\text{and}\qquad \xi_{P_a} =
  \frac{\d}{\d x^a} +  \tfrac12 \epsilon_{ab} x^b \frac{\d}{\d t}.
\end{equation}
One checks that $[\xi_{P_a}, \xi_{P_b}] = - \epsilon_{ab} \xi_H$, as
expected.

\subsubsection{Soldering form and canonical connection}
\label{sec:sold-form-canon-2da}

Since $D(z) = 1 - \frac12 z + O(z^2)$, equation~\eqref{eq:MC-pullback}
says that the connection one-form $\omega = 0$ and the soldering form
is given by
\begin{equation}
  \theta = (dt + \tfrac12 \epsilon_{ab} dx^a x^b) H + d\x \cdot \P,
\end{equation}
which is clearly everywhere invertible.  The torsion of the canonical
connection is given by
\begin{equation}
  \Theta = - \tfrac12 \epsilon_{ab} dx^a \wedge dx^b H.
\end{equation}
The vielbein is given by
\begin{equation}
  E_H = \frac{\d}{\d t} \qquad\text{and}\qquad E_{P_a} = \frac{\d}{\d
    x^a} - \tfrac12 \epsilon_{ab} x^b \frac{\d}{\d t}.
\end{equation}

\section{Symmetries of the spacetime structure}
\label{sec:symm-spac-struct}

In this section we investigate the (conformal) symmetries of the
carrollian and galilean spacetimes and their respective invariant
structures. A carrollian structure $(\kappa, b)$ consists of a spatial
metric $b$ and a so-called carrollian vector field $\kappa$, whereas a
galilean structure $(\tau,h)$ consists of a spatial co-metric $h$ and
a clock-one form $\tau$. Let us remark that some authors would add the
invariant connection as part of the structure, but we will not do so
in the following. This means that, in the terminology of
\cite{Duval:2014lpa}, we treat the ``weak'' rather than the ``strong''
structures.

The calculations in this section are motivated by the intriguing
connection between conformally carrollian
symmetries~\cite{Duval:2014uva,Duval:2014lpa} and the symmetries of
asymptotic flat spacetimes~\cite{Bondi:1962px,Sachs:1962zza} in $3+1$
dimensions. This connection is given by an isomorphism between the Lie
algebra of infinitesimal conformal transformations of a carrollian
structure~\cite{Duval:2014uva} and the Lie algebra of the
Bondi--Metzner--Sachs (BMS) group~\cite{Bondi:1962px,Sachs:1962zza}.

Similarly, the infinitesimal conformal symmetries of the galilean and
carrollian structures of the homogeneous kinematical spacetimes will
turn out to be infinite-dimensional and one might hope this has
interesting consequences. It should be mentioned that were one to add
the invariant connection as part of the data of the homogeneous
carrollian or galilean structure, the symmetry algebra would be
typically cut down to the (finite-dimensional) transitive kinematical
Lie algebra.

Let $(M,\tau,h)$ be a galilean spacetime.  We say that a vector field
$\xi \in \eX(M)$ is a \textbf{galilean Killing vector field} if it
generates a symmetry of the galilean structure:
\begin{equation}
  \label{eq:g-kv}
  \L_\xi \tau = 0 \qquad\text{and}\qquad \L_\xi h = 0,
\end{equation}
whereas we say that it is a \textbf{galilean conformal Killing vector
  field at level $N \in \NN$} if it generates a conformal symmetry (at
level $N$) of the galilean structure:
\begin{equation}
  \label{eq:g-ckv}
  \L_\xi \tau =   -\frac{\lambda}{N} \tau \qquad\text{and}\qquad
  \L_\xi h = \lambda h,
\end{equation}
for some $\lambda \in C^\infty(M)$.  Similarly, if $(M,\kappa,b)$ is a
carrollian spacetime, we say that $\xi \in \eX(M)$ is a
\textbf{carrollian Killing vector field} if it generates a symmetry of
the carrollian structure:
\begin{equation}
  \label{eq:c-kv}
  \L_\xi \kappa = 0 \qquad\text{and}\qquad \L_\xi b = 0,
\end{equation}
whereas we say that it is a \textbf{carrollian conformal Killing
  vector field at level $N \in \NN$} if it generates a conformal
symmetry (at level $N$) of the carrollian structure:
\begin{equation}
  \label{eq:c-ckv}
  \L_\xi \kappa = -\frac{\lambda}{N} \kappa \qquad\text{and}\qquad \L_\xi b = \lambda b,
\end{equation}
for some $\lambda \in C^\infty(M)$.  These definitions agree (modulo
notation) with the ones in \cite{Duval:1993pe} and
\cite{Duval:2014uva,Duval:2014lpa}.  The set of galilean/carrollian
Killing vector fields close under the Lie bracket of vector fields to
give rise to Lie algebras.  The same is true for the set of
galilean/carrollian conformal Killing vector fields of a given fixed
level $N$.  In this section we will determine the structure of these
Lie algebras for the homogeneous carrollian and galilean spacetimes.

The calculations in this section are easier to perform if we change
coordinates from the exponential coordinates
$\sigma: \RR^{D+1} \to M$, with
$\sigma(t,\x) = \exp(t H + \x \cdot \P) \cdot o$, that we have been
using until now to modified exponential coordinates
$\sigma': \RR^{D+1} \to M$, with
$\sigma'(t,\x) = \exp(t H) \exp(\x \cdot \P) \cdot o$.
Appendix~\ref{app:modexp} discusses these coordinates further.  In
many of the calculations we require knowledge of the Lie algebra of
conformal Killing vector fields on the simply-connected riemannian
symmetric spaces $\EE$, $\SS$ and $\HH$.  In
Appendix~\ref{sec:conf-kill-vect} we collect a few standard results in
low dimension.

\subsection{Symmetries of the carrollian structure ($\zC$)}
\label{sec:symm-carr-struct}

We start by determining the carrollian Killing vector fields for the
(flat) carrollian spacetime $\hyperlink{S13}{\zC}$
(as has already been done in, e.g., \cite{Duval:2014uva}). Since $H$
and $\P$ commute in this spacetime, the exponential and modified
exponential coordinates agree. The invariant carrollian structure on
the spacetime parametrised by $(t, x^a) \in \RR^{D+1}$, with
$a=1,\ldots,D$, is given by $\kappa = \frac{\d}{\d t}$ and
$b = \delta_{ab} dx^a dx^b$. Let
$\xi = \xi^{0} \frac{\d}{\d t} + \xi^a \frac{\d}{\d x^a}$ be a
carrollian Killing vector field of $(\kappa, b)$, so
that it satisfies equation~\eqref{eq:c-kv}. Then,
$\L_{\xi}\kappa=[\xi,\kappa] = 0$ says that $T := \xi^{0}$ and $\xi^a$
are $t$-independent. The condition $\L_\xi b = 0$, says that
\begin{equation}
  0 \stackrel{!}{=} \L_\xi b = 2 (\L_\xi dx^a) dx^a = 2 d (\L_\xi x^a)
  dx^a = 2 d \xi^a dx^a = 2 \frac{\d \xi^a}{\d x^b} d x^b dx^a
  \implies \frac{\d \xi^a}{\d x^b} + \frac{\d \xi^b}{\d x^a} = 0.
\end{equation}
This says that $\xi^a(\x) \frac{\d}{\d x^a}$ is a Killing vector field
of euclidean space.  In summary, the most general carrollian Killing
vector field of $(\kappa,b)$ is given by
\begin{equation}
  \xi = T(\x) \frac{\d}{\d t} + \xi_X~,
\end{equation}
for some $X \in \e$, the euclidean Lie algebra of $\EE^D$, and some
``supertranslations'' $T \in C^\infty(\EE^D)$.  As a vector space,
then, the Lie algebra $\a$ of carrollian Killing vector fields is
given by $C^\infty(\EE^D) \oplus \e$, but as a Lie algebra it is a
semidirect product
\begin{equation}
\a^{\zC} \cong \e \ltimes C^\infty(\EE^D) ,
\end{equation}
where the action of $\e$ on
$C^\infty(\EE^D)$ is via the Lie derivative.  In other words, we have
a split exact sequence
\begin{equation}
  \label{eq:c-ses}
  \begin{tikzcd}
    0 \arrow[r] & C^\infty(\EE^D) \arrow[r] &  \a^{\zC} \arrow[r] & \e \arrow[r] & 0.
  \end{tikzcd}
\end{equation}
The carrollian algebra is embedded here by considering the subalgebra
of $C^\infty(\EE^D)$ consisting of polynomial functions of degree at
most $1$: with the constant function $1$ corresponding to $H$ and the
linear function $x^a$ corresponding to $B_a$. When we identify
$J_{ab}$ and $P_{a}$ in $\e$ in the obvious way we recover
\eqref{eq:carrvec}.

Let us now determine the carrollian conformal Killing vector fields. Let
$\xi = \xi^0 \frac{\d}{\d t} + \xi^a \frac{\d}{\d x^a}$ satisfy
equation~\eqref{eq:c-ckv} where $(\kappa,b)$ is again the invariant
carrollian structure on $\hyperlink{S13}{\zC}$:
$\kappa = \frac{\d}{\d t}$ and $b = \delta_{ab} dx^a dx^b$. The
condition $\L_\xi \kappa = -\frac{\lambda}{N} \kappa$ imposes
\begin{equation}
  \label{eq:xiaxi0car}
  \frac{\d \xi^a}{\d t} = 0 \qquad\text{and}\qquad \lambda = N
  \frac{\d \xi^0}{\d t}.
\end{equation}
The condition $\L_\xi b = \lambda b$ says that
\begin{equation}
  \label{eq:confkill}
  \frac{\d \xi^a}{\d x^b} +  \frac{\d \xi^b}{\d x^a} = \lambda
  \delta_{ab},
\end{equation}
so that $\xi^a\frac{\d}{\d x^a}$ is a conformal Killing vector of
$\EE^D$. Since $\xi^{a}$ is independent of time, so is
$\lambda = \frac{2}{D} \frac{\d \xi^a}{\d x^a}$, which we can now use
to solve for $\xi^0$ in \eqref{eq:xiaxi0car}:
\begin{equation}
  \xi^0 = T(\x) + \frac{2 t}{N D} \frac{\d \xi^a}{\d x^a},
\end{equation}
for some ``supertranslations'' $T \in C^\infty(\EE^D)$. The carrollian
conformal symmetries vary with respect to the space dimension $D$.

Let $D \geq 3$. Thus we see that, as a vector space, the Lie algebra
$\c^{\zC}$ of carrollian conformal Killing vector fields of
$\hyperlink{S13}{\zC}$ is isomorphic to $\so(D+1,1) \oplus
C^\infty(\EE^D)$, where $\so(D+1,1)$ is the Lie algebra of conformal
Killing vectors on $\EE^D$ which we denote by $\xi_{X}$. In summary we
have the vector field
\begin{align}
  \xi=
  \xi_X
  +  \frac{2 t}{N D} \div \xi_{X}  \frac{\d}{\d t}
  + T(\x)  \frac{\d}{\d t} .
\end{align}
for $X \in \so(D+1,1)$ and $T \in C^\infty(\EE^D)$, and
$\div \xi = \frac{\d \xi^a}{\d x^a}$. The vector space isomorphism is
then given by
\begin{equation}
  X \mapsto \xi_X + \frac{2t}{N D} \div \xi_X \frac{\d}{\d t}
  \qquad\text{and}\qquad T \mapsto T \frac{\d}{\d t}.
\end{equation}

As Lie algebras, $\c^{\zC}$ is a semidirect product. Indeed,
\begin{equation}\label{eq:semidir}
  \left[ \xi_X + \frac{2t}{N D} \div \xi_X \frac{\d}{\d t}, T
    \frac{\d}{\d t} \right] =
  \left(\xi_X(T) - \frac2{N D}\div \xi_X T \right)\frac{\d}{\d t},
\end{equation}
so that $T$ does not actually transform as a function but as a section
of $\Lbdl^{\frac2N}$ where $\Lbdl$ is the density line bundle,
normalised so that the spatial metric $b$ is a section of
$S^2T^*M \otimes \Lbdl^2$.

It may help to spell this out. A conformal metric is a section of
$S^2T^*M \otimes \Lbdl^2$ and a conformal Killing vector field is one
which preserves the conformal metric. Now if $\zeta$ is a conformal
Killing vector field for $(M,g)$, then
\begin{equation}
  \L_\zeta g = \frac{2}{D} \div \zeta g \iff \left(\L_\zeta -
    \frac{2}{D} \div \zeta \right) g = 0.
\end{equation}
If we interpret this as the invariance of $g$ under the action of
$\zeta$ on sections of $S^2T^*M \otimes \Lbdl^2$, we see that the
action of $\xi_X$ on $T$, which is given in
equation~\eqref{eq:semidir} by
\begin{equation}
  T \mapsto \left(\L_{\xi_X} - \frac2{N D}\div \xi_X \right) T,
\end{equation}
says that $T$ is a section of $\Lbdl^{\frac2N}$, as claimed.  In
particular, if $N =2$, $T$ has conformal weight $1$ in agreement with
\cite{Ashtekar:2018lor}.

In summary, for $D\geq 3$, $\c^{\zC}$ is isomorphic to a split
extension
\begin{equation}
  \begin{tikzcd}
    0 \arrow[r] & \Gamma(\Lbdl^{\frac2N}) \arrow[r] & \c^{\zC}_{D\geq
      3} \arrow[r] & \so(D+1,1) \arrow[r] & 0,
  \end{tikzcd}
\end{equation}
a result first derived in \cite{Duval:2014uva}. We notice that
comparing to the Lie algebra of carrollian Killing vector fields in
equation~\eqref{eq:c-ses}, all that has happened is that the Lie
algebra $\e$ of euclidean isometries gets enhanced to the Lie algebra
$\so(D+1,1)$ of euclidean conformal symmetries, under which the
``supertranslations'' transform not as functions, but as sections of a
(trivial) line bundle with conformal weight $2/N$ (in conventions
where the metric scales with weight $2$). We did not see this when we
calculated the carrollian Killing vector fields because the Lie
algebra $\e$ does not contain the generator of dilatations and cannot
tell the weight.

Now let $D=2$.  In this case, as reviewed in
Appendix~\ref{sec:conf-kill-vect}, the Lie algebra of conformal Killing
vector fields on $\EE^2$ is enhanced to the Lie algebra
$\mathscr{O}(\CC)$ of entire functions on the complex plane with the
wronskian Lie bracket:
$[f,g] = f \d g - g \d f$.
Hence for $D=2$, $\c^{\zC}$ is isomorphic to a split extension
\begin{equation}
  \begin{tikzcd}
    0 \arrow[r] & \Gamma(\Lbdl^{\frac2N}) \arrow[r] & \c^{\zC}_{D=2} \arrow[r]
    & \mathscr{O}(\CC) \arrow[r] & 0.
  \end{tikzcd}
\end{equation}
The vector field is given explicitly by
\begin{equation}
    \xi=
  \xi_f
  +  \frac{t}{N} \div \xi_{f} \frac{\d}{\d t}
  + T(z)  \frac{\d}{\d t} \qquad\text{where}\qquad   \xi_{f} = f(z) \d + \overline{f(z)}\, \dbar .
\end{equation}

Finally, if $D=1$, every vector field on $\EE^1$ is conformal Killing
and hence now $\c^{\zC}$ is isomorphic to
\begin{equation}
  \begin{tikzcd}
    0 \arrow[r] & \Gamma(\Lbdl^{\frac2N}) \arrow[r] & \c^{\zC}_{D=1} \arrow[r]
    & C^\infty(\RR) \arrow[r] & 0,
  \end{tikzcd}
\end{equation}
where the vector field is given by
\begin{equation}
  \xi= \xi(x) \frac{\d}{\d x} + \frac{2t}{N} \xi'(x) \frac{\d}{\d t} + T(x)\frac{\d}{\d t} .
\end{equation}

The last two results were already obtained in Section IV of
\cite{Duval:2014lpa}, to which we refer for further information.

\subsection{Symmetries of the (anti) de~Sitter carrollian structure ($\zdSC$ and $\zAdSC$)}
\label{sec:symm-carr-struct-1}

We now investigate the symmetries of the (anti) de~Sitter carrollian
spacetimes ($\hyperlink{S14}{\zdSC}$ and $\hyperlink{S15}{\zAdSC}$)
with their carrollian structure. They can be embedded as null surfaces
of the (anti) de~Sitter spacetime. Unlike the carrollian space
$\hyperlink{S13}{\zC}$, the invariant connection on these spacetimes
is not flat. The carrollian structure becomes much more transparent if
we work in modified exponential coordinates, as described in
Appendix~\ref{app:modexp}. In order to be able to treat both cases at
once, let us introduce the functions
\begin{equation}
  C(r) :=
  \begin{cases}
    \cos(r) & \text{for $\hyperlink{S14}{\zdSC}$}\\
    \cosh(r) & \text{for $\hyperlink{S15}{\zAdSC}$}
  \end{cases}
  \qquad
  S(r) := C'(r)  \qquad\text{and}\qquad
  G(r) := \frac{S(r)}{C(r)},
\end{equation}
with the understanding that $r\in(0,\frac\pi2)$ for $\hyperlink{S14}{\zdSC}$  and $r>0$
for $\hyperlink{S15}{\zAdSC}$.  In those coordinates, the invariant
carrollian structures are given by
\begin{equation}
  \kappa = C(r)^{-1} \frac{\d}{\d t} \qquad\text{and}\qquad b = dr^2 +
  S(r)^2 g_{\SS^{D-1}}.
\end{equation}
The metric $b$ defines the round metric on the sphere $\SS^D$ for
$\hyperlink{S14}{\zdSC}$ and the hyperbolic metric on $\HH^D$ for
$\hyperlink{S15}{\zAdSC}$.  Although the coordinates only cover a
hemisphere of $\SS^D$, we proved in
\cite[§4.2.5]{Figueroa-OFarrill:2018ilb} that $\hyperlink{S14}{\zdSC}$
is diffeomorphic to $\RR \times \SS^D$ for $D \geq 2$ and to $\RR^2$
for $D =1$.

Now let $\xi = \xi^0 \frac{\d}{\d t} + \xi^a \frac{\d}{\d x^a}$ be a
carrollian Killing vector field, so that
$\L_\xi \kappa = 0$ and $\L_\xi b = 0$.  We calculate
\begin{equation}
  [\xi, \kappa] = -C(r)^{-1} \left( \left( \frac{\d \xi^0}{\d t} + \x \cdot \bxi
    \frac{G(r)}{r}\right) \frac{\d}{\d t} + \frac{\d \xi^a}{\d t}
  \frac{\d}{\d x^a}\right) \stackrel{?}{=} 0,
\end{equation}
which is solved by
\begin{equation}
  \xi^a = \xi^a(\x) \qquad\text{and}\qquad \xi^0 = T(\x) - t \x \cdot
  \bxi \frac{G(r)}{r},
\end{equation}
for some $t$ independent ``supertranslations'' $T(\x)$ and where we
have introduced the shorthand notation
$\x \cdot \bxi = \delta_{ab} x^a \xi^b$. Therefore,
\begin{equation}
  \xi = \left( T(\x) - \frac{G(r)}{r} t \x \cdot \bxi \right)
  \frac{\d}{\d t} + \xi^a(\x) \frac{\d}{\d x^a}.
\end{equation}
Now we impose $\L_\xi b = 0$.  We observe that this does not
constrain the $\frac{\d}{\d t}$ component of $\xi$, so it is only a
condition on $\xi^a(\x) \frac{\d}{\d x^a}$.  But in the submanifolds
of constant $t$, $b$ defines a metric and $\L_\xi b = 0$ says that
$\xi^a(\x) \frac{\d}{\d x^a}$ is a Killing vector.  Therefore, we have
\begin{equation}
  \xi = \left( T(\x)  - \frac{G(r)}{r} t \x \cdot \bxi_X \right)
  \frac{\d}{\d t} + \xi_X^a(\x) \frac{\d}{\d x^a} \qquad\text{for } X \in
  \begin{cases}
    \so(D+1), & \text{for $\hyperlink{S14}{\zdSC}$}\\
    \so(D,1), & \text{for $\hyperlink{S15}{\zAdSC}$}.
  \end{cases}
\end{equation}
In summary, the Lie algebra of carrollian Killing vector fields is
isomorphic to
\begin{equation}
  \a^{\zdSC} \cong \so(D+1) \ltimes
  C^\infty(\SS^D)\qquad\text{and}\qquad   \a^{\zAdSC} \cong \so(D,1)
  \ltimes C^\infty(\HH^D),
\end{equation}
where the action of $\so$ on $C^\infty$ is given by
\begin{equation}
  [X,T] = \xi_X T + \frac{G(r)}{r} \x \cdot \bxi_X T.
\end{equation}
If we define $T \mapsto \widehat{T} := -C(r) T$ then it follows that
\begin{equation}
  \label{eq:act}
  \widehat{[X, T]} = \xi_X \widehat{T},
\end{equation}
so the action of $\so$ on $C^\infty$ is just a ``dressed'' version of
the standard action of vector fields on functions.\footnote{%
  Alternatively, we may view this ``dressing''  as a change of
  coordinates to a new rescaled time $t' = -t g(r)$.}
In this way, we  may identify the finite-dimensional transitive kinematical
Lie algebras as the subalgebras
\begin{equation}
  \so(D+1) \ltimes C_{\leq 1}^\infty(\SS^D)\qquad\text{and}\qquad   \so(D,1)
  \ltimes C_{\leq 1}^\infty(\HH^D),
\end{equation}
respectively, where $C^\infty_{\leq 1}$ denotes the functions $T(\x)$
which are polynomial of degree $\leq 1$ in $\x$. Comparing with
Table~\ref{tab:spacetimes}, one can see that the $\so$ factors are the
span of $\J$ and $\P$, whereas $C^\infty_{\leq 1}$ are spanned by $H$
and $\B$, which do indeed commute.

Let us now consider the carrollian conformal Killing vector fields.
Let $\xi = \xi^0 \frac{\d}{\d t} + \xi^a \frac{\d}{\d x^a}$ satisfy
equation~\eqref{eq:c-ckv}.  The condition $\L_\xi \kappa =
[\xi,\kappa] = - \frac{\lambda}{N} \kappa$ is satisfied provided that
\begin{equation}
  \frac{\d \xi^a}{\d t} = 0 \qquad\text{and}\qquad \lambda = N
  \left(\frac{\d \xi^0}{\d t} + \frac{G(r)}{r} \x \cdot \bxi\right),
\end{equation}
where $\x \cdot \bxi := x^a \xi^a$.  The condition
$\L_\xi b = \lambda b$ says that $\xi^a\frac{\d}{\d x^a}$ is a
conformal Killing vector field of the metric $b$ with
$\lambda = \frac2D \nabla_a \xi^a$, with $\nabla$ the Levi-Civita
connection for $b$, which is the round metric on $\SS^D$ for
$\hyperlink{S14}{\zdSC}$, and the metric on hyperbolic space $\HH^D$
for $\hyperlink{S15}{\zAdSC}$.

Let $D\geq 3$. Both $\SS^D$ and $\HH^D$ are conformally flat, so their
Lie algebras of conformal Killing vector fields are isomorphic, and
indeed isomorphic to that of $\EE^D$: namely, $\so(D+1,1)$.

Solving for $\xi^0$ we find
\begin{equation}
  \xi^0 = T(\x) + t \left( \frac{2}{ND} \div \xi -
    \frac{G(r)}{r} \x \cdot \bxi \right),
\end{equation}
where $\div\xi := \nabla_a \xi^a$ and where $T$ is a smooth
function on $\SS^D$ or $\HH^D$ depending on whether we are in
$\hyperlink{S14}{\zdSC}$ or $\hyperlink{S15}{\zAdSC}$,
respectively. As vector spaces, the Lie algebras $\c^{\zdSC}$
(resp.\ $\c^{\zAdSC}$) of conformal symmetries of
$\hyperlink{S14}{\zdSC}$ (resp.\ $\hyperlink{S15}{\zAdSC}$) are
isomorphic to $C^\infty(\SS^D) \oplus \so(D+1,1)$ (resp.
$C^\infty(\HH^D) \oplus \so(D+1,1)$), with the isomorphism given by
\begin{equation}
  X \mapsto \xi_X + \left( \frac{2}{ND} \div \xi_X + \frac{G(r)}{r} \x
    \cdot \bxi_X \right) t \frac{\d}{\d t}
  \qquad\text{and}\qquad T \mapsto T \frac{\d}{\d t},
\end{equation}
for $X \in \so(D+1,1)$ and $T$ a smooth function in the relevant
space.

As Lie algebras, $\c^{\zdSC}$ and $\c^{\zAdSC}$ are again semidirect
products. Indeed, if $X \in \so(D+1,1)$ and $f \in C^\infty$, then we
find
\begin{equation}
  [X,T] = \xi_X(T) + \frac{G(r)}{r} \x \cdot \bxi_X T
  - \frac{2}{ND} \div \xi_X T.
\end{equation}
If we again define $T \mapsto \widehat{T} = -C(r) T$, then
\begin{equation}
  \widehat{[X,T]} = \xi_X(\widehat{T}) - \frac{2}{ND} \div \xi_X
\widehat{T},
\end{equation}
so that $\widehat{T}$ is a section of the line bundle
$\Lbdl^{\frac2N}$. In summary, just as in the case of the flat
carrollian spacetime $\hyperlink{S13}{\zC}$, we find that the Lie
algebras $\c^{\zdSC}$ and $\c^{\zAdSC}$ are split extensions
\begin{equation}
   \begin{tikzcd}
      0 \arrow[r] & \Gamma(\Lbdl^{\frac2N}) \arrow[r] & \c_{D\geq 3}^{\mathsf{(A)dSC}}
      \arrow[r] & \so(D+1,1) \arrow[r] & 0,
    \end{tikzcd}
\end{equation}
where $\Lbdl$ is the density bundle on $\SS^D$ or $\HH^D$ for
$\hyperlink{S14}{\zdSC}$ or $\hyperlink{S15}{\zAdSC}$, respectively.
So again we see that in going from the Lie algebras of symmetries to
the Lie algebras of conformal symmetries, all that happens is that the
isometries enhance to conformal symmetries and what earlier were
thought (after the ``dressing'') to be functions are actually sections
of $\Lbdl^{\frac2N}$.

Now let $D=2$. Here the situation differs. As reviewed in
Appendix~\ref{sec:conf-kill-vect}, the case of
$\hyperlink{S14}{\zdSC}$ is just as for $D\geq 3$, whereas for
$\hyperlink{S15}{\zAdSC}$, the Lie algebra of conformal Killing vector
fields on $\HH^2$ is enhanced to $\mathscr{O}(\HH)$, the holomorphic
functions on the upper half-plane with the wronskian Lie bracket
$[f,g] = f\d g - g \d f$. Therefore we have
\begin{equation}
   \begin{tikzcd}
      0 \arrow[r] & \Gamma(\Lbdl^{\frac2N}) \arrow[r] & \c_{D=2}^{\zdSC}
      \arrow[r] & \so(3,1) \arrow[r] & 0,
    \end{tikzcd}
\end{equation}
but
\begin{equation}
   \begin{tikzcd}
      0 \arrow[r] & \Gamma(\Lbdl^{\frac2N}) \arrow[r] & \c_{D=2}^{\zAdSC}
      \arrow[r] & \mathscr{O}(\HH) \arrow[r] & 0.
    \end{tikzcd}
\end{equation}

For $D=1$ again every vector field is conformal Killing and their Lie
algebra is isomorphic to the Lie algebra of smooth functions on the
real line or the circle with the wronskian Lie bracket:
\begin{equation}
   \begin{tikzcd}
      0 \arrow[r] & \Gamma(\Lbdl^{\frac2N}) \arrow[r] & \c_{D=1}^{\zdSC}
      \arrow[r] & C^\infty(S^1) \arrow[r] & 0,
    \end{tikzcd}
\end{equation}
but
\begin{equation}
   \begin{tikzcd}
      0 \arrow[r] & \Gamma(\Lbdl^{\frac2N}) \arrow[r] & \c_{D=1}^{\zAdSC}
      \arrow[r] & C^\infty(\RR) \arrow[r] & 0.
    \end{tikzcd}
\end{equation}

Let us restrict the discussion to $N=2$. Then the conformal symmetries
of the dS carrollian structure are (at least in $3+1$ dimension)
isomorphic to the BMS symmetries~\cite{Bondi:1962px,Sachs:1962zza}
(for a definition of the BMS algebra in higher dimension see, e.g.,
\cite{Barnich:2006av}). This could have been anticipated since the dS
carrollian structure is, up to a rescaling of time, the same as in
\cite{Duval:2014uva}. It should however not be forgotten that
$\hyperlink{S14}{\zdSC}$ is a null surface in de Sitter spacetime and
has nowhere vanishing curvature. For $D=2$, if we allow for conformal
Killing vector fields on the sphere which are not everywhere smooth,
then we may extend $\sl(2,\CC)$ to
``superrotations''~\cite{Barnich:2009se,Barnich:2010eb} (see also
\cite{Banks:2003vp}). For $D=1$, the superrotations are built in from
the start, which again is in agreement with the BMS group for $2+1$
``bulk'' dimensions~\cite{Ashtekar:1996cd,Barnich:2006av}.

Let us also observe that we find for the AdS carrollian spacetime in
$D=2$, a null surface of AdS in $3+1$ dimensions, an infinite
dimensional enhancement with ``superrotations'', in addition to the
supertranslations.

\subsection{Symmetries of the carrollian light cone ($\zLC$) }
\label{sec:symm-carr-struct-2}

These were already determined in \cite{Duval:2014uva}, but we present
it here for completeness. To determine the symmetries of the
carrollian structure of $\hyperlink{S16}{\zLC}$, it is convenient to
change coordinates.

Let $D\geq 2$. As shown in
\cite{Duval:2014uva,Figueroa-OFarrill:2018ilb},
$\hyperlink{S16}{\zLC}$ can be embedded as the future light cone in
$(D+2)$-dimensional Minkowski spacetime $\MM^{D+2}$ in such a way that
the carrollian structure is the one induced by the Minkowski metric on
that null hypersurface. We may parametrise the future light cone in
$\MM^{D+2}$ by $\x \in \RR^{D+1}\setminus\{0\}$ and the map
$i : \RR^{D+1}\setminus\{0\} \to \MM^{D+2}$ is given by
$i(\x) = (r,\x)$, where $r = \|\x\|>0$. The carrollian structure
$(\kappa,b)$ is given by $\kappa = r\frac{\d}{\d r}$ and $b = i^* g$,
where $g$ is the Minkowski metric:
\begin{equation}
  g = \eta_{\mu\nu} dX^\mu dX^\nu = -(dX^0)^2 + \sum_i (dX^i)^2, 
\end{equation}
where the $X^\mu$ are the affine coordinates on $\MM^{D+2}$.  On the future
light cone, $X^0 = r$ and $X^i = x^i$.  Therefore, we see that
\begin{equation}
  b = i^*g = - dr^2 + (dr^2 + r^2 g_{\SS^D}) = r^2 g_{\SS^D}.
\end{equation}
In terms of the coordinates $\x$, we have that $\kappa = x^a
\frac{\d}{\d x^a}$ and
\begin{equation}
  b = \left(\delta_{ab} - \frac{x^ax^b}{r^2}\right) dx^a
  dx^b.
\end{equation}

Now let $\xi = \xi^a \frac{\d}{\d x^a}$ be a symmetry of the
carrollian structure $(\kappa,b)$. Then
$\L_{\xi}\kappa = [\xi,\kappa] = 0$ and $\L_\xi b = 0$. We find it
more convenient to write
\begin{equation}
  \xi =\xi^r \frac{\d}{\d r} + \zeta,
\end{equation}
where $\xi^r \in C^\infty(\RR^{D+1}\setminus\{0\})$ and $\zeta$ is a possibly
$r$-dependent vector field tangent to the spheres of constant $r$;
that is, $\zeta r = 0$.  The condition $[\kappa,\xi] = 0$ results in
\begin{equation}
  0 \stackrel{!}{=}[\kappa,\xi] = \left[r \frac{\d}{\d r}, \xi^r
  \frac{\d}{\d r} + \zeta\right] = \left(r \frac{\d \xi^r}{\d r} -
    \xi^r\right) \frac{\d}{\d r} + r \frac{\d \zeta}{\d r}.
\end{equation}
This implies that $\xi^r = r F$, where $F \in C^\infty(\SS^D)$, so
that $\frac{\d F}{\d r} = 0$, and $\zeta$ is independent of $r$. The
condition $\L_\xi b = 0$ results in
\begin{equation}
  0 \stackrel{!}{=} \L_\xi (r^2 g_{\SS^D}) = 2 r^2 F g_{\SS^D} + r^2
  \L _{\zeta} g_{\SS^D},
\end{equation}
so that $\zeta$ is a conformal Killing vector on $\SS^D$ and
$F = -\frac{1}{D} \div \zeta$, where $\div \zeta$ is the intrinsic
divergence of $\zeta$ on the sphere relative to the round metric, but
which agrees with $\frac{\d \zeta^a}{\d x^a}$ in this case.
Therefore, the symmetry algebra of the carrollian structure on
$\hyperlink{S16}{\zLC}$ is isomorphic to $\so(D+1,1)$, even for $D=2$
as shown in Appendix~\ref{sec:conf-kill-vect}, which is the transitive
kinematical Lie algebra. It is an intriguing result that among the
homogeneous carrollian spacetimes, it is precisely the non-reductive
one whose symmetry algebra is finite-dimensional.

For $D=1$, $\zLC$ is the universal cover of the future light cone in
three-dimensional Minkowski spacetime. One can model $\zLC$ as the
submanifold of $\RR^3$ with points
\begin{equation}
  \zLC  = \left\{ (r\cos\theta, r\sin\theta, \theta) ~\middle |~
    r>0,~\theta\in\RR\right\},
\end{equation}
with the covering map from $\zLC$ to the future light cone in $\MM^3$
given by
$(r\cos\theta, r\sin\theta,\theta) \mapsto (r, r\cos\theta,
r\sin\theta)$. Notice that the non-contractible circles of constant
$r$ in the light cone lift to contractible helices in $\zLC$. The
transitive kinematical Lie algebra is isomorphic to $\sl(2,\RR)$ and
is spanned by the vector fields
\begin{equation}
  \frac{\d}{\d\theta},\qquad \cos\theta\frac{\d}{\d\theta} + r \sin\theta
  \frac{\d}{\d r}\qquad\text{and}\qquad
  \sin\theta\frac{\d}{\d\theta} - r \cos\theta \frac{\d}{\d r}.
\end{equation}
Since they are periodic in $\theta$ with period $2\pi$, they descend
to tangent vector fields to the future light cone. The carrollian
structure is given by $\kappa = r\frac{\d}{\d r}$ and
$b = r^2 d\theta^2$, except that $\theta$ is not angular in $\zLC$.
It is straightforward to work out the Lie algebra of carrollian
Killing vector fields and obtain that it is isomorphic to
$C^\infty(\RR_\theta)$ with the wronskian Lie bracket.  Indeed, if
$f \in C^\infty(\RR_\theta)$, the corresponding vector field is
\begin{equation}
  \xi_f = f(\theta) \frac{\d}{\d \theta} - f'(\theta) r \frac{\d}{\d r}
\end{equation}
and the Lie bracket is given by
\begin{equation}
  [\xi_f, \xi_g] = \xi_h \qquad\text{with}\qquad h = fg'-f'g.
\end{equation}
For the (non-simply connected) future light cone, we must consider
periodic functions, so that the Lie algebra of carrollian Killing
vector fields is $C^\infty(S^1)$ with the wronskian Lie bracket.

Let us now consider the carrollian conformal Killing vector fields.
Again we first consider $D\geq 2$.  This was treated already
in~\cite{Duval:2014uva}, but we write it here for completeness.  As
before we work in embedding coordinates where the carrollian structure
on $\hyperlink{S16}{\zLC}$ is given by
\begin{equation}
  \kappa = r \frac{\d}{\d r} \qquad\text{and}\qquad b = r^2
  g_{\SS^D}
\end{equation}
and let $\xi = \xi^r \frac{\d}{\d r} + \zeta$, with $\zeta r = 0$,
satisfy equation~\eqref{eq:c-ckv}. The condition
$\L_\xi \kappa = -\frac{\lambda}{N} \kappa$ results in
\begin{equation}
  \frac{\d \zeta}{\d r} = 0 \qquad\text{and}\qquad r \frac{\d
    \xi^r}{\d r} - \xi^r = \frac{\lambda}{N} r,
\end{equation}
whereas the condition $\L_\xi b = \lambda b$ results in
\begin{equation}
  \L_\zeta g_{\SS^D} = \left(\lambda - \frac{2 \xi^r}{r}\right) g_{\SS^D},
\end{equation}
so that $\zeta$ is a conformal Killing vector field on $\SS^D$ with
divergence
\begin{equation}
  \div \zeta = \frac{D}{2} \left(\lambda - \frac{2 \xi^r}{r}\right).
\end{equation}
Solving for $\xi^r$ we find
\begin{equation}
  \xi^r = r \left(r^{\frac{N}2} T - \frac{1}{D}\div \zeta\right),
\end{equation}
for some $T \in C^\infty(\SS^D)$. Therefore, as a vector space, the
Lie algebra $\c^{\zLC}$ of carrollian conformal Killing vector fields of
$\hyperlink{S16}{\zLC}$ is isomorphic to
$C^\infty(\SS^D) \oplus \so(D+1,1)$, with the isomorphism given by
\begin{equation}
  X \mapsto \zeta_X - \frac{1}{D}\div \zeta_X r \frac{\d}{\d r}
  \qquad\text{and}\qquad T \mapsto r^{\frac2N} T r \frac{\d}{\d r},
\end{equation}
for $X \in \so(D+1,1)$ and $T \in C^\infty(\SS^D)$.

As a Lie algebra, $\c^{\zLC}$ is a semi-direct product with
\begin{equation}
  [X,T] = \zeta_X(T) - \frac{2}{ND} \div \zeta_X T,
\end{equation}
so that $T$ is actually a section of $\Lbdl^{\frac2N}$. In summary,
$\c^{\zLC}$ is a split extension
\begin{equation}
   \begin{tikzcd}
      0 \arrow[r] & \Gamma(\Lbdl^{\frac2N}) \arrow[r] & \c^{\zLC}
      \arrow[r] & \so(D+1,1) \arrow[r] & 0
    \end{tikzcd},
\end{equation}
which shows that there is an isomorphism $\c^{\zLC} \cong
\c^{\zdSC}$.

For $D=1$, analogous to the case of carrollian Killing vector fields,
we find that now the Lie algebra of carrollian conformal Killing
vector fields is larger.  The carrollian conformal Killing vector
fields at level $N$ are given by
\begin{equation}
  f(\theta) \frac{\d}{\d \theta} - f'(\theta) r \frac{\d}{\d r} +
  r^{\frac2N} g(\theta) r \frac{\d}{\d r},
\end{equation}
for some $f,g \in C^\infty(\RR_\theta)$. The Lie algebra structure is
now a semidirect product of the wronskian Lie algebra
$C^\infty(\RR_\theta)$ of carrollian Killing vector fields and the
abelian ideal of sections of $\Lbdl^{\frac2N}$:
\begin{equation}
   \begin{tikzcd}
      0 \arrow[r] & \Gamma(\Lbdl^{\frac2N}) \arrow[r] & \c_{D=1}^{\zLC}
      \arrow[r] & C^\infty(\RR_\theta) \arrow[r] & 0
    \end{tikzcd},
\end{equation}
where under the isomorphism
$\Lbdl^{\frac2N} \cong C^\infty(\RR_\theta)$, to a function
$g \in C^\infty(\RR_\theta)$ there corresponds the vector field
$\zeta_g = r^{\frac2N} g(\theta) r \frac{\d}{\d r}$, so that with
$\xi_f = f(\theta) \frac{\d}{\d\theta} - f'(\theta) r \frac{\d}{\d
  r}$, we have
\begin{equation}
  [\xi_f, \zeta_g] = \zeta_h \qquad\text{with}\qquad h = f g' -
  \tfrac2N f' g.
\end{equation}

\subsection{Symmetries of galilean structures}
\label{sec:symm-galil-spac}

In this section, we will work out the Lie algebra of galilean Killing
vector fields for the homogeneous galilean spacetimes.  This Lie
algebra has been termed the \emph{Coriolis algebra} of a
galilean spacetime in \cite{Duval:1993pe}.  In the modified exponential
coordinates of Appendix~\ref{app:modexp}, the invariant galilean
structure takes the same form in all the homogeneous spacetimes
$\hyperlink{S7}{\zG}$, $\hyperlink{S8}{\zdSG}$,
$\hyperlink{S10}{\zAdSG}$, $\hyperlink{S9}{\ztdSG_\gamma}$,
$\hyperlink{S11}{\ztAdSG_\chi}$ and
$\hyperlink{S12}{\text{\twodgal}_{\gamma,\chi}}$: the clock one-form
is given by $\tau = dt$ and the inverse spatial metric by
$h = \delta^{ab} \frac{\d}{\d x^a} \otimes \frac{\d}{\d x^b}$.

Let $\xi = \xi^0 \frac{\d}{\d t} + \xi^a \frac{\d}{\d x^a}$ satisfy
equation~\eqref{eq:g-kv}.  The condition that $\xi$ preserves the
clock-one form says
\begin{equation}
  0 \stackrel{!}{=} \L_\xi \tau = \L_\xi dt = d \L_\xi t = d \xi^0
  \implies \xi^0~\text{is constant.}
\end{equation}
The condition that $\L_\xi h = 0$ says that
\begin{equation}\label{eq:Lxh=0}
  0 \stackrel{!}{=} \L_\xi h = -  (\d_a \xi_b + \d_b \xi_a)  \frac{\d}{\d x^a} \otimes
  \frac{\d}{\d x^b}  \implies \d_a \xi_b + \d_b \xi_a = 0.
\end{equation}
This equation says that $\xi^a\d_a$ is a (possibly) $t$-dependent
Killing vector field of the $D$-dimensional euclidean space $\EE^D$,
so that
\begin{equation}
  \xi^a(x,t) = f^a(t) + \Lambda^a{}_b(t) x^b,
\end{equation}
where $\Lambda_{ab} = - \Lambda_{ba}$.  In other words,
\begin{equation}
  \xi = \xi^0 \frac{\d}{\d t} + f^a(t) \frac{\d}{\d x^a} +
    \Lambda(t)^a{}_b x^b \frac{\d}{\d x^a},
\end{equation}
so that, as a vector space, the Lie algebra $\a$ of vector fields
which preserve the galilean structure $(\tau,h)$, is isomorphic to
$\a \cong \RR \oplus C^\infty(\RR_t, \e)$, with $\e$ the euclidean Lie
algebra and $\RR_t$ the real line with coordinate $t$. As a Lie
algebra,
\begin{equation}
  \a \cong \RR \ltimes C^\infty(\RR_t, \e)
\end{equation}
has the structure of a semidirect product or,
equivalently, a split extension
\begin{equation}
  \label{eq:a-gal}
  \begin{tikzcd}
    0 \arrow[r] &  C^\infty(\RR_t, \e) \arrow[r] &  \a \arrow[r] &  \RR \arrow[r] &  0,
  \end{tikzcd}
\end{equation}
where the splitting $\RR \to \a$ is given by sending $1\in\RR$ to
$\frac{\d}{\d t}$, corresponding to the action of $H$.  This was
originally worked out in \cite{Duval:1993pe}, who
named it the Coriolis algebra.

We will now determine the Lie algebra $\c$ of conformal symmetries of
the galilean structure and we will see that it has a very similar
structure to $\a$ in equation~\eqref{eq:a-gal}, except that $\RR$ gets
enhanced to a non-abelian Lie algebra structure on $C^\infty(\RR_t)$.

Let $\xi = \xi^0 \frac{\d}{\d t} + \xi^a \frac{\d}{\d x^a}$ satisfy
equation~\eqref{eq:g-ckv}. The condition
$\L_\xi \tau = -\frac{\lambda}{N}\tau$ results in
\begin{equation}
  \frac{\d \xi^0}{\d x^a} = 0 \qquad\text{and}\qquad \frac{\d
    \xi^0}{\d t} = -\frac{\lambda}{N} \implies \lambda = \lambda(t).
\end{equation}
The condition $\L_\xi h = \lambda h$ results in
\begin{equation}
  \frac{\d \xi^a}{\d x^b} +  \frac{\d \xi^b}{\d x^a} = - \lambda \delta_{ab},
\end{equation}
so that $\xi^a \frac{\d}{\d x^a}$ is a (possibly) $t$-dependent
conformal Killing vector field on $\EE^D$, but since
$\lambda = \lambda(t)$, we see that that $\xi^a \frac{\d}{\d x^a}$ is
either Killing or homothetic.  In other words, we can write
\begin{equation}
  \xi^a = f^a(t) + \Lambda^a_b(t) x^b + g'(t) x^a,
\end{equation}
where we have found it convenient to think of the homothetic component
as the derivative of a smooth function $g \in C^\infty(\RR_t)$. Doing
so, we may solve for $\xi^0$ to arrive at
\begin{equation}
  \xi^0 =-\frac{2}{N D} g(t),
\end{equation}
so that
\begin{equation}
  \xi = \left(f^a(t) + \Lambda^a_b(t) x^b \right) \frac{\d}{\d x^a} +
  \left( -\frac{2}{N D} g(t) \frac{\d}{\d t} + g'(t) x^a \frac{\d}{\d
      x^a}\right),
\end{equation}
in agreement with \cite[eq.\ (3.12)]{Duval:2009vt} and \cite[eq.\
(III.5)]{Duval:2014lpa}, who worked out the case of
$\hyperlink{S7}{\zG}$.

Thus we see that, as a vector space, the Lie algebra $\c$ of conformal
symmetries of the galilean spacetime is isomorphic to
$C^\infty(\RR_t, \e) \oplus C^\infty(\RR_t)$, with the isomorphism
such that $g \in C^\infty(\RR_t)$ is sent to the vector field
\begin{equation}
  g(t) \mapsto g'(t) x^a \frac{\d}{\d x^a} - \frac{2}{N D} g(t) \frac{\d}{\d t}.
\end{equation}
In particular, the Lie algebra structure on $C^\infty(\RR_t)$ is not
abelian, but rather if $f,g \in C^\infty(\RR_t)$, their Lie bracket is
a multiple of the wronskian:
\begin{equation}
  [f,g] = \frac{-2}{N D} (f g' - f' g).
\end{equation}

As a Lie algebra, $\c$ is a semidirect product, where $f \in
C^\infty(\RR_t)$ acts on $(\v(t),\Lambda(t)) \in C^\infty(\RR_t, \e)$
by
\begin{equation}
  [f,(\v,\Lambda)] = \left( \tfrac{-2}{ND} f \v' + f' \v, \tfrac{-2}{ND}
  f \Lambda'\right).
\end{equation}
In summary, the Lie algebra $\c$ is a split extension
\begin{equation}
  \begin{tikzcd}
    0 \arrow[r] &  C^\infty(\RR_t, \e) \arrow[r] &  \c \arrow[r] &  C^\infty(\RR_t) \arrow[r] &  0,
  \end{tikzcd}
\end{equation}
so that in going from the symmetries to the conformal symmetries, the
abelian Lie algebra $\RR$ has been enhanced to the non-abelian
``wronskian'' Lie algebra $C^\infty(\RR_t)$.

It is intriguing that the galilean spacetimes, despite admitting
non-isomorphic transitive kinematical Lie algebras, have isomorphic
conformal symmetry Lie algebras. It would be interesting to
investigate how the transitive Lie algebras relate via their
embeddings in $\c$.

\section{Conclusions}
\label{sec:conclusions}

The main results of this and our previous paper
\cite{Figueroa-OFarrill:2018ilb} are
\begin{enumerate}
\item the classification of simply-connected spatially isotropic
  homogeneous spacetimes, recorded in Tables~\ref{tab:spacetimes} and
  \ref{tab:aristotelian};
\item the proof that the boosts act with generic non-compact orbits on
  all spacetimes in Table~\ref{tab:spacetimes} except for the
  riemannian symmetric spaces, and
\item the determination of the Lie algebra of infinitesimal
  (conformal) symmetries of these structures.
\end{enumerate}
The second point is an important physical requirement, already
mentioned in \cite{Bacry:1968zf}. We also discussed the subtle
interplay between the kinematical Lie algebras and their
spacetimes~\cite{Figueroa-OFarrill:2018ilb}. Among them is the
intriguing connection between the anti de Sitter carrollian and
Minkowski spacetime, which are different homogeneous spacetimes, but
based on the same Lie algebra.

In addition, we also determined the invariant affine connections on
these homogeneous spacetimes and calculated their torsion and
curvature.  These connections allow us to define geodesics, which we
hope to study in future work.

Table~\ref{tab:spacetimes-props} summarises the basic geometric
properties of the spacetimes.  This table makes it clear that the bulk
of the spacetimes do not admit an invariant metric and hence
that there is a very rich landscape beyond lorentzian geometry, even
if we remain within the realm of homogeneous spaces with space
isotropy.

Another aspect of this work was the analysis of the, generically
infinite dimensional, (conformal) symmetries of the carrollian and
galilean structures. One observation is that the Lie algebra of
infinitesimal conformal symmetries of carrollian (anti) de Sitter
spacetime, which embeds as a null hypersurface of (anti) de Sitter
spacetime, is infinite dimensional and reminiscent of the BMS
algebra. It is tempting to speculate that this might be relevant for
BMS physics (memory effect,
$\ldots$)~\cite{Strominger:2017zoo,Ashtekar:2018lor} on these non-flat
backgrounds (see also \cite{Morand:2018tke}).

Some of the above results were made possible by the introduction of
local coordinates.  We chose to consider exponential coordinates;
although admittedly these are not always the simplest coordinates for
calculations.  We have found modified exponential coordinates to be
quite useful as well, particularly for the determination of the
infinitesimal (conformal) symmetries of the spacetimes.  We expressed
the kinematical vector fields -- that is, the infinitesimal generators
of rotations, boosts and translations -- in terms of exponential
coordinates, and we did the same for the invariant structures (if
any).  This was particularly useful in order to determine their
infinitesimal (conformal) symmetries.

There are a number of possible directions for future research
departing from our results.

One open problem we did not address is to exhibit the galilean
spacetimes as null reductions of lorentzian spacetimes in one 
higher dimension.  This would complement the description of the
carrollian spacetimes as null hypersurfaces in an ambient lorentzian
manifold.

We showed that all of the galilean spacetimes in this paper
($\hyperlink{S7}{\zG}$, $\hyperlink{S8}{\zdSG}$,
$\hyperlink{S10}{\zAdSG}$, $\hyperlink{S9}{\ztdSG_\gamma}$,
$\hyperlink{S11}{\ztAdSG_\chi}$ and
$\hyperlink{S12}{\text{\twodgal}_{\gamma,\chi}}$) have isomorphic Lie
algebras of infinitesimal conformal symmetries.  We did not determine
how the transitive kinematical Lie algebras are embedded in these
infinite-dimensional Lie algebras.  Perhaps studying those embeddings
might teach us something about how the kinematical Lie algebras relate
to each other.

It would be interesting to promote the homogeneous spacetimes to
Cartan geometries and hence study the possible theories based on them.
For a discussion in $2+1$ dimensions see \cite{Matulich:2019cdo}.

Another intriguing direction is to explore the applications of these
geometries to non-AdS holography. It is not inconceivable that some of
these homogeneous geometries might play a similar rôle in non-AdS
holography to that played by anti~de Sitter spacetime in the AdS/CFT
correspondence~\cite{Maldacena:1997re}. One particularly interesting
property of a non-zero cosmological constant is that acts as an
infrared regulator (often paraphrased as ``AdS is like a box'') and it
would be interesting to investigate if this persists in the
non-relativistic or ultra-relativistic limits to
$\hyperlink{S10}{\zAdSG}$ or $\hyperlink{S15}{\zAdSC}$, respectively.

\section*{Acknowledgments}
\label{sec:acknowledgments}

During the embryonic stages of this work, JMF and SP were
participating at the MITP Topical Workshop ``Applied Newton--Cartan
Geometry'' (APPNC2018), held at the Mainz Institute for Theoretical
Physics, to whom we are grateful for their support, their hospitality
and for providing such a stimulating research atmosphere. We are
particularly grateful to Eric Bergshoeff and Niels Obers for the
invitation to participate. We are grateful to Yvonne Calò for checking
some calculations in the paper. JMF would like to acknowledge helpful
conversations with Jelle Hartong, James Lucietti and Michael Singer.
SP is grateful to Glenn Barnich, Carlo Heissenberg, Marc Henneaux,
Yegor Korovin, Javier Matulich, Arash Ranjbar, Jan Rosseel, Romain
Ruzziconi and Jakob Salzer for useful discussions.

The research of JMF is partially supported by the grant ST/L000458/1
``Particle Theory at the Higgs Centre'' from the UK Science and
Technology Facilities Council. The research of SP is partially
supported by the ERC Advanced Grant ``High-Spin-Grav'' and by
FNRS-Belgium (convention FRFC PDR T.1025.14 and convention IISN
4.4503.15). SP acknowledges support from the Erwin Schrödinger
Institute during his stay at the ``Higher Spins and Holography''
workshop.

SP wants to dedicate this work to his ``kleine Oma'' Amelie Prohazka.

\appendix

\section{Modified exponential coordinates}
\label{app:modexp}

In this appendix we revisit the local geometry of the homogeneous
carrollian and galilean spacetimes, but this time in modified
exponential coordinates.

\subsection{Carrollian spacetimes}
\label{sec:carr-spac}

\subsubsection{Carrollian (anti) de Sitter spacetimes}
\label{sec:carollian-anti-de}

Let $\sigma'(t,\x) = \exp(t H) \exp(\x \cdot \P) \cdot o$. We
calculate the soldering form by pulling back the left-invariant
Maurer--Cartan one-form $\vartheta$ on the Lie group:
\begin{equation}
  \sigma'^*(\vartheta) = \theta + \omega = \exp(-\ad_A) H dt +
  D(\ad_A) (d\x \cdot \P),
\end{equation}
where $A = \x \cdot \P$.  We find
\begin{equation}
  \ad_A H = \varepsilon \x \cdot \B \qquad\text{and}\qquad \ad^2_A H =
  -\varepsilon x^2 H,
\end{equation}
so that
\begin{equation}
  \exp(-\ad_A) H = \cosh(x_+) H - \frac{\sinh(x_+)}{x_+} \varepsilon
  \x \cdot \B,
\end{equation}
where $x_+^2 = - \varepsilon x^2$ and $x_- = - x_+$.  Also, we find
\begin{equation}
  \begin{split}
    \ad_A P_a &= \varepsilon J_{ab} x^b \\
    \ad^2_A P_a &= -\varepsilon x^2 P_a + \varepsilon x^a A \\    
    \implies \ad^3_A P_a &= -\varepsilon x^2 \ad_A P_a.
  \end{split}
\end{equation}
Therefore,
\begin{equation}
  D(\ad_A) P_a = P_a + D^- \varepsilon x^b J_{ab} +
  \frac{1}{x_+^2}(D^+ - 1)(-\varepsilon x^2 P_a + \varepsilon x^a \x
  \cdot \P),
\end{equation}
where $D^- = \frac{1}{2x_+}(D(x_+) - D(x_-))$ and $D^+ = \frac12
(D(x_+) + D(x_-))$.  In summary,
\begin{equation}
  \theta = \cosh(x_+) dt H + D^+ d\x\cdot \P +
  \frac{\varepsilon}{x_+^2} (D^+ -1 ) \x \cdot d\x\, \x \cdot \P.
\end{equation}
Using that $D^+ = \frac{sinh(x_+)}{x_+}$, we find
\begin{equation}
  \theta = \cosh(x_+) dt H + \frac{\sinh(x_+)}{x_+} d\x \cdot \P +
  \varepsilon \frac{\sinh(x_+)-x_+}{x_+^3} \x\cdot d\x \, \x \cdot \P.
\end{equation}
The carrollian structure is given by $\kappa = E_H =
\sech(x_+)\frac{\d}{\d t}$ and $b = \pi^2(\theta,\theta)$, which
expands to
\begin{equation}
  b = -\frac{\varepsilon}{x^2} \sinh^2(x_+) d\x \cdot d \x +
  \frac{\varepsilon}{x^4}(\sinh^2(x_+) - x_+^2) (\x \cdot d\x)^2.
\end{equation}
If $\varepsilon=1$, $x_+ = i r$, where $r = |\x|$ and hence
$\sinh^2 x_+ = - \sin^2 r$, so that
\begin{equation}
  b = \frac{\sin^2 r}{r^2} \left(dr^2 + r^2 g_{\SS^{D-1}}\right) -
  \frac{(\sin^2 r - r^2)}{r^2} dr^2 = dr^2 + \sin^2 r\, g_{\SS^{D-1}},
\end{equation}
which is the round metric on $\SS^D$.  The coordinate system is good
provided that $r \in (0,\frac\pi2)$.  On the other hand, if
$\varepsilon = -1$, $x_+ = r$ and, therefore,
\begin{equation}
  b = \frac{\sinh^2 r}{r^2} \left(dr^2 + r^2 g_{\SS^{D-1}}\right) -
  \frac{(\sinh^2 r - r^2)}{r^2} dr^2 = dr^2 + \sinh^2 r\, g_{\SS^{D-1}},
\end{equation}
which is the metric on hyperbolic space $\HH^D$ and the coordinate
system is good for all $r>0$. In summary, the carrollian structures in
these coordinate systems are given by
\begin{equation}
  \begin{aligned}[m]
    \kappa^{\zdSC} &= \sec(r) \frac{\d}{\d t}\\
    b^{\zdSC} &= dr^2 + \sin^2 r\, g_{\SS^{D-1}}
  \end{aligned}
  \qquad\text{and}\qquad
  \begin{aligned}[m]
    \kappa^{\zAdSC} &= \sech(r) \frac{\d}{\d t}\\
    b^{\zAdSC} &= dr^2 + \sinh^2 r\, g_{\SS^{D-1}}.
  \end{aligned}
\end{equation}

\subsection{Galilean spacetimes}
\label{sec:galilean-spacetimes}

The transitive kinematical Lie algebras for the homogeneous galilean
spacetimes (with the exception of
\hyperlink{S12}{$\text{\twodgal}_{\gamma,\chi}$}, which will be
treated separately below) has additional brackets of the form
\begin{equation}
  [H,\B] = - \P \qquad\text{and}\qquad [H,\P] = \alpha \B + \beta \P~,
\end{equation}
for some $\alpha,\beta \in \RR$.  In other words, $\ad_H$ is
represented by a matrix of the form $\begin{pmatrix}  0 &  \alpha \\
  -1 & \beta\end{pmatrix}$.  We define
\begin{equation}
  M(t) = \exp \left( t
    \begin{pmatrix}
      0 &  \alpha \\
      -1 & \beta
    \end{pmatrix}
\right).
\end{equation}

We introduce modified exponential coordinates $(t,\x)$ by acting with
$L(t,\x) := \exp(t H) \exp(\x \cdot \P)$ on the origin $o$. Relative
to them $\xi_H = \frac{\d}{\d t}$ and $\xi_{J_{ab}}$ are as in
exponential coordinates. We will determine $\xi_{B_a}$ and then
calculate $\xi_{P_a} = [\xi_H,\xi_{B_a}]$.

Let $s \in (-\varepsilon, \varepsilon)$ and consider
\begin{equation}
  \exp(s \v \cdot \B) L(t,\x)\cdot o = L(\tau(s),\y(s)) \cdot o~,
\end{equation}
where $\tau(0) = t$ and $\y(0) = \x$.  This is equivalent to
\begin{equation}
  \exp(s \v \cdot \B) L(t,\x) = L(\tau(s) , \y(s) ) \exp( \w(s) \cdot \B),
\end{equation}
where $\w(0) = 0$, which we may re-write yet again as
\begin{equation}
  \exp(\tau(s) H) \exp(\y(s) \cdot \P)  = \exp(s \v \cdot \B) L(t,\x)
  \exp(- \w(s) \cdot \B).
\end{equation}
We now differentiate with respect to $s$ at $s=0$ to obtain (in the
notation of matrix groups)
\begin{equation}
(\tau'(0) H)
  L(t,\x) + L(t,\x) (\y'(0)\cdot \P) = (\v \cdot \B) L(t,\x) - L(t,\x)
  (\w'(0)\cdot \B),
\end{equation}
which implies that $\tau'(0) = 0$ and
\begin{equation}
  \y'(0) \cdot \P + \w'(0)\cdot \B = M(-t) \v \cdot \B
  \qquad\text{or}\qquad
  \begin{pmatrix}
    \w'(0)\\\y'(0)
  \end{pmatrix} = M(-t)
  \begin{pmatrix}
    \v \\ \bzero
  \end{pmatrix}.
\end{equation}
We now proceed to treat the different galilean spacetimes in turn, but
first we simply comment on the fact that the galilean structure is
formally identical in all cases. Indeed,
\begin{equation}
  L(t,\x)^{-1}dL(t,\x) = H dt + (\beta dt x^a + d x^a) P_a + \alpha dt
  x^a B_a,
\end{equation}
where $[H,\P] = \alpha \B + \beta \P$ defines $\alpha$ and $\beta$. It
follows from this that the soldering form is given by
\begin{equation}
  \theta^H = dt \qquad\text{and}\qquad \theta^{P_a} = dx^a + \beta x^a dt ,
\end{equation}
the invariant canonical connection by
\begin{equation}
  \omega = \alpha dt\, \x \cdot \B 
\end{equation}
and the vielbein is
\begin{equation}
  E_H = \frac{\d}{\d t} - \beta x^a \frac{\d}{\d x^a}
  \qquad\text{and}\qquad E_{P_a} = \frac{\d}{\d x^a}.
\end{equation}
The galilean structure is given by the clock one-form
\begin{equation}
  \eta(\theta) = dt
\end{equation}
and the inverse spatial metric
\begin{equation}
  \delta^{ab} E_{P_a} \otimes E_{P_b} = \delta^{ab} \frac{\d}{\d x^a}
  \otimes \frac{\d}{\d x^b}.
\end{equation}
The torsion and curvature are, respectively
\begin{equation}
  \Theta=- \beta dt \wedge d\x \cdot \P
\qquad\text{and}\qquad
  \Omega= - \alpha dt \wedge  d\x \cdot \B .
\end{equation}

We now work out the expressions of the fundamental vector fields
$\xi_{B_a}$ and $\xi_{P_a}$ in each case.

\subsubsection{Galilean spacetime}
\label{sec:G-modexp}

For the galilean spacetime $\hyperlink{S7}{\zG}$,
\begin{equation}
  M(t) = \exp \left(t \begin{pmatrix} 0 & 0 \\ -1 & 0 \end{pmatrix} \right) =
  \begin{pmatrix}
    1 & 0 \\ -t & 1
  \end{pmatrix},
\end{equation}
and hence
\begin{equation}
    \begin{pmatrix}
    \w'(0)\\\y'(0)
  \end{pmatrix} =
  \begin{pmatrix}
    \v \\ t \v
  \end{pmatrix},
\end{equation}
from where we read off
\begin{equation}
  \xi_{B_a} = t \frac{\d}{\d x^a} \qquad\text{and hence}\qquad
  \xi_{P_a} = \frac{\d}{\d x^a}.
\end{equation}

\subsubsection{Galilean de~Sitter spacetime}
\label{sec:dSG-modexp}

For the galilean de~Sitter spacetime $\hyperlink{S8}{\zdSG}$,
\begin{equation}
  M(t) = \begin{pmatrix}
    \cosh t & -\sinh t \\ -\sinh t & \cosh t
  \end{pmatrix},
\end{equation}
and hence
\begin{equation}
    \begin{pmatrix}
    \w'(0)\\\y'(0)
  \end{pmatrix} =
  \begin{pmatrix}
    \v \cosh t\\ \v\sinh t
  \end{pmatrix},
\end{equation}
from where we read off
\begin{equation}
  \xi_{B_a} = \sinh t \frac{\d}{\d x^a} \qquad\text{and hence}\qquad
  \xi_{P_a} = \cosh t \frac{\d}{\d x^a}.
\end{equation}

\subsubsection{Torsional galilean de~Sitter spacetime}
\label{sec:tdSG-modexp}

For the torsional galilean de~Sitter spacetime $\hyperlink{S9}{\ztdSG_\gamma}$,
\begin{equation}
  M(t) = \frac1{1-\gamma} \begin{pmatrix}
    e^{t\gamma} - \gamma e^t & \gamma(e^t - e^{t\gamma})\\
    e^{t\gamma} - e^t & e^t - \gamma e^{t\gamma}\\
  \end{pmatrix},
\end{equation}
and hence
\begin{equation}
    \begin{pmatrix}
    \w'(0)\\\y'(0)
  \end{pmatrix} =
  \frac1{1-\gamma}
  \begin{pmatrix}
    (e^{t\gamma} - \gamma e^t) \v \\ (e^{t\gamma} - e^t) \v
  \end{pmatrix},
\end{equation}
from where we read off
\begin{equation}
  \xi_{B_a} = \frac{e^{t\gamma} - e^t}{1-\gamma} \frac{\d}{\d x^a} \qquad\text{and hence}\qquad
  \xi_{P_a} =  \frac{\gamma e^{t\gamma} - e^t}{1-\gamma}  \frac{\d}{\d x^a}.
\end{equation}

\subsubsection{Galilean anti de~Sitter spacetime}
\label{sec:adSG-modexp}

For the galilean anti de~Sitter spacetime $\hyperlink{S10}{\zAdSG}$,
\begin{equation}
  M(t) = \begin{pmatrix}
    \cos t & \sin t \\ -\sin t & \cos t
  \end{pmatrix},
\end{equation}
and hence
\begin{equation}
    \begin{pmatrix}
    \w'(0)\\\y'(0)
  \end{pmatrix} =
  \begin{pmatrix}
    \v \cos t\\ \v\sin t
  \end{pmatrix},
\end{equation}
from where we read off
\begin{equation}
  \xi_{B_a} = \sin t \frac{\d}{\d x^a} \qquad\text{and hence}\qquad
  \xi_{P_a} = \cos t \frac{\d}{\d x^a}.
\end{equation}

\subsubsection{Torsional galilean anti de~Sitter spacetime}
\label{sec:tAdSG-modexp}

For the torsional galilean anti de~Sitter spacetime $\hyperlink{S11}{\ztAdSG_\chi}$,
\begin{equation}
  M(t) = \begin{pmatrix}
    e^{t\chi} (\cos t - \chi \sin t) & e^{t\chi}(1 + \chi^2)\sin t\\
    - e^{t\chi}(1+\chi^2)\sin t & e^{t\chi} (\cos t + \chi \sin t)\\
  \end{pmatrix},
\end{equation}
and hence
\begin{equation}
    \begin{pmatrix}
    \w'(0)\\\y'(0)
  \end{pmatrix} =
  \begin{pmatrix}
    e^{-t\chi} (\cos t + \chi \sin t) \v \\ e^{-t\chi} \sin t\, \v
  \end{pmatrix},
\end{equation}
from where we read off
\begin{equation}
  \xi_{B_a} = e^{-t\chi}\sin t \frac{\d}{\d x^a} \qquad\text{and hence}\qquad
  \xi_{P_a} =  e^{-t\chi} (\cos t - \chi \sin t)  \frac{\d}{\d x^a}.
\end{equation}

\subsubsection{Spacetime $\text{\twodgal}_{\gamma,\chi}$}
\label{sec:twodgal}

For spacetime $\hyperlink{S12}{\text{\twodgal}_{\gamma,\chi}}$, the
expression for the fundamental vector fields $\xi_{B_a}$ and
$\xi_{P_a}$ are not particularly transparent in modified exponential
coordinates, so we will not give them here. We will show, however,
that the galilean structure is formally identical to that of all the
other homogeneous galilean spacetimes.

The transitive Lie algebra in this case is defined by the following
brackets
\begin{equation}
  [H,B_a] = - P_a \qquad\text{and}\qquad [H,P_a] = (1+\gamma) P_a +
  \gamma B_a - \chi \epsilon_{ab} (B_b + P_b).
\end{equation}
Letting $L(t,\x) = \exp(t H) \exp(\x \cdot \H)$, we find
\begin{equation}
  L(t,\x)^{-1} d L(t,\x) = H dt + (1+\gamma) x^a P_a - \chi
  \epsilon_{ab} x^a P_b + \gamma x^a B_a - \chi \epsilon_{ab} x^a B_b
  + dx^a P_a,
\end{equation}
so that the soldering form has components
\begin{equation}
  \theta^H = dt \qquad\text{and}\qquad \theta^{P_a} = dx^a + f^a(x) dt,
\end{equation}
where $f^a(x) := (1+\gamma) x^a + \chi \epsilon_{ab} x^b$.  The
vielbein has components
\begin{equation}
  E_H = \frac{\d}{\d t} - f^a(x) \frac{\d}{\d x^a}
  \qquad\text{and}\qquad
  E_{P_a} = \frac{\d}{\d x^a}.
\end{equation}
Therefore, the invariant galilean structure has clock one-form
\begin{equation}
  \eta(\theta) = dt
\end{equation}
and inverse spatial metric
\begin{equation}
  \delta^{ab} E_{P_a} \otimes E_{P_b} = \delta^{ab} \frac{\d}{\d x^a}
  \otimes \frac{\d}{\d x^b}.
\end{equation}

\section{Conformal Killing vectors in low dimension}
\label{sec:conf-kill-vect}

In this appendix we collect some results concerning the conformal
Killing vectors of euclidean space $\EE^D$, round sphere $\SS^D$ and
hyperbolic space $\HH^D$ for $D \leq 2$. We have used these results in
determining the infinitesimal (conformal) symmetries of the carrollian
spacetimes.

For $D= 1$, every smooth vector field is conformal Killing. For
example, the ``metric'' on $\EE^1$ is given by $g = dx^2$ relative to
the global coordinate $x$. Since the tangent bundle is trivial, we may
identify smooth vector fields with smooth functions globally, so
$\xi = f(x)\frac{d}{d x}$ for some $f \in C^\infty(\RR)$. Then we see
that $\L_\xi g = 2 df dx = 2 f' g$. Similar considerations apply to
$\SS^1$ and $\HH^1$, with conformal Killing vector fields being in
bijective correspondence with the smooth functions $C^\infty(\SS^1)$
and $C^\infty(\RR)$, respectively.

In all cases, the Lie algebra of conformal Killing vector fields is
isomorphic to the Lie algebra of smooth functions under the wronskian
Lie bracket:
\begin{equation}
  [f,g] =f g' - g f'.
\end{equation}

Things are more interesting for $D=2$. Let us first consider euclidean
space with metric $g = dx^2 + dy^2$ relative to global coordinates
$(x,y)$. Every vector field is of the form
$\xi = u(x,y) \frac{\d}{\d x} + v(x,y) \frac{\d}{\d y}$ for
$u,v \in C^\infty(\RR^2)$. Then the conformal Killing condition
\begin{equation}
  \L_\xi g = 2 \frac{\d u}{\d x} dx^2 + 2 \frac{\d u}{\d y} dy^2 + 2
  \left(\frac{\d u}{\d y} + \frac{\d v}{\d x}\right) dx dy \stackrel{!}{=}
  \lambda \left(dx^2 + dy^2\right)
\end{equation}
is equivalent to
\begin{equation}
  \frac{\d u}{\d y} + \frac{\d v}{\d x} = 0 \qquad\text{and}\qquad \frac{\d
    u}{\d x} = \frac{\d v}{\d y} = \frac{\lambda}2.
\end{equation}
This says that $u$ and $v$ obey the Cauchy--Riemann equations and,
since they are smooth, that $w = u(x,y) + i v(x,y)$ is a holomorphic
function $f(z)$, say, of $z = x + i y$. In other words, every
conformal Killing vector field on $\EE^2$ is given by
\begin{equation}
  \xi = f(z) \d + \overline{f(z)}\, \dbar
\end{equation}
for some entire function $f : \CC \to \CC$. The Lie algebra of
conformal Killing vector fields on $\EE^2$ is therefore isomorphic to
the Lie algebra $\mathscr{O}(\CC)$ of entire functions relative to the
``wronskian'' Lie bracket:
\begin{equation}
  [f,g] = f \d g - g \d f.
\end{equation}

The round sphere $\SS^2$ is the one-point compactification of $\EE^2$.
A conformal Killing vector field on $\SS^2$ takes the form
$\xi = f(z)\d + \overline{f(z)}\, \dbar$ away from the North pole,
say. But demanding that $f(z) \d$ extends to a holomorphic vector
field at the North Pole, says that if $\zeta = 1/z$, then
$-\zeta^2 f(1/\zeta)$ should be holomorphic at $\zeta = 0$ and this
requires $f(z) = a_0 + a_1 z + a_2 z^2$, for some
$a_0,a_1,a_2 \in \CC$. This is the well-known result that the
(everywhere smooth) conformal Killing vector fields on $\SS^2$ define
a real Lie algebra isomorphic to $\sl(2,\CC) \cong \so(3,1)$. Indeed,
the wronskian Lie bracket of the polynomials of degree $\leq 2$ is
given by
\begin{equation}
  [1,z] = 1, \qquad [1,z^2] = 2 z \qquad\text{and}\qquad [z,z^2] =
  z^2.
\end{equation}

Finally, let us consider hyperbolic space $\HH^2$, which we model as
the upper half-plane $\{(x,y) \in \RR^2 ~\mid~ y>0\}$ with metric
\begin{equation}
  g = \frac{dx^2 + dy^2}{y^2}.
\end{equation}
The tangent bundle is trivial so that we can write any smooth vector
field as $\xi = u(x,y) \frac{\d}{\d x} + v(x,y) \frac{\d}{\d y}$ for
some $u,v \in C^\infty(\RR^2)$.  The conformal Killing condition
\begin{equation}
  \L_\xi g = \frac2{y^2} \left( \frac{\d u}{\d y} + \frac{\d v}{\d x}
    \right) dx dy + \frac2{y^2}\left( \frac{\d u}{\d x} -
      \frac{v}{y}\right) dx^2+ \frac2{y^2} \left( \frac{\d v}{\d y} -
      \frac{v}{y}\right) dy^2 \stackrel{!}{=}
    \lambda \frac{dx^2 + dy^2}{y^2}
\end{equation}
results in
\begin{equation}
  \frac{\d u}{\d y} + \frac{\d v}{\d x} = 0 \qquad\text{and}\qquad \frac{\d
    u}{\d x} = \frac{\d v}{\d y} = \frac{\lambda}2 + \frac{v}{y}.
\end{equation}
In particular, $u,v$ satisfy the Cauchy--Riemann equations and hence
again $w = u + i v = f(z)$, where $f$ is a holomorphic function of
$z = x + i y$ in the upper half-plane. The Schwarz reflection
principle says that if $f$ extends continuously to $y=0$ then it
extends to an entire function on the whole complex plane such that
$f(z) = \overline{f(\zbar)}$ for $z$ in the lower half-plane. But of
course $f$ may develop singularities as $y \to 0$ and hence there are
more holomorphic functions on the upper half-plane than can be
obtained by restricting entire functions.

In summary, the Lie algebra of conformal Killing vector fields is
isomorphic to the Lie algebra $\mathscr{O}(\HH)$ of holomorphic
functions on the upper half-plane relative to the ``wronskian'' Lie
bracket $[f,g] = f \d g - g \d f$.

\providecommand{\href}[2]{#2}\begingroup\raggedright\endgroup


\end{document}